\documentclass[useAMS,usenatbib,usegraphicx]{mn2e}

\usepackage{times}

\usepackage{amsmath}

\usepackage{color}
\usepackage{soul}
\usepackage{hyperref}

\usepackage[letterpaper,margin=0.8in]{geometry}
\title[Significance of cross-correlations for uneven sampling]{A method for the estimation of the significance of cross-correlations in unevenly sampled red-noise time series}

\author[W. Max-Moerbeck et al.]{
W.~Max-Moerbeck$^{1,2}$\thanks{E-mail: wmax@nrao.edu},  
J.~L.~Richards$^{3}$,
T.~Hovatta$^{1,4}$,
V.~Pavlidou$^{1,5,6}$,
T.~J.~Pearson$^{1}$,\newauthor
A.~C.~S.~Readhead$^{1}$\\
$^{1}$Cahill Center for Astronomy and Astrophysics, California Institute of Technology, Pasadena, CA 91125, USA\\
$^{2}$National Radio Astronomy Observatory (NRAO), P.O. Box 0, Socorro, NM 87801, USA\\
$^{3}$Department of Physics, Purdue University, West Lafayette, IN 47907, USA\\
$^{4}$Aalto University Mets\"ahovi Radio Observatory, Mets\"ahovintie 114, 02540 Kylm\"al\"a, Finland\\
$^{5}$Max-Planck-Institut f\"ur Radioastronomie, Auf dem H\"ugel 69, 53121 Bonn, Germany\\
$^{6}$Department of Physics, University of Crete / Foundation for Research and Technology - Hellas, Heraklion 71003, Greece
}

\begin{document}

\date{Accepted 2014 August 19. Received 2014 August 18; in original form 2014 June 20}

\pagerange{\pageref{firstpage}--\pageref{lastpage}} \pubyear{0000}

\maketitle

\label{firstpage}

\begin{abstract}
We present a practical implementation of a Monte Carlo method to estimate the significance of cross-correlations in unevenly sampled time series of data, whose statistical properties are modeled with a simple power-law power spectral density. This implementation builds on published methods, we introduce a number of improvements in the normalization of the cross-correlation function estimate and a bootstrap method for estimating the significance of the cross-correlations. A closely related matter is the estimation of a model for the light curves, which is critical for the significance estimates. We present a graphical and quantitative demonstration that uses simulations to show how common it is to get high cross-correlations for unrelated light curves with steep power spectral densities. This demonstration highlights the dangers of interpreting them as signs of a physical connection. We show that by using interpolation and the Hanning sampling window function we are able to reduce the effects of red-noise leakage and to recover steep simple power-law power spectral densities. We also introduce the use of a Neyman construction for the estimation of the errors in the power-law index of the power spectral density. This method provides a consistent way to estimate the significance of cross-correlations in unevenly sampled time series of data.
\end{abstract}


%--------------------------------------------------------------------------------------------------------------------------------
% Introduction
%--------------------------------------------------------------------------------------------------------------------------------
\section{Introduction}

Studies of the variability in astronomical sources can reveal aspects that are not accessible to imaging, which is limited by the angular resolution of current instruments. For example, variability can be used to set limits on the sizes of the emitting regions through causality arguments \citep[e.g.,][]{abdo+2011}, to determine the size of the broad line region in active galactic nuclei \citep[e.g.,][]{peterson+1988}, or to detect extrasolar planets \citep[e.g.,][]{charbonneau+2000} among many other applications. In this paper we describe the practical implementation of a cross-correlation technique to determine the location of the gamma-ray emission site in blazars, by studying the relation between the variability in the radio and gamma-ray bands. For this purpose we are carrying out a blazar monitoring program with the Owens Valley Radio Observatory (OVRO) 40 meter telescope \citep[][]{richards+2011} and the Large Area Telescope (LAT) on board of the \emph{Fermi Gamma-ray Space Telescope} \citep[\emph{Fermi},][]{atwood+2009}. Our approach is to search for correlated variability between these two energy bands, which would enable us to determine the location of the gamma-ray emission regions relative to the radio emission regions. The study of cross-correlations between two energy bands presents a number of challenges from the data analysis and statistical point of view: among these are uneven sampling, non-equal error bars, and short time duration of the light curves. The techniques we develop here should be useful for other applications.  

Related methods have been presented in the literature, for example the study of cross-correlations with unevenly sampled light curves has an extensive literature about its application to reverberation mapping \citep[e.g.,][]{peterson1993}. These methods present a detailed treatment of the estimation of cross-correlations and time lags, but not of the estimation of significance of the observed correlations, a critical aspect for the interpretation of cross-correlation results.  The literature abounds with claims of statistically significant correlations that are not backed up by rigorous statistical analyses.

This paper presents a detailed discussion of the methods used for our investigation of time-correlation between radio and gamma-ray activity in blazars, which is discussed in \citet[][]{max-moerbeck+2014}. Here we present a description of the Monte Carlo method used to estimate the significance of cross-correlations between unevenly sampled time series using the method of \citet[][]{edelson_1988}. In order to estimate the distribution of cross-correlations in two uncorrelated data streams we need a model for the light curves. A commonly used model for time variability in blazars and other AGNs is a simple power-law power spectral density (PSD $\propto 1/\nu^\beta$), as has been measured for a small number of sources at various wavelengths \citep[e.g.,][]{hufnagel+1992, edelson+1995, uttley+2003, arevalo+2008, chatterjee_2008, abdo_variability_2010}. The results presented in \citet[][]{abdo_variability_2010} are of particular interest for the OVRO blazar monitoring program. In their paper, they find a value of $\beta_{\gamma} = 1.4 \pm 0.1$ for bright BL Lacs and  $\beta_{\gamma} = 1.7 \pm 0.3$ for bright FSRQs in the gamma-ray band. In the radio band a number of publications have measured $\beta_{\rm radio}$. It has been found that $\beta_{\rm radio} = 2.3 \pm 0.5$ for 3C279 at 14.5 GHz \citep[][]{chatterjee_2008} using a fit to the PSD for an 11 year light curve. Additional indirect estimates for the PSD power-law index are obtained by \citet[][]{hufnagel+1992} using structure function fits. For five sources, they obtain values of $\alpha = 0.4 \pm 0.2$ to $1.5 \pm 0.1$, where $\alpha$ is the exponent on the structure function ${\rm SF}(\tau) \propto \tau^{\alpha}$. The same method is used for 51 sources by \citet[][]{hughes+1992} who found that most values of $\alpha$ lie between 0.6 and 1.8, while a couple are closer to 0. However, the often assumed relation between the exponents of the PSD and the structure function ($\beta  = \alpha + 1$) is only valid under special conditions, not necessarily found in real data sets \citep[][]{paltani1999, emmanoulopoulos+2010}. The structure function has been widely used in blazar variability studies but its interpretation is not straightforward, as has been recently discussed by \citet[][]{emmanoulopoulos+2010}. These authors used simulations to demonstrate that many of the features in the structure function are associated with the length and sampling patterns of the light curves rather than anything of statistical significance. For these reasons, values obtained from the structure function can only be taken as a rough measure of the properties of the time series, and therefore we do not use them here. Instead we fit the PSDs directly.

We start by giving a brief description of the data sets used (Section \ref{observations}), and then provide detailed descriptions of the methods in Sections \ref{psd_estimation_method} and \ref{cross_corr_method}. In Section \ref{psd_estimation_method} we describe our approach to the critical problem of estimating a model for the light curves to use with the Monte Carlo significance estimate. Here we describe an implementation of the method of \citet[][]{uttley+2002} that contains some important modifications. In Section \ref{cross_corr_method} we provide a description of a number of modifications we propose to common methods used to estimate the significance of cross-correlations. We give a justification for the use of the local normalization \citep[][]{welsh_1999} in the \citet[][]{edelson_1988} method, demonstrate the strong dependence of the significance estimate on the model light curves and introduce a bootstrap method to estimate the error in the cross-correlation significance estimates. We close the paper with a summary of our main findings and recommendations for the use of this and related techniques (Section \ref{conclusions}).

An important aspect of this work is the use of a statistically well-defined data set, where long light curves are used independent of the flaring state of the object. A fatal trap that many authors fall into is that of "cherry picking" the data by selecting small intervals of data. This approach can produce spurious levels of significance for the cross-correlations, and hence cannot be used to draw conclusions about blazar populations, or the long term behaviour of individual sources.

%--------------------------------------------------------------------------------------------------------------------------------
% Observations
%--------------------------------------------------------------------------------------------------------------------------------
\section{The parameters of the observations}
\label{observations}

The methods we discuss in this paper can be adapted to use with any data set, but since we are describing a particular implementation, our simulated data sets are generated making some choices related to the intended application. These choices are motivated by the data sets associated with our blazar monitoring program in the radio and gamma-ray bands. These data sets are described in detail in \citet[][]{max-moerbeck+2014} and here we only summarize their main properties.

A radio observation for each one of the monitored blazars is attempted twice per week, but because of the effects of weather and other technical problems we obtain unevenly sampled light curves \citep[][]{richards+2011}. \emph{Fermi} observes the whole sky once every three hours \citep[][]{atwood+2009}, but because of the highly varying nature of blazars in gamma-rays, sources may sometimes fall below that detection threshold, resulting in upper limits for a given integration period. We consider gamma-ray light curves with a time binning of one week, which allows us to detect about 100 sources most of the time. We have upper limits for about 30\% of the data. At this level we find that treating the upper limits as non-observations does not have important effects in the measured time lags or significance of cross-correlations for the cases with interesting values of the cross-correlation significance \citep[][]{max-moerbeck+2014}. This behaviour could be a result of the particular properties of the light curves we considered in that study, such as long time scales for the variability compared to the gaps created by ignoring the upper limits or the power spectral densities, and it might not hold in other situations. We thus obtain unevenly sampled gamma-ray light curves as the ones we discuss here.

%--------------------------------------------------------------------------------------------------------------------------------
% Power spectral density fit
%--------------------------------------------------------------------------------------------------------------------------------
\section{Power spectral density estimation for unevenly sampled time series of short duration} 
\label{psd_estimation_method}

We begin with a brief summary of the standard methods used for the estimation of the PSD and then move to the uneven sampling and short time series cases. This discussion is based on the method presented in \citet[][]{uttley+2002} which is modified to suit our dataset and the range of PSDs we fit. Additional justification of the need for binning and interpolation of the light curves is given in Section \ref{rebin_interp_reasons}. We also present an example of the application of our method to a simulated light curve, and a number of tests using real data sampling for simulated light curves that demonstrate the accuracy of the fitting procedure under different conditions. The real data sampling is based on the data set presented in \citet[][]{max-moerbeck+2014}, which have 4 year 15 GHz radio light curves from the OVRO 40 meter telescope blazar monitoring program, and 3 year gamma-ray light curves from the LAT on board of \emph{Fermi}. A study of the effect of increasing the number of simulations in the fitting procedure is performed to guide our choice of parameters for the data analysis. A summary of the method, with emphasis on the improvements we add to the original formulation is given in Section \ref{conclusions}.

\subsection{The basics of power spectral density estimation}

We define a time series as a time ordered sequence of triplets $(t_i, f_i, e_i)$, where $t_i$ is the observation time, $f_i$ is the measured value of the quantity of interest (e.g., flux density, photon flux, etc.), and $e_i$ is an estimate of the observational error associated with the measurement. We assume that the time series is sorted in time and $i=1, ..., N$. \footnote{In what follows we use $\nu$ for the frequency and $f_i$ for time series data, e.g., flux density, photon flux, etc.}

An estimation of the PSD can be obtained through the periodogram, which is conventionally defined as the squared modulus of the discrete Fourier transform:
\begin{equation}
P(\nu_k) = \left [\sum_{i=1}^{N} f_i \cos(2 \pi \nu_k t_i) \right]^2 + \left [\sum_{i=1}^{N} f_i \sin(2 \pi \nu_k t_i) \right]^2
\end{equation}
where the periodogram is evaluated at the discrete set of frequencies $\nu_k = k / T$ for $k = 1, ..., N / 2$ for $N$ even, or $k = 1, ..., (N - 1) / 2$ for $N$ odd, $\nu_{\rm Nyq} = N / 2T$ is the Nyquist frequency and $T = N (t_N - t_1) / (N - 1)$, see footnote\footnote{This choice of $T$ is consistent with the definition of the discrete Fourier transform \citep[][]{brigham1988} and allows us to make use of the Fast Fourier Transform algorithm to increase the speed of the computations.}.

Estimating the PSD in this way requires sampling a continuous time series at discrete times for a finite amount of time. The sampling operation is equivalent to multiplication of the time series by a Dirac comb, while sampling for a finite time corresponds to a multiplication by a rectangular observing window. These two multiplications appear as convolutions in frequency space: the original spectrum is convolved with the Fourier transform of the Dirac comb and of the rectangular window. As a final step we only look at a discrete set of frequencies which is equivalent to multiplication by a Dirac comb in frequency space.\footnote{A graphical representation of these operations can help the reader understand their effect. See Figure 6.1 in \citet[][]{brigham1988} or elsewhere.}

Ignoring the effect of sampling with a Dirac comb in the frequency domain, and omitting normalization factors, we find that the periodogram is given by
\begin{equation}
P(\nu) = |W(\nu) * {\rm III}_{\frac{1}{\Delta t}}(\nu) * F(\nu)|^2,
\end{equation}
where $F(\nu)$ is the Fourier transform of the time series $(t_i, f_i)$,  ${\rm III}_{\frac{1}{\Delta t}}(\nu)$ is the Fourier transform of the Dirac comb with sampling interval $\Delta t$, and $W(\nu)$ is the Fourier transform of the sampling window function, which is by default a rectangular window, and $*$ denotes convolution.

As a result of the convolution with the Dirac comb, we do not have access to the original spectrum but a modified version that repeats periodically. Another distortion comes from the convolution with the sampling window function, which modifies the shape of the original spectrum, and finally we only look at discrete set of frequencies. All these factors have to be taken into account when analysing data and interpreting the results. The periodic repetition of the spectrum gives rise to aliasing, in which high frequency components are mistaken as low frequency components. Convolution with a window function can be a serious problem when the sidelobes of the frequency window function lie on regions of the spectrum where the power is much higher than at the frequency of interest -- this is the origin of the red-noise leakage problem. Having the spectrum sampled at a number of discrete frequencies can be problematic if we are searching for narrow spectral components which can be smeared or missed.

For the case of evenly sampled time series, PSD estimation amounts to using the discrete Fourier transform (DFT) along with periodogram or frequency averaging to decrease the noise which is distributed as a $\chi^2_2$ for a single frequency component. Each of these averaging processes can reduce the variance at the price of reduced spectral resolution. For example, in the case of frequency or periodogram averaging of $M$ components the resulting distribution is  $\chi^2_{2 M}$, which reduces the variance by a factor of $1 / M$ with respect to the non-averaging case.

The application of these methods is straightforward in the case of long time series, where a good estimate of the PSD can be obtained at the expense of reduced frequency resolution. Nonetheless, problems of aliasing and red-noise leakage can still complicate the analysis of broadband signals like the simple power-law PSDs we fit to our data ($P(\nu) \propto 1/\nu^\beta$), for the reasons outlined below. For relatively flat spectra ($\beta$ from 0 to 2) aliasing can be a problem as high frequency power above the Nyquist frequency contaminates low frequencies. This problem is less serious for steep spectra ($\beta \ge 2$), that have relatively small amounts of power at high frequencies. But in this case red-noise leakage can flatten the high frequency part of the spectrum: power from low frequencies contaminates the low amplitude high frequency parts of the spectrum through sidelobes on the sampling window functions. To reduce the effects of these problems a combination of filters and sampling window functions can be used \citep[e.g.,][]{brigham1988, press+1992, shumway+2011}.

\subsection{Power spectral density estimation for unevenly sampled data and short time series}

When working with time series data, problems often arise because the time series is unevenly sampled and relatively short. Uneven sampling requires the use of a different estimate of the periodogram: the best known alternatives are the Deeming periodogram \citep[][]{deeming1975} and the Lomb-Scargle periodogram \citep[][]{scargle1982}. The Lomb-Scargle periodogram is well suited to the detection of periodic signals in white noise, because its statistical properties are well understood. For the analysis of broadband signals, the Deeming periodogram is often used for reasons that are mainly historical as it does not present any real advantages. These two methods allow us to obtain an estimate of the periodogram for unevenly sampled time series directly, but do not provide a way to correct for distortions produced by the sampling window functions, which can modify the shape of the periodogram significantly as explained below.

\subsubsection{Description of the method \label{psd_method_description}}

The method we present here was originally developed and described in detail in \cite{uttley+2002}. We describe the main steps here to highlight the differences between theirs and our implementation.

\begin{enumerate}
\item
Obtain the periodogram for the light curve and bin it in frequency to reduce scatter. The periodogram is given by a frequency binned version of the following expression
\begin{equation}
\label{periodogram_equation}
P(\nu_k) = \frac{2 T}{N^2} \left( \left [\sum_{i=1}^{N} f_i \cos(2 \pi \nu_k t_i) \right]^2 + \left [\sum_{i=1}^{N} f_i \sin(2 \pi \nu_k t_i) \right]^2 \right )
\end{equation}
where the frequencies are $\nu_k = k / T$ for $k = 1, ..., N/2$ for $N$ even, or $k = 1, ..., (N - 1)/2$ for $N$ uneven. The minimum frequency is $\nu_{\rm min} = 1 / T$, the maximum frequency is the Nyquist frequency $\nu_{\rm Nyq} = (N / 2)(1 /  T)$, and $T = N (t_N - t_1) / (N - 1)$. The multiplicative factor is a normalization, such that the integral from $\nu_i$ to $\nu_f$ is equal to the variance contributed to the light curve in this frequency range. The evenly sampled time series $(t_i, f_i)$ is obtained from the original one by interpolation onto a regular grid. This interpolated time series is first multiplied by an appropriate sampling window in order to reduce red-noise leakage. A justification of these steps is given in Sections \ref{rebin_interp_reasons} and \ref{spectral_window}.

\item
Choose a PSD model to test against the data. In this case we are fitting power-laws of the form $P(\nu) \propto 1/\nu^\beta$ but this can be generalized to any functional dependence. For the given model simulate $M$ time series, where $M$ is a large number that allows us to represent a variety of possible realizations of this PSD model. These and all the simulated light curves used in this work are generated using the method described in \citet{timmer+1995}, which randomizes both the amplitude and phase of the Fourier transform coefficients consistent with the statistical properties of the periodogram.

\item
For each simulated light curve apply the same sampling, add observational noise, and interpolate into the same even grid. Calculate the periodogram for each one. From these $M$ periodograms determine the mean periodogram and its associated error as the scatter at each frequency bin.

\item
Using the mean periodogram and errors obtained in step (iii) construct a $\chi^2$-like test, defined by 
\begin{equation}
\label{chi2_definition}
\chi^2_{\rm obs} = \sum_{\nu = \nu_{\rm min}}^{\nu_{\rm max}} \frac{[\overline{P_{\rm sim}}(\nu) - P_{\rm obs}(\nu)]^2}{\Delta \overline{P_{\rm sim}}(\nu)^2}
\end{equation}
where $P_{\rm obs}(\nu)$ is the periodogram of the observed light curve, $\overline{P_{\rm sim}}(\nu)$ and $\Delta \overline{P_{\rm sim}}(\nu)$ are the mean and scatter in the periodogram obtained from the simulated light curves, and $\chi^2_{\rm obs}$ is the $\chi^{2}$ of the observed light curve when compared to the simulations for a given PSD model. This $\chi^2_{\rm obs}$ is then compared to the simulated distribution of $\chi^2$ for which we can obtain $M$ samples, $\chi^2_{{\rm sim},\,i}$, by replacing the $P_{\rm obs}(\nu)$ term by the periodogram of the simulated light curves, $P_{{\rm sim},\,i}(\nu)$, in Equation \ref{chi2_definition}. The fraction of the distribution for which $\chi^2_{{\rm sim},\,i} > \chi^2_{\rm obs}$ is the significance level at which the tested PSD model can be rejected, also known as the $p$-value. Thus a high value of this percentage represents a good fit, while a low one corresponds to a poor fit.
\end{enumerate}

The process described above can be repeated for a number of models with different parameters. The final step consists of selecting the best model fit as the one with the highest value of $p$. As with any statistical procedure, a measurement of the uncertainty in the parameters of the model needs to be given. In this point we depart from the original formulation and provide uncertainties based on Monte Carlo simulations of the model fitting process (as described in Section \ref{uncertainty_estimate}).

The most significant differences with the original implementation are the use of sampling window functions to reduce red-noise leakage and the Monte Carlo estimation of fitting uncertainties. Another difference is that we simulate the effects of aliasing by simulating light curves with high frequency components with a sampling period of 1 day, instead of adding a constant noise term to the PSD of the simulated light curves as in the original formulation. The high frequency cut at 1 day$^{-1}$ is justified in our implementation by the small amount of power seen at higher frequencies specially in the radio band. Intraday variable sources show variability in short time scales \citep[][]{wagner+1995}, but even in these cases the amplitude of the variability is only a few percent for most sources \citep[][]{quirrenbach+1992}. At gamma-rays this is not necessarily true as fast variability has been observed, but given that gamma-ray photon fluxes correspond to mean values of long integrations of at least a week for most blazars, the effects of fast variability are less important as they are averaged out. Other applications where fast variability is expected might require a higher frequency cut, making this approach impractical. Another important difference, although less important conceptually, is the use of the Fast Fourier Transform to perform the computations, which substantially decreases computing time. Further discussions of the most important elements of the method are given below.

\subsubsection{The necessity for rebinning and interpolation of the light curves\label{rebin_interp_reasons}}

\begin{figure*}
\begin{center}
\includegraphics[width=12cm, trim=20 10 20 20]{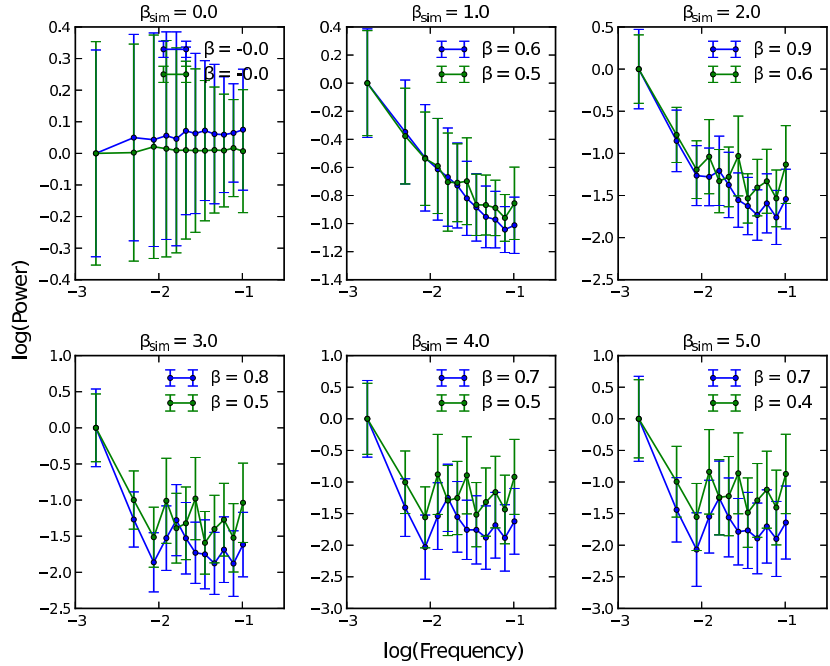}
\caption{Effect of the use of window functions for \textbf{uneven sampling} cases using the rectangular (blue) and Hanning window (green). Each figure shows the result of simulating 1000 light curves with a given simple power-law PSD $\propto 1/\nu^{\beta}$, with $\beta$ given in each figure title. The data points are the mean PSD and the error is the standard deviation in the simulation, while the units of power (vertical axis) and frequency (horizontal axis) are arbitrary. Also included are direct fits of the slopes of the mean PSDs for the simulated data for each window (individual panels legend). Notice how the linear fits can hardly discriminate between different slopes and how all the estimated PSDs look very similar.}
\label{window_uneven}
\end{center}
\end{figure*}

This step is very important when estimating steep PSDs. It is easy to get mislead by intuition developed from the behavior of window functions for evenly sampled time series, but it turns out that window functions for unevenly sampled data do not behave in the same way. An example is presented in Figure \ref{window_uneven}, where we show the frequency response of a uneven sampling pattern with rectangular and Hanning windows, for the periodogram of power-law PSDs with different values of $\beta$ from 0 to 5. These window functions are 
\begin{equation}
w_{\rm rect}(t)=
\begin{cases}
1, & 0 \leq t \leq T \\
0, & {\rm otherwise}
\end{cases}
\label{rectangular_window}
\end{equation}

\begin{equation}
w_{\rm Hanning}(t)=
\begin{cases}
\cos (\pi \frac{(t - T / 2)}{T})^2, & 0 \leq t \leq T \\
0, & {\rm otherwise}
\end{cases}
\label{hanning_window}
\end{equation}

% Window function for uneven and even sampling cases 

\begin{figure*}
\begin{center}
\includegraphics[width=12cm, trim=20 10 20 20]{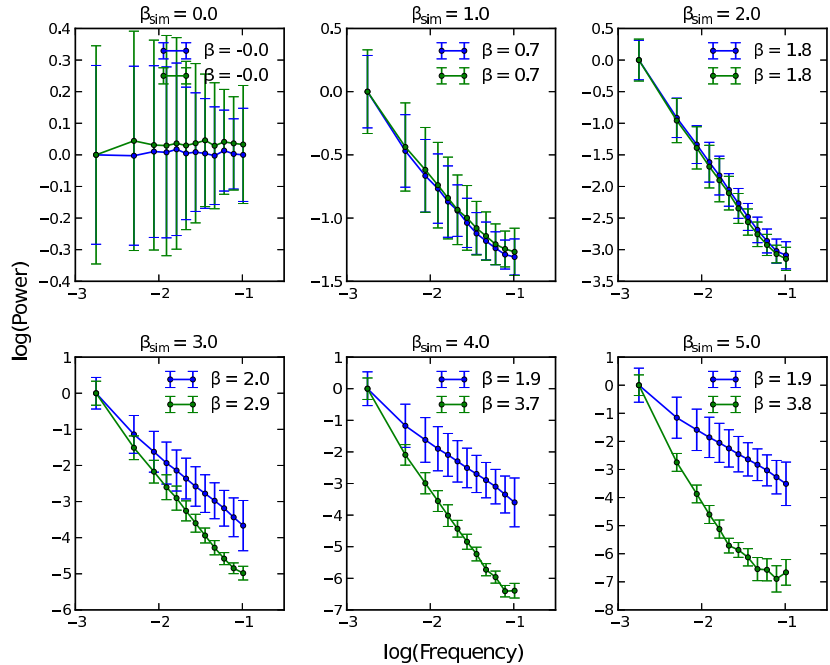}
\caption{Effect of the use of window functions for \textbf{even sampling} cases using the rectangular (blue) and Hanning window (green). Each figure shows the result of simulating 1000 light curves with a given simple power-law PSD $\propto 1/\nu^{\beta}$, with $\beta$ given in each figure title. The data points are the mean PSD and the error bar is the standard deviation in the simulation, while the units of power (vertical axis) and frequency (horizontal axis) are arbitrary. Also included are direct fits of the slopes of the mean PSDs for the simulated data for each window (individual panels legend). In this case, the shape of the PSDs is less noisy and the estimated PSDs for steep cases look different from each other. Even in this case, direct linear fitting of the PSD produces biased results.}
\label{window_even}
\end{center}
\end{figure*}

From Figure \ref{window_uneven} it can be seen that even though we can calculate the periodogram directly for an unevenly sampled time series the results we obtain are very noisy and do not vary much among different values of $\beta$. The main problem is that all the PSDs with $\beta \ge 1$ look very similar, showing almost the same slope when fitted with a linear function after a log-log transformation. This is problematic as the fitting procedure relies on the differences between different PSD power-law indices to choose the best model. 

Doing the same exercise for a time series with the same time length and number of data points but with even sampling we obtain the results shown in Figure \ref{window_even}. In this case the results are much less noisy and the estimated PSDs look different from each other even for very steep PSDs. This allows for better discrimination and is required to find an upper limit to the source power-law exponent of the PSD. 

The problems associated with the window functions become evident when trying to apply the fitting method using unevenly sampled data, and show up as an inability to find an upper limit to the power-law exponent $\beta$ due to the lack of difference between the estimated PSDs for the simulated data. This problem can be solved by the use of interpolation and an appropriate window function, a subject that is discussed in Section \ref{spectral_window}.

Figures \ref{window_uneven} and \ref{window_even} illustrate the limited use we can make of direct PSD fitting, even for the case of long time series. In this case, red-noise leakage makes it impossible to recover the right power-law index for steep PSDs.

The subject of windowing of unevenly sampled data is briefly discussed in \citet[][]{scargle1982}. In particular figure 3 in \citet[][]{scargle1982} shows a few example window functions for the cases of even and uneven sampling using the classic periodogram. That figure illustrate the very different sidelobe structure that is obtained for the uneven sampling case, which is at the root of the problem described here. 

\begin{figure*}
\begin{center}
\includegraphics[width=12cm, trim=20 10 20 20]{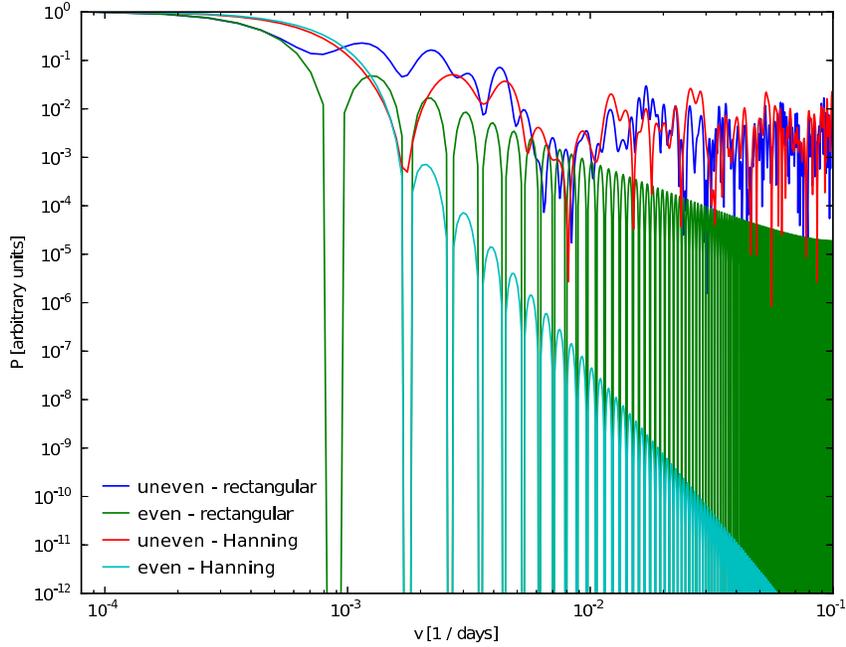}
\caption{Spectral window functions in the cases of even and uneven sampling. In the uneven sampling case the rectangular (blue curve) and Hanning (red curve) windows have a response with a relatively high sidelobe level, that does not decay as the frequency increases. For the even sampling case with the same time length and number of data points we see that the rectangular (green curve) and Hanning window (cyan curve) behave as expected in the usual case, with a regular sidelobe structure whose amplitude decreases as the frequency increases. }
\label{window_even_uneven_example}
\end{center}
\end{figure*}

To clarify this point, we also include the window functions for our test data along with the results of applying the Hanning window. An examination of Figure \ref{window_even_uneven_example} helps us understand the results described below. In conventional Fourier analysis, window functions change the frequency response of the sampling, changing the sidelobe structure and thus helping mitigate the effects of red-noise leakage and aliasing. This behavior can be seen when using evenly sampled data sets, where the sidelobe structure is regular and decays as frequency increases. The case for uneven sampling is very different: the shapes of the window functions explains the strong red-noise leakage seen in the simulations and the increased noise. In the case of even sampling we recover the results of conventional Fourier analysis, with all the known properties of window functions. 

%\clearpage

For the reasons described above we use linear interpolation and rebinning to interpolate the unevenly sampled light curves to a regular grid, thus allowing for the PSD fitting.

\subsubsection{Spectral window function \label{spectral_window}}

One fundamental difference between the implementation of the method of \cite{uttley+2002} and ours is that we use window functions to reduce the effects of red-noise leakage. We found that this is necessary when dealing with steep power spectral densities, like those found in blazar studies. In our first attempts to fit the PSDs we found that with a rectangular window we were not able to set an upper limit to the value of $\beta$ and were only able to set a lower limit. The upper limit on $\beta$ is necessary to constrain the significance of cross-correlations, as will be described in Section \ref{cross_corr_method}. In this section we explain the origin of that problem and the solution we implemented.

For broadband time series a big problem is the leakage of power through far sidelobes of the spectral window response. This problem is evident when dealing with high dynamic range PSDs, such as steep power-laws. For these power-law PSDs, it is seen as a flattening of the high frequency part of the periodogram due to power leaking from low frequency part which has much higher power. In practical terms, it means that after some critical value of the power-law index all the periodograms have a flat slope which does not depend strongly on the PSD (Figures \ref{window_uneven} and \ref{window_even}). Most of this high frequency power is actually coming from low frequencies through sidelobes of the window function. One way to deal with this problem is by using window functions with low level sidelobes; some details about their application to our data set are presented below.

\paragraph*{Spectral window functions for our data sets \label{best_window}}

There is a great variety of window functions, which differ mainly in the width of their main lobe, the maximum level, and the fall-off rate of the sidelobes. The ideal window function will depend on the application and some experimentation might be necessary. Properties of various window functions can be found elsewhere \citep[e.g.,][]{harris1978}

We tried a number of them and compared their performance in recovering steep PSDs. We found that among the ones tested the most suitable one was the Hanning window, which is able to recover a steep spectrum in a range that allows us to fit our light curves. Among the special characteristics of this window are its low sidelobe level, more than $32$ dB below main lobe, and the fast fall-off at $-18$ dB/decade. As a downside the Hanning window has a broader main lobe at 3 dB ($1.44 \cdot 1 / T$) when compared to the rectangular window ($0.89 \cdot 1 / T$), where $T$ is the length of the time series.
 
\begin{figure*}
\begin{center}
\includegraphics[width=12cm, trim=20 10 20 20]{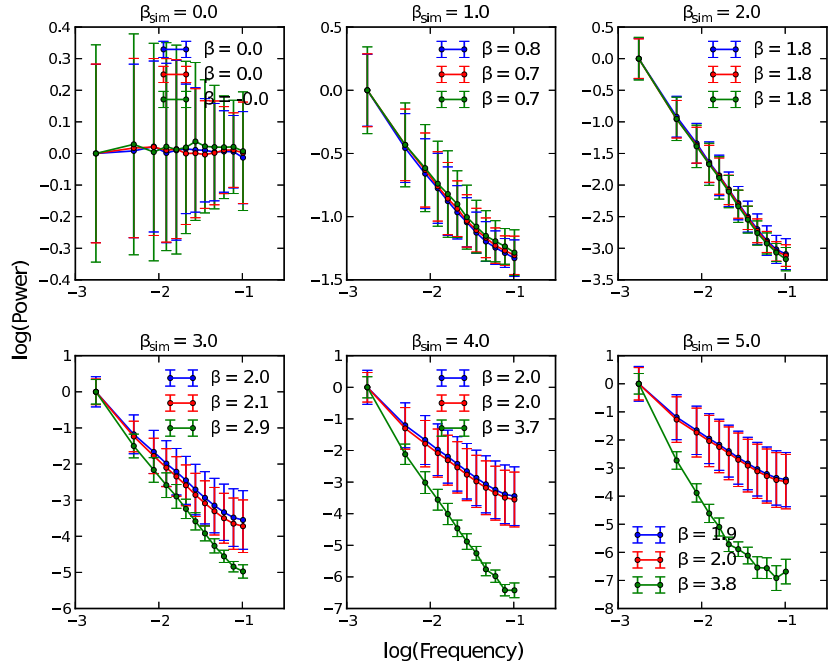}
\caption{Comparison of windowed periodogram for power-law PSDs for evenly sampled data. Each figure shows the result of simulating 1000 light curves with a given simple power-law PSD $\propto 1/\nu^{\beta}$, with $\beta$ given in each figure title. The data points are the mean PSD and the error is the standard deviation in the simulation, while the units of power (vertical axis) and frequency (horizontal axis) are arbitrary. Also included are direct fits of the slopes of the mean PSDs for the simulated data in each case using a rectangular (blue), triangular (red) and Hanning (green) windows.}
\label{window_steep_psds}
\end{center}
\end{figure*}

The effect of different windows is illustrated in Figure \ref{window_steep_psds}, which shows the periodogram for a series of steep PSDs. From the figure it is also clear why other window functions fail to distinguish between steep PSDs, and thus are not suitable to use with this method. 

% Window functions and steep PSDs

The results of Figure \ref{window_steep_psds} can be understood by comparing the properties of the window functions shown in Table \ref{windows_properties} \citep[][]{harris1978}. The reduction of the red-noise leakage when using the Hanning window allows us to discriminate between different steep power-law indices of the PSD, and is due to the low level and fast fall-off its sidelobes.

\begin{table}
\caption{Properties of selected window functions}
\begin{center}
\begin{tabular}{c c c c}
\hline
Window & Sidelobe Level & Sidelobe Fall-Off & 3-dB BW \\
 & (dB) & (dB/oct) & (bins) \\ 
 \hline
Rectangular & $-13$ & $-6$ & $0.89$ \\
Triangle or Bartlett & $-27$ & $-12$ & $1.28$ \\
$\cos^2(x)$ or Hanning & $-32$ & $-18$ & $1.44$ \\
\hline
\end{tabular}
\end{center}
\label{windows_properties}
\end{table}

Windowing is good for fitting a featureless PSD, but it can be a source of problems if the goal is to find narrow spectral components. This is because of the well-known trade-off between resolution and sidelobe level: tapering the window function in order to decrease the sidelobe level must reduce the resolution. This has to be considered when searching for periodic components, a case which is outside of the scope of the current analysis.

\subsubsection{Filtering}

The windowing technique is able to solve the problem with red-noise leakage, but another method that can be used is filtering in the time domain followed by a correction to the frequency domain result. The goal of filtering is to eliminate the low-frequency components that produce the red-noise leakage before computing the periodogram. Since this changes the spectrum of the time series, it has to be compensated in the final periodogram by the application of a frequency filter. 

One of these techniques is called pre-whitening and post-darkening by first differencing. In this case, the original time series $(t_i, f_i)$ with even sampling is transformed to $(t_i, g_i \equiv f_i - f_{i-1})$. In the frequency domain this is equivalent to filtering with $|H(\nu)|^2 = 2 [1 - \cos(2 \pi \nu)]$. Higher order filtering is possible, for example by the application of first order differencing multiple times \citep[][]{shumway+2011}.

% Filtering to fit steep PSDs
\begin{figure*}
\begin{center}
\includegraphics[width=12cm, trim=20 10 20 20]{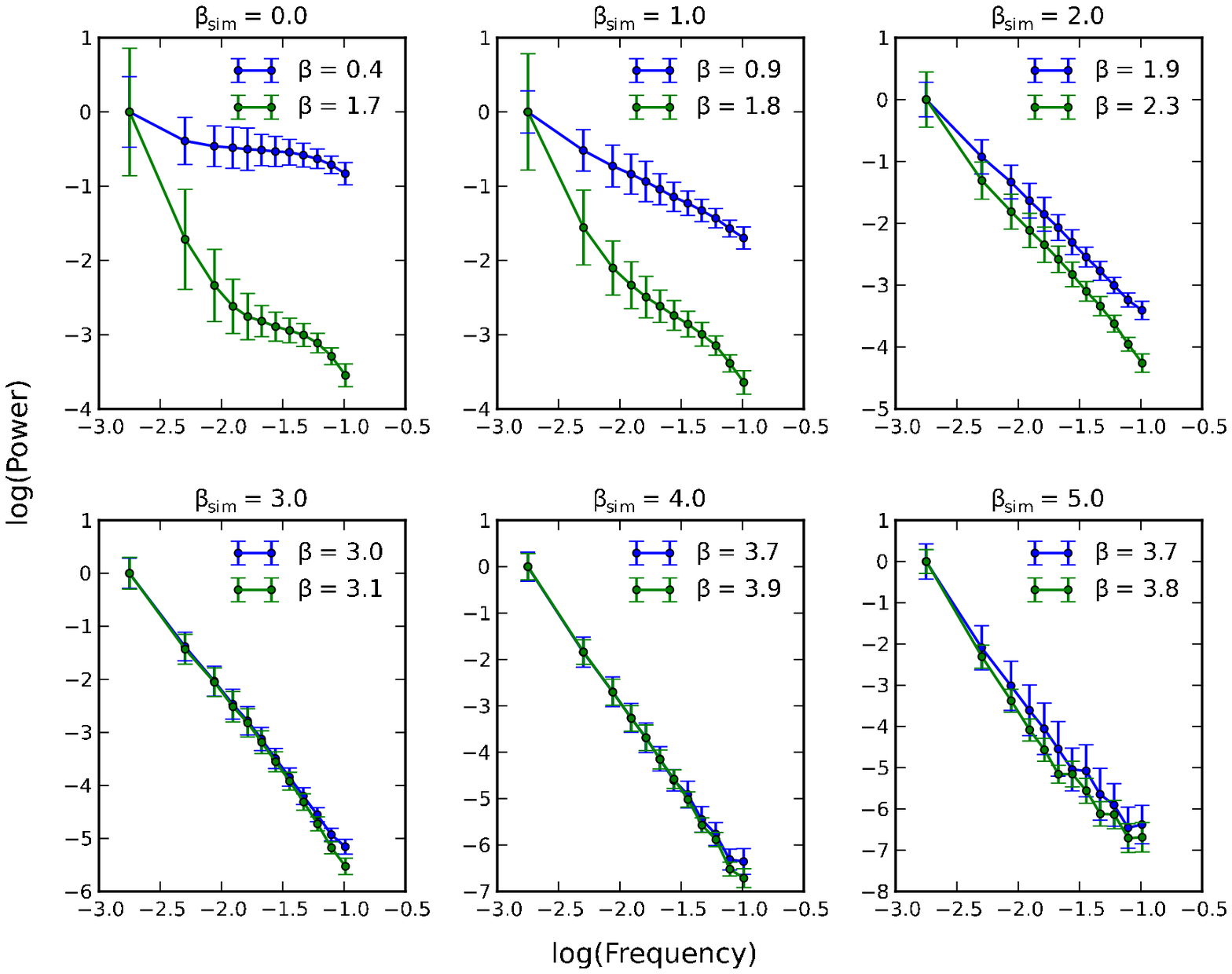}
\caption{Effect of the pre-whitening and post-darkening filters in the PSD of evenly sampled time series. Each figure shows the result of simulating 1000 light curves with a given simple power-law PSD $\propto 1/\nu^{\beta}$, with $\beta$ given in each figure title. The data points are the mean PSD and the error is the standard deviation in the simulation, while the units of power (vertical axis) and frequency (horizontal axis) are arbitrary. Also included are direct fits of the slopes of the mean PSDs for the simulated data in each case using first difference (blue curve) and second difference (green curves).}
\label{filter_steep_psds}
\end{center}
\end{figure*}
%\clearpage

Figure \ref{filter_steep_psds} shows the result of applying this procedure to simulated data with even sampling and a range of power-law slopes of the PSD. It can be seen that this method has problems recovering flat PSDs with $\beta \le 2$ and very steep PSDs with $\beta \ge 4$. We also tested it with the sampling of the OVRO data set and found that in a large number of cases it was not able to provide good upper limits for $\beta$ and was outperformed by windowing with the Hanning window. We therefore use Hanning windowing for the data analysis.

\subsubsection{Adding noise to simulated light curves}

A final issue is the addition of noise to simulated light curves, a necessary step to consider the effect of observational uncertainties in our ability to measure the PSD. This is not a serious problem for the radio light curves, which in most cases have very high signal-to-noise ratio. But it is important for most gamma-ray light curves, which have moderate signal-to-noise ratios.

In order to add the observational noise to the light curves we first need to normalize the simulated data to match the observations. One way to obtain an approximate normalization is by using Parseval's theorem, which with the normalization we use implies that
\begin{equation}
\sigma^2  = \sum_{\nu_{\rm min}}^{\nu_{\rm max}} P(\nu) \Delta \nu
\end{equation}

We can estimate the variance for the observations and the simulations and use a constant factor to make them equal, thus getting an approximate normalization of the PSD. One problem is that the data already contain observational noise added to the signal, so for each data point we have $d_i = s_i  + n_i$, where $d$ is the data, $s$ the signal and $n$ the noise. We estimate the variance to obtain $\sigma^2_{\rm d} = \sigma^2_{\rm s} + \sigma^2_{\rm n}$, under the assumption that the noise and signal are uncorrelated.

The variance of the noise can be obtained from the observational uncertainty by $\sigma^2_{\rm n} \approx \bar{e_i^2}$, where $e_i$ is the $1\sigma$ observational uncertainty associated with the $i$-th measurement. The final normalization equation is
\begin{equation}
 \sigma^2_{\rm sim} = A^2 (\sigma^2_{\rm d} - \bar{e_i^2})
\label{sim_normalization}
\end{equation}

We multiply the originally arbitrarily normalized simulated data by $A^{-1}$, to get a normalization equivalent to the one in the observations. In practice, we use $A$ to transfer the observational error bars to the simulations, to which we add Gaussian observational noise to the time domain signal such that $e_{\rm{sim}, i} = A \, e_i$. In the original formulation the noise is applied to the periodogram, but we choose to apply it directly to the time series in order to account for the different magnitudes of the observational uncertainties. The assumption of Gaussian error bars is only approximate for the gamma-ray data, which have a Poisson distribution. Since in this analysis we are only considering highly significant gamma-ray detections, we usually have at least 5 photons in each integration and in most cases many more. In this regime, the difference between Poisson and Gaussian distributed errors is negligible.

\subsubsection{Estimation of the uncertainty in the model parameters\label{uncertainty_estimate}}

\citet[][]{uttley+2002} defined the region of confidence for the fitted model parameters as the region for which $p(\hat{\theta}) > p_{\rm conf}$, where $p(\hat{\theta})$ is the $p$-value for a given set of parameters $\hat{\theta}$. For example a 68.3\% confidence interval has $p_{\rm conf} = 0.317$, while a 95.5\% confidence interval has $p_{\rm conf} = 0.045$. One problem with this rule is that it is not possible to get 68.3\% confidence intervals for fits in which $p < 0.317$. This contrast with the usual approach to measure uncertainties from $\chi^2$ fits that defines a 68.3\% (or any other level) confidence interval by the region of parameter space for which $\chi^2(\theta) - \chi^2_{\rm min} \leq \Delta \chi^2$, where $\Delta \chi^2$ depends on the number of interesting parameters being fit and the confidence level \citep[][]{avni_1976, press+1992, wall+2003}. In this widely used method a confidence interval can be obtained independently of the value of $\chi^2_{\rm min}$ for the fit, thus effectively decoupling the goodness of fit estimate from the estimation of confidence intervals.

For these reasons we decided to estimate frequentist best fit confidence intervals by using the method of the Neyman construction. These intervals are constructed to include the true value of the parameter with a probability greater than a specified level, as demonstrated in \citet[][]{beringer+2012} and  \citet[][]{james_2006}. Here we only describe the mechanics of obtaining confidence intervals and refer the reader to the references for a formal demonstration. The procedure requires that we know the probability of a given experimental result $\beta_{\rm fit}$, as a function of the value of the unknown parameter $\beta$. The distribution for the fitted value of the power-law index for a given value of $\beta$ is estimated with a Monte Carlo simulation resulting in the distribution of $\beta_{\rm fit}$ as a function of $\beta$. At each value of $\beta$ we can construct a confidence interval at a desired level, the results of these confidence intervals are summarized in a figure with $\beta$ in the vertical axis and $\beta_{\rm fit}$ in the horizontal axis (see the lower panel of Figure \ref{fit_example_results} for an example). For each value of $\beta$ we can draw a horizontal line going from the lower to the upper limit of the confidence interval. We join the confidence intervals for a range of values of $\beta$ to get a confidence band. We then fit the PSD to the data, draw a vertical line at $\beta_{\rm fit}$, and determine the confidence interval as the intersection of the vertical line and confidence band. This procedure requires that for each PSD fit we run a large number of fits to simulated data, which increases the computational time. This is feasible when fitting a single power-law index but it can be prohibitive when fitting a larger number of parameters. In principle we are required to have a confidence interval for each possible value of $\beta$, but to reduce the computational time we discretize a reasonable parameter range and use linear interpolation to fill the gaps in the confidence band.

An example of the application of this method is presented in Section \ref{example_of_psd_method}.

\subsection{Implementation}

This section starts with an example of the application of the method to a simulated light curve of known PSD. We present four tests intended to validate the procedure by fitting a large number of simulated data with known PSD, using sampling patterns taken from the OVRO program, and with observational errors consistent with our data. This section ends with a study of the effects of changing the number of simulated light curves ($M$ as defined in Section \ref{psd_method_description}), when fitting simulated data in one of them, and an example light curve from the OVRO program in the other. The goal is to get an indication of the associated uncertainties by changing $M$, as it can have a large impact on the computational time.

\subsubsection{An example of the application of the method \label{example_of_psd_method}}

\begin{figure}
\begin{center}
\includegraphics[width=7cm, trim=20 10 20 0]{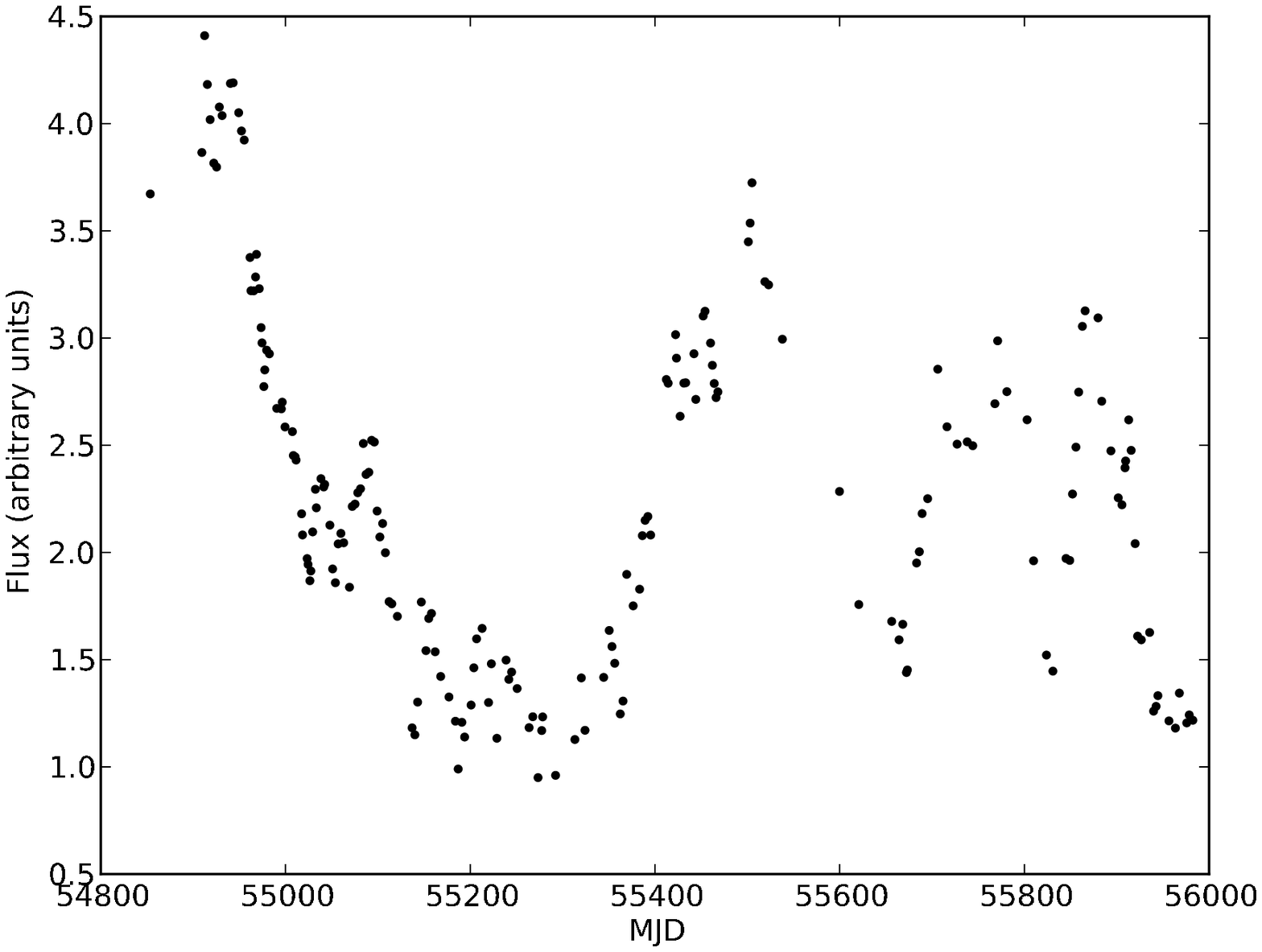}
\includegraphics[width=7cm, trim=20 10 20 0]{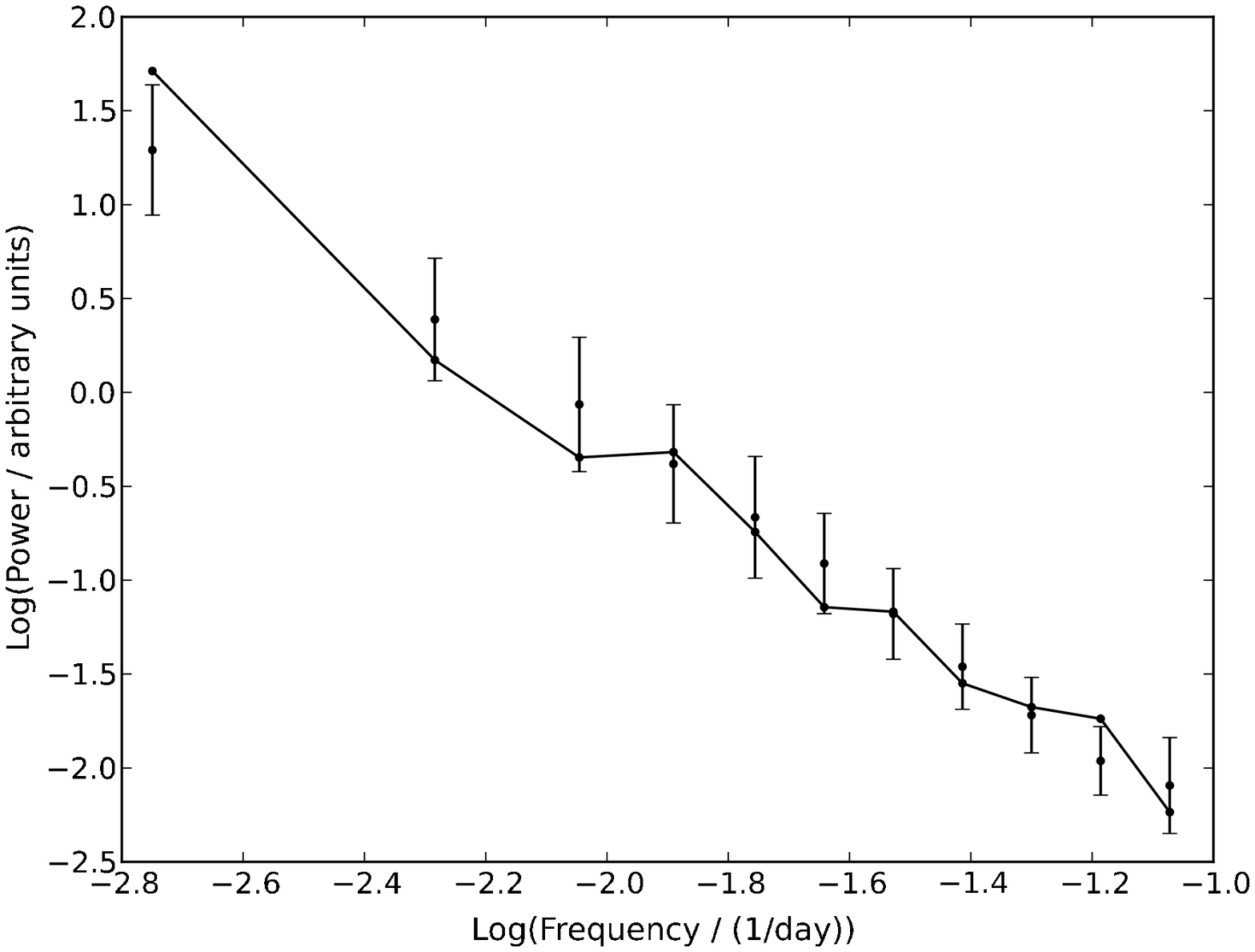}
\caption{Example of the PSD fit method applied to simulated data. Upper panel is the simulated light curve with a PSD $\propto 1/\nu^2$ and no noise. Lower panel is the data periodogram binned in frequency (black line) and the mean PSD and scatter for the best fit with $\beta = 1.85 \pm 0.2$ (black dots and error bars).}
\label{fit_example_lc}
\end{center}
\end{figure}

A simulated light curve with a power-law PSD with $\beta=2.0$, no observational noise, and sampled in the same way as the source J1653+3945 is shown in Figure \ref{fit_example_lc}; along with the periodogram and best fit.

% Simulated light curve with sampling for J1653+3945

% Fit results for simulated light curve with sampling for J1653+3945
\begin{figure}
\begin{center}
\includegraphics[width=7cm, trim=20 10 20 0]{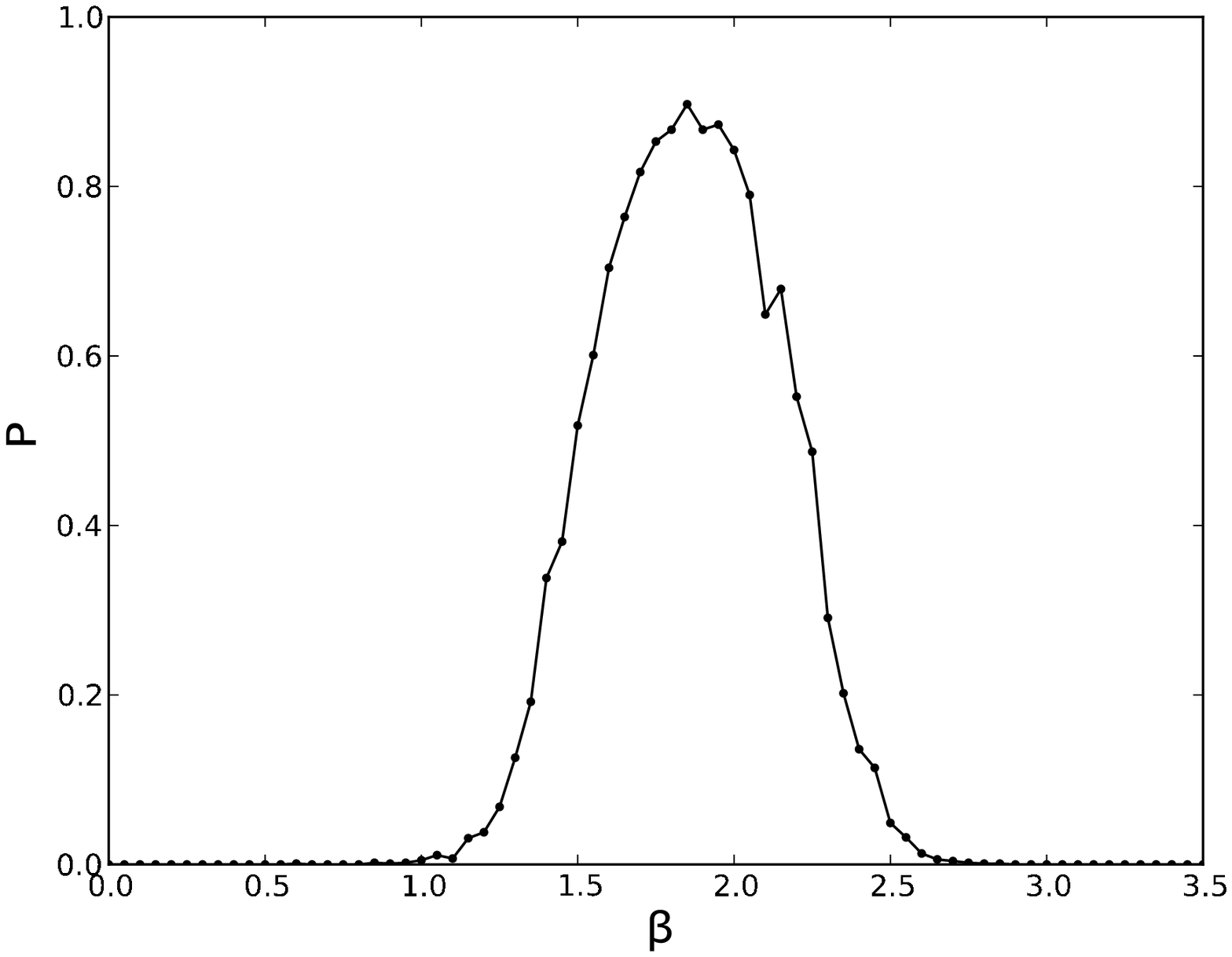}
\includegraphics[width=7cm, trim=20 10 20 0]{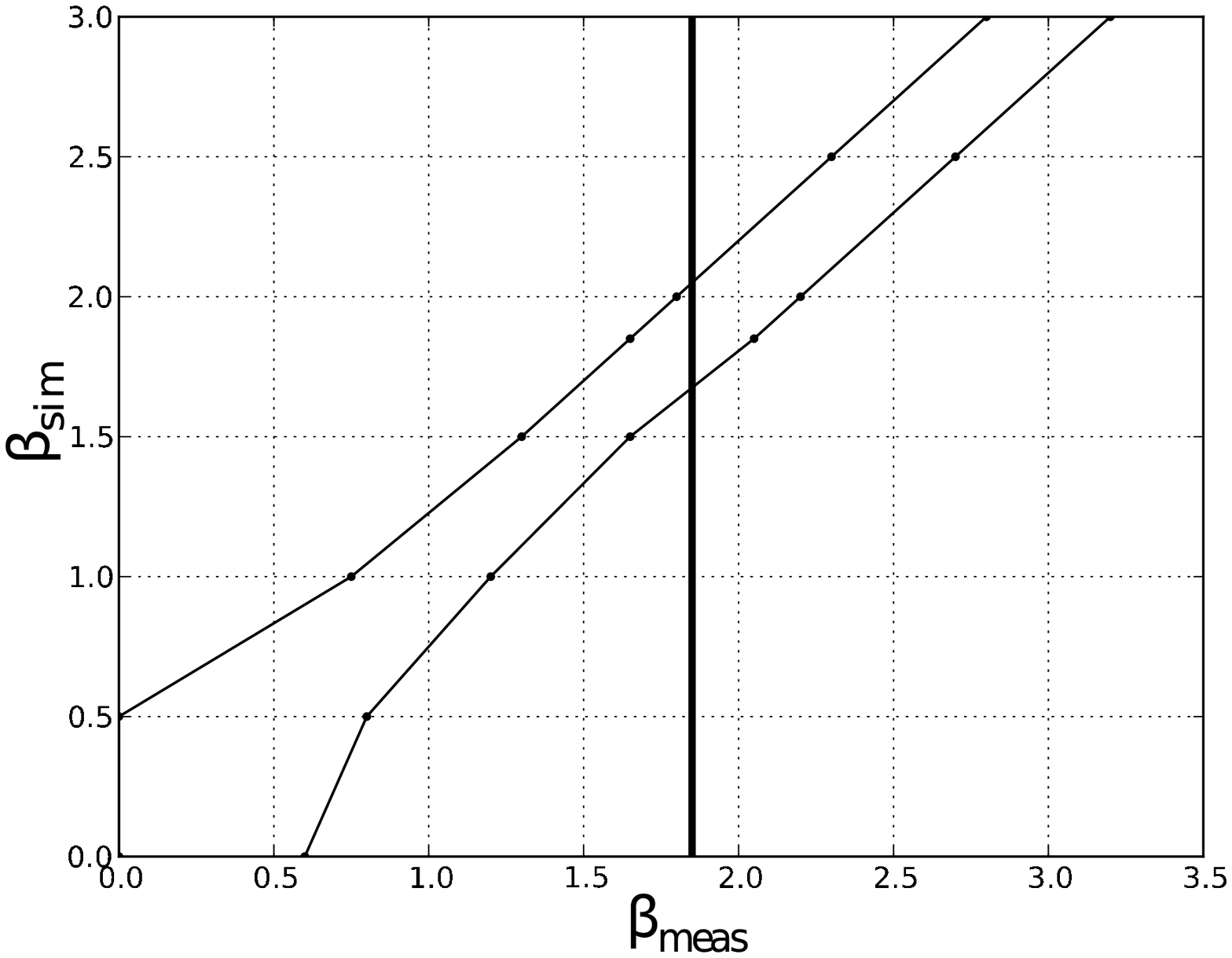}
\caption{Example of the fitting method applied to simulated data of known PSD. Upper panel is $p$ versus $\beta$ for the different model power-law exponents we tested. The peak at 1.85 indicates the best fit. The error on the fit is obtained from the confidence band which is shown in the lower panel. The intersection of the vertical line with the confidence band give us $\beta = 1.85 \pm 0.2$.}
\label{fit_example_results}
\end{center}
\end{figure}
%\clearpage

The results of the fitting procedure are summarized by a plot of $p$ vs $\beta$ (Figure \ref{fit_example_results}). The best fit corresponds to $\beta = 1.85 \pm 0.2$, where the errors were obtained with a Neyman construction, whose resulting 68.3\% confidence band is also shown in the figure. In what follows all the errors are obtained in this way. This error can be compared with the original error prescription, which in this case produces a value of $\pm 0.5$, more than twice the value estimated from the simulations.

\subsubsection{Validation of the implementation with simulated data sets}

In order to validate the implementation we tested it with simulated data sets of known PSD. Typical sampling patterns and various relative amounts of noise are considered to investigate the behavior of the method under different conditions. In each of the tests we use $M=1000$ to get the mean PSD and scatter at each trial value of $\beta$. We use trial values of $\beta$ from 0.0 to 3.5 in steps of 0.05. Besides validating the method, these tests illustrate how our ability to measure the PSD and the associated error in the power-law exponent depend on the sampling and noise for the particular light curves.

\paragraph*{OVRO sampling pattern 1 and no noise}

\begin{figure}
\begin{center}
\includegraphics[width=8cm, trim=40 0 40 0]{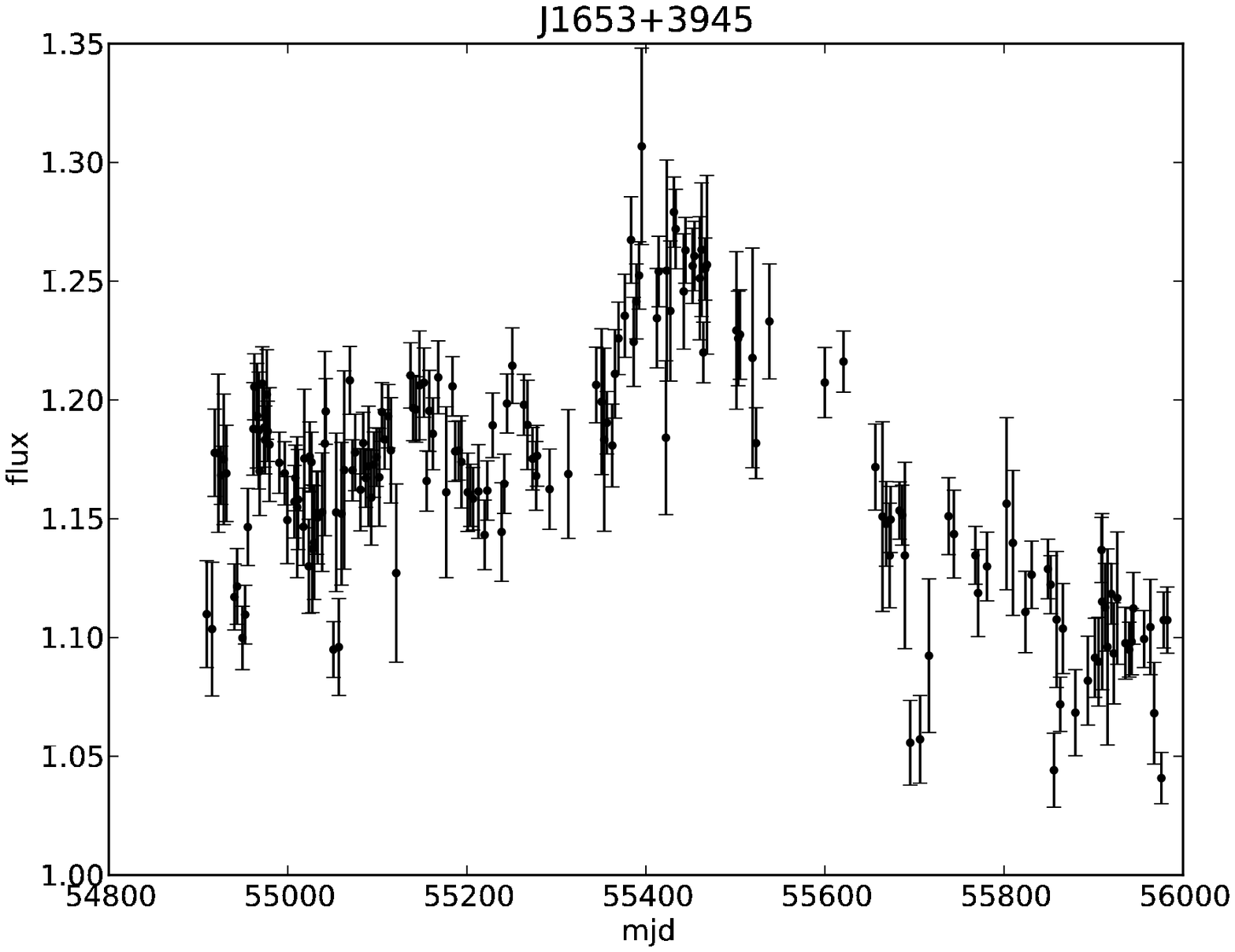}\\
\includegraphics[width=4cm, trim=30 5 20 0]{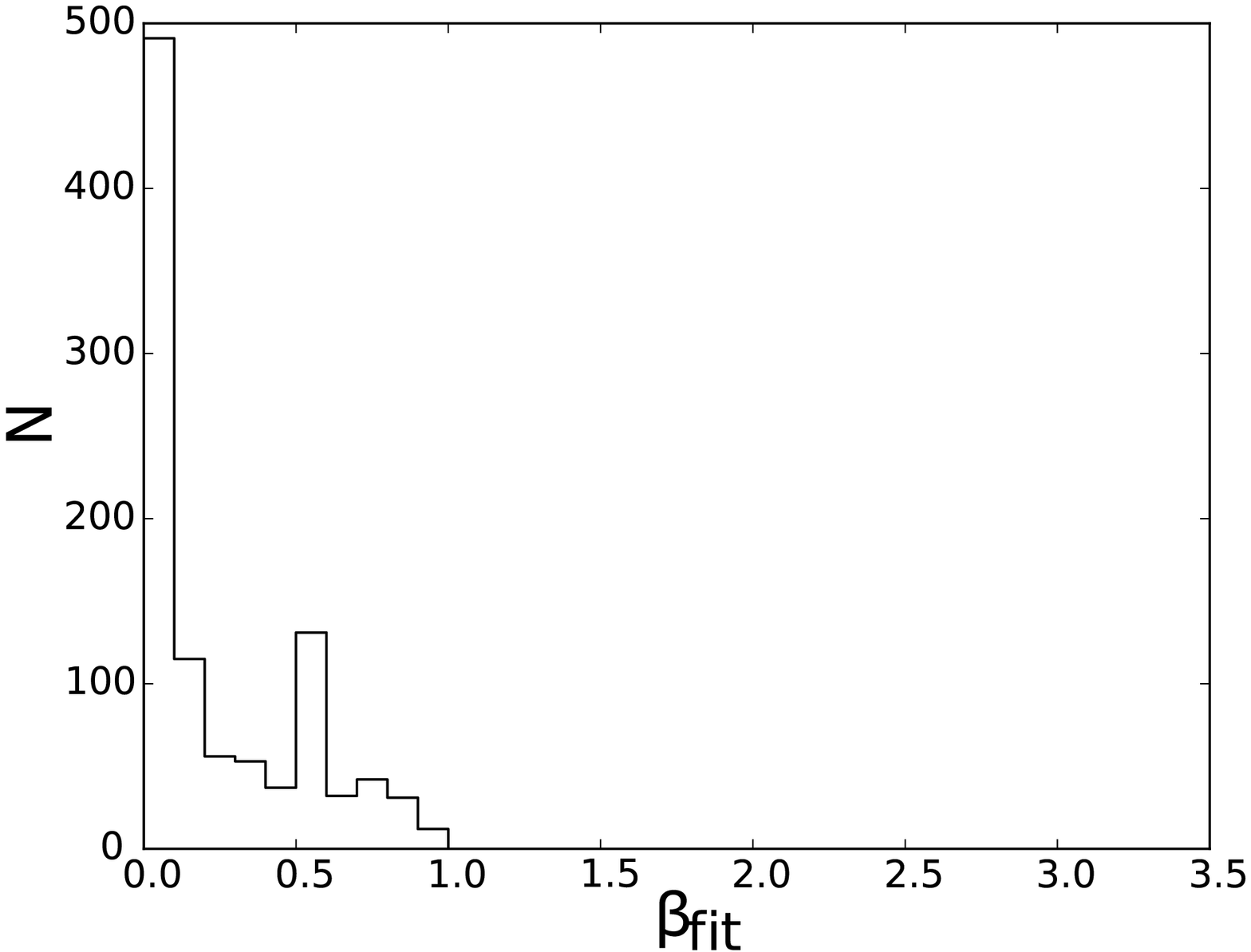}
\includegraphics[width=4cm, trim=30 5 20 0]{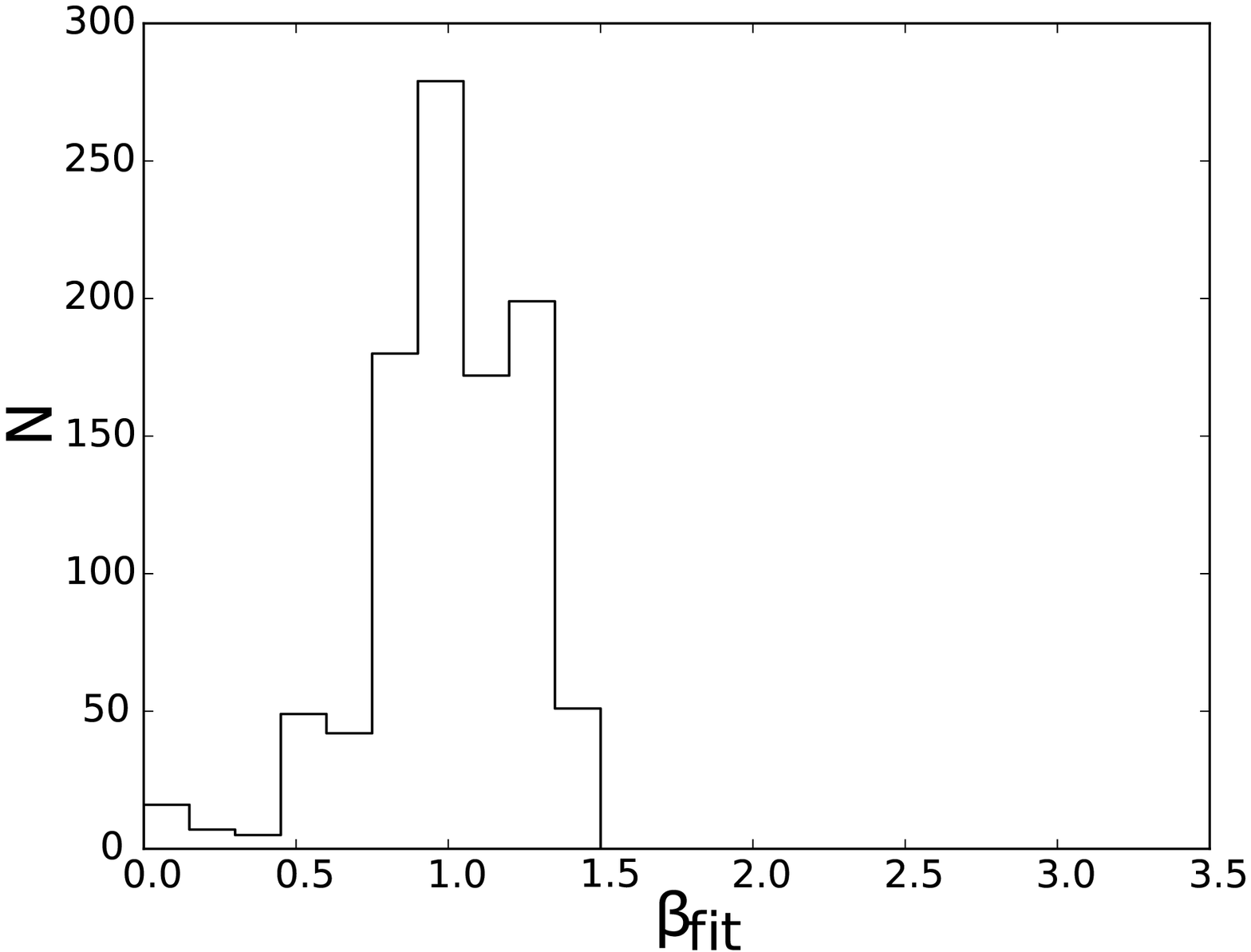}\\
\includegraphics[width=4cm, trim=30 5 20 0]{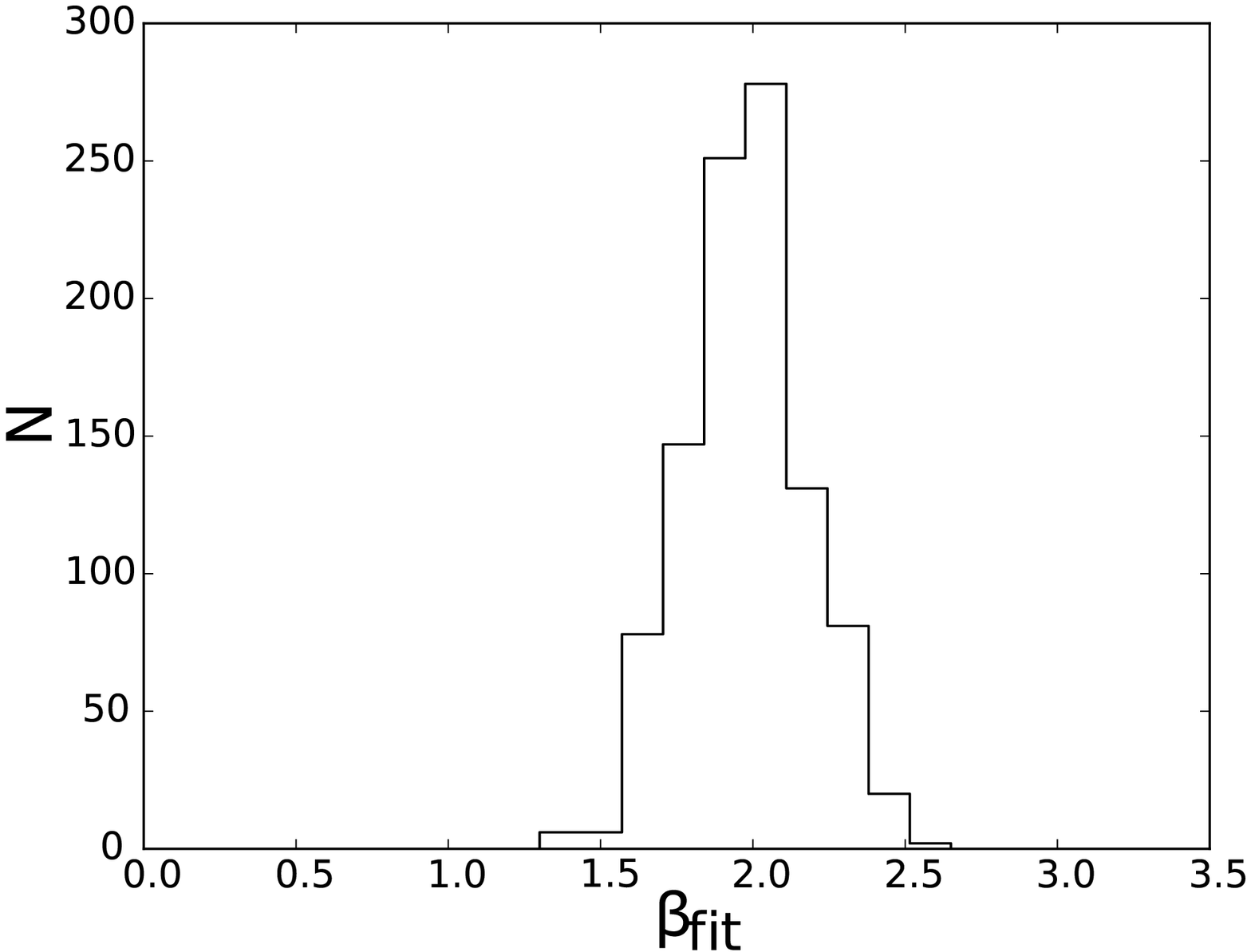}
\includegraphics[width=4cm, trim=30 5 20 0]{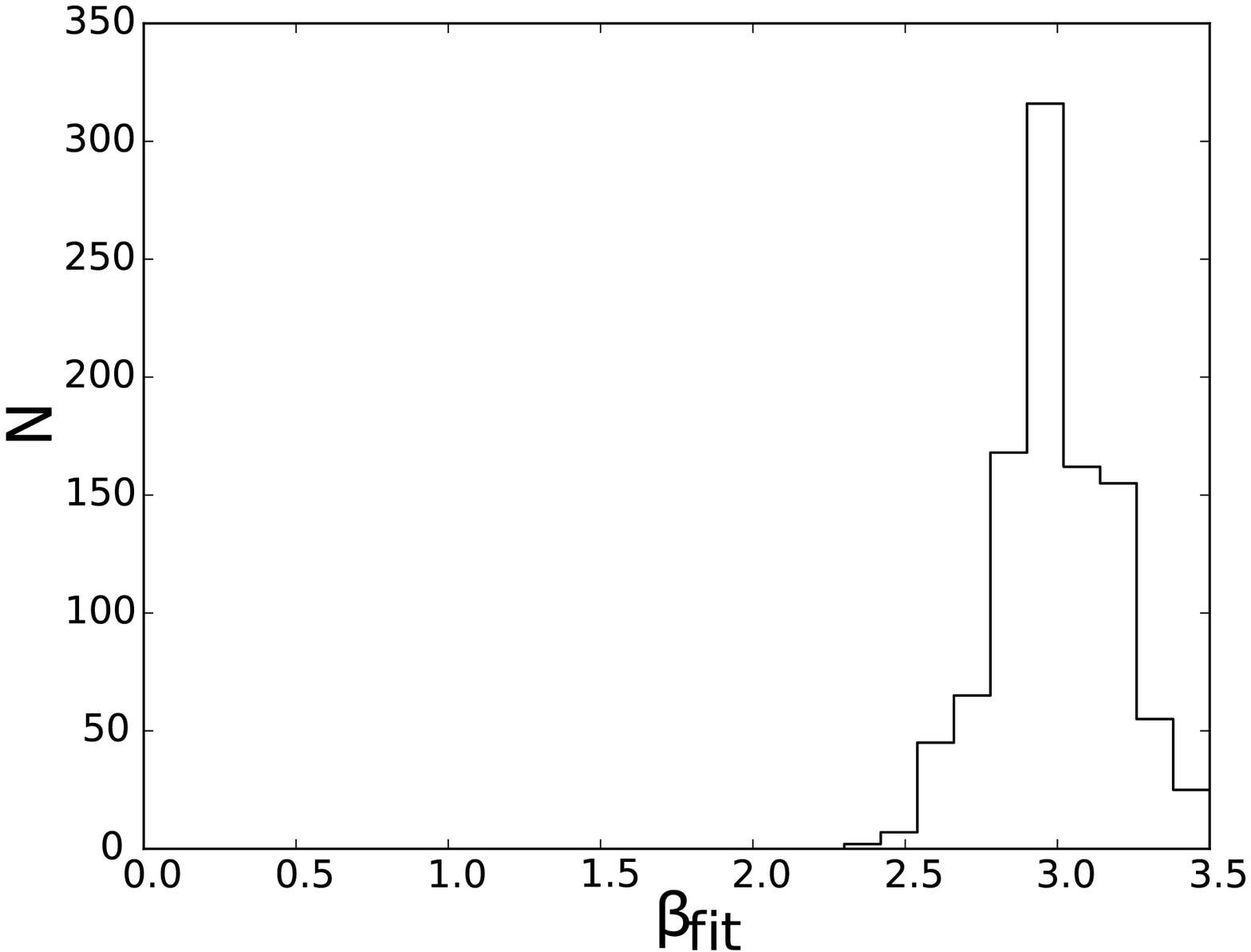}
\caption{Upper panel shows the OVRO data used to get the time sampling. The flux density uncertainties are not used in this test and we assume a perfect measurement. Four lower panels are the distribution of best fit values for 1000 simulated light curves in each case. In each case this distribution gives an estimation of the error on the fit and is used to construct the confidence band. Top left is for $\beta_{\rm sim}  = 0.0$ and $\beta_{\rm fit} = 0.0^{+0.3}_{-0.0}$, top right is for $\beta_{\rm sim}  = 1.0$ and $\beta_{\rm fit} = 1.0 \pm 0.2$, lower left is for $\beta_{\rm sim}  = 2.0$ and $\beta_{\rm fit} = 2.0^{+0.15}_{-0.2}$, and lower right is for $\beta_{\rm sim}  = 3.0$  and $\beta_{\rm fit} = 3.0^{+0.2}_{-0.15}$. In the case of $\beta_{\rm sim} = 0.0$ we report the mode and dispersion about that value. All the other cases use the median and dispersion.}
\label{val_ex_1_fit}
\end{center}
\end{figure}

In this test we use the sampling pattern for the source J1653+3945, the OVRO data are shown in Figure \ref{val_ex_1_fit} as reference. Note that for all the tests in this section, the fitted data are simulations, with only the sampling taken from the OVRO observations. The results of the fit for simulated data as a distribution of best fit values are shown in Figure \ref{val_ex_1_fit}. We find that in all cases we are able to recover the true $\beta$ with a typical uncertainty of 0.2.

% PSD fit for simulated data for J1653+3945 sampling and no noise
%\clearpage

\paragraph*{OVRO sampling pattern 1 and noise}

\begin{figure}
\begin{center}
\includegraphics[width=8cm, trim=40 0 40 0]{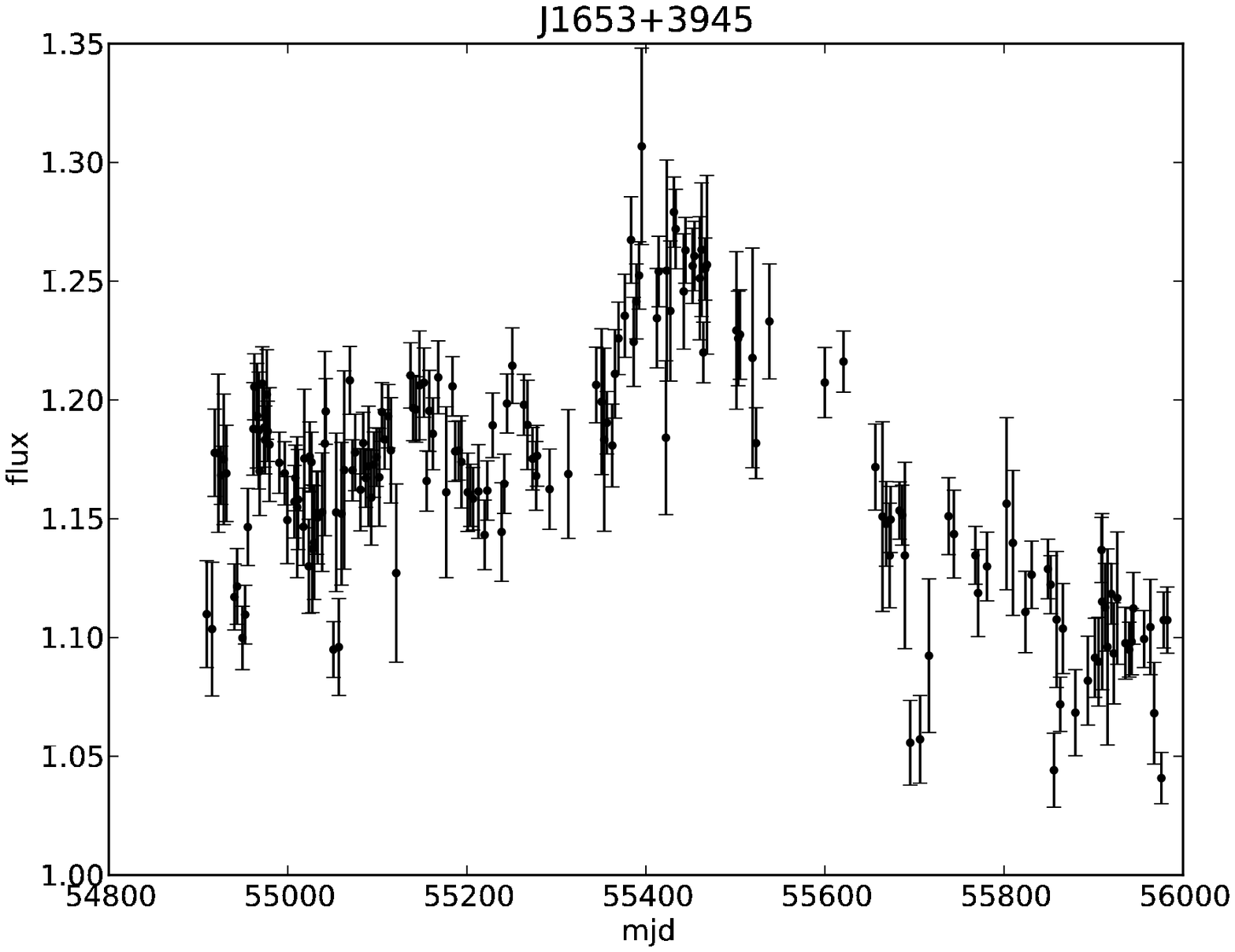}\\
\includegraphics[width=4cm, trim=30 5 20 0]{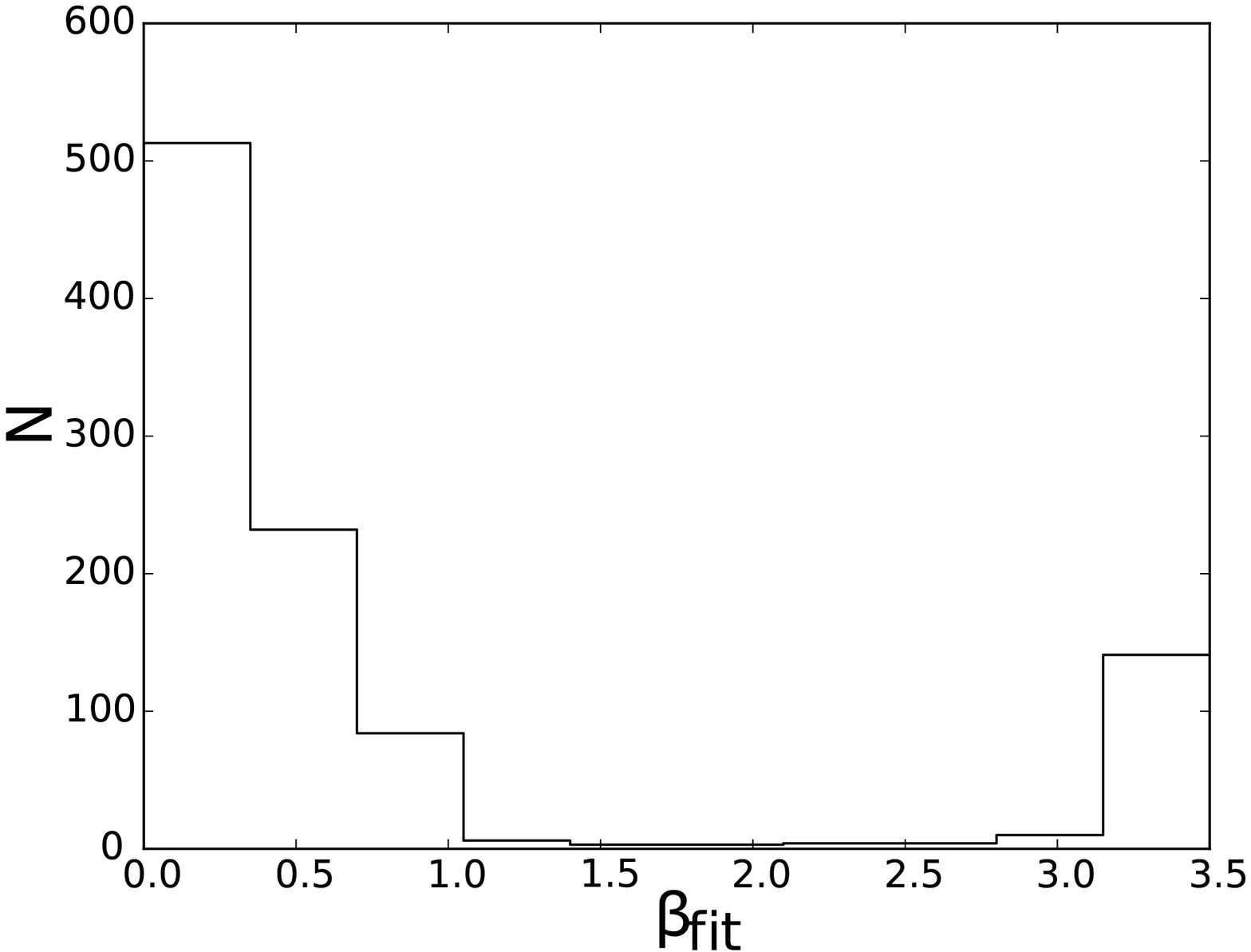}
\includegraphics[width=4cm, trim=30 5 20 0]{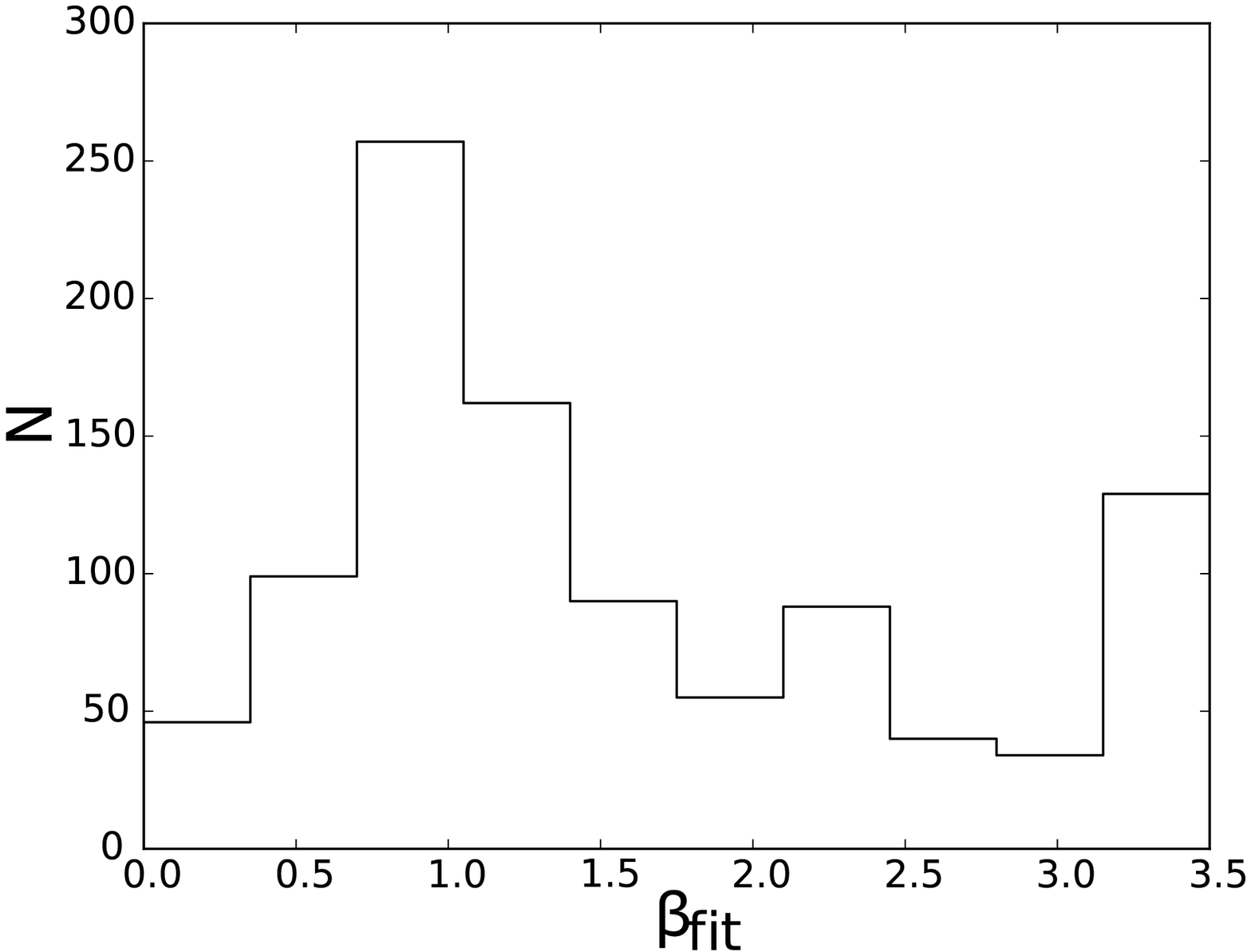}\\
\includegraphics[width=4cm, trim=30 5 20 0]{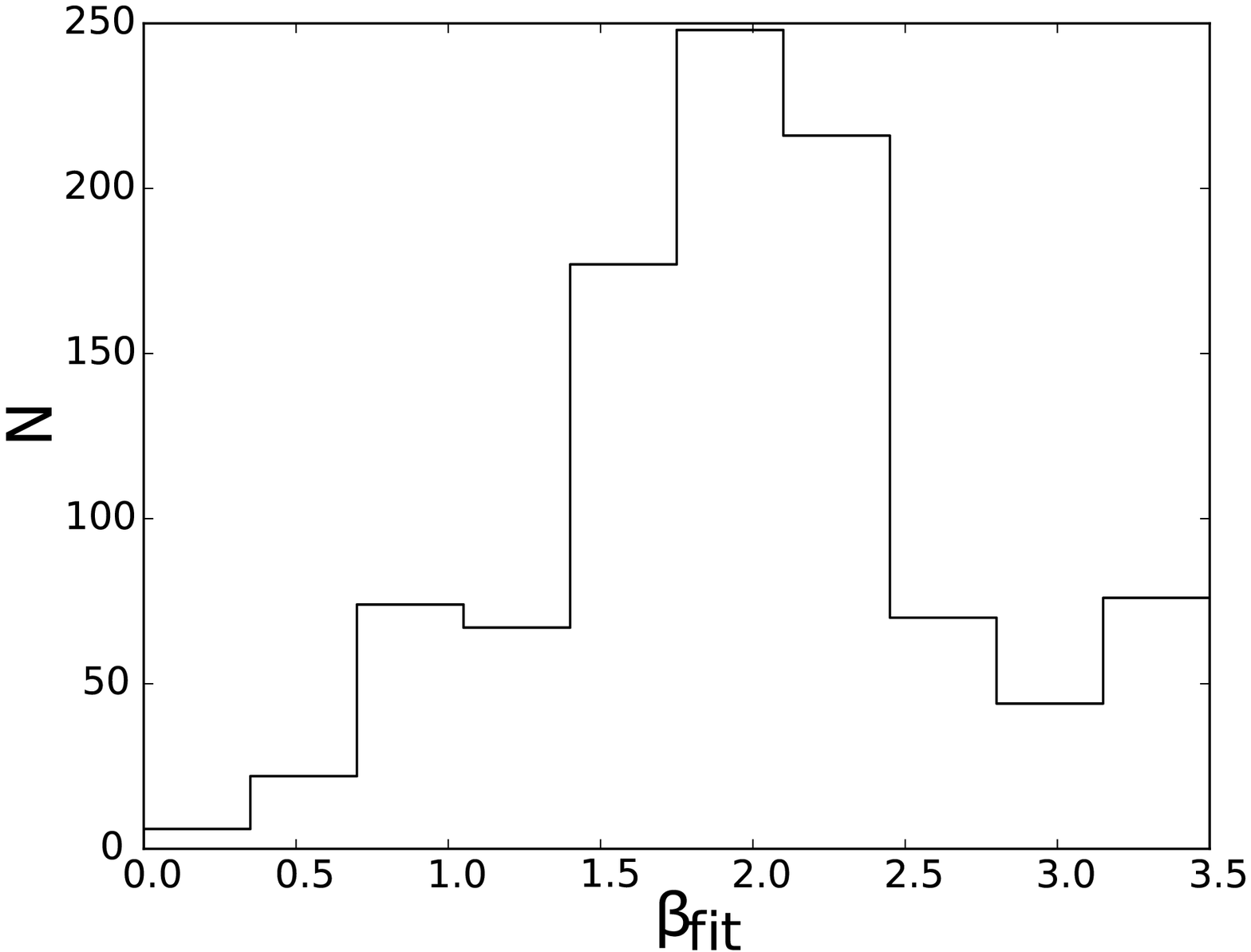}
\includegraphics[width=4cm, trim=30 5 20 0]{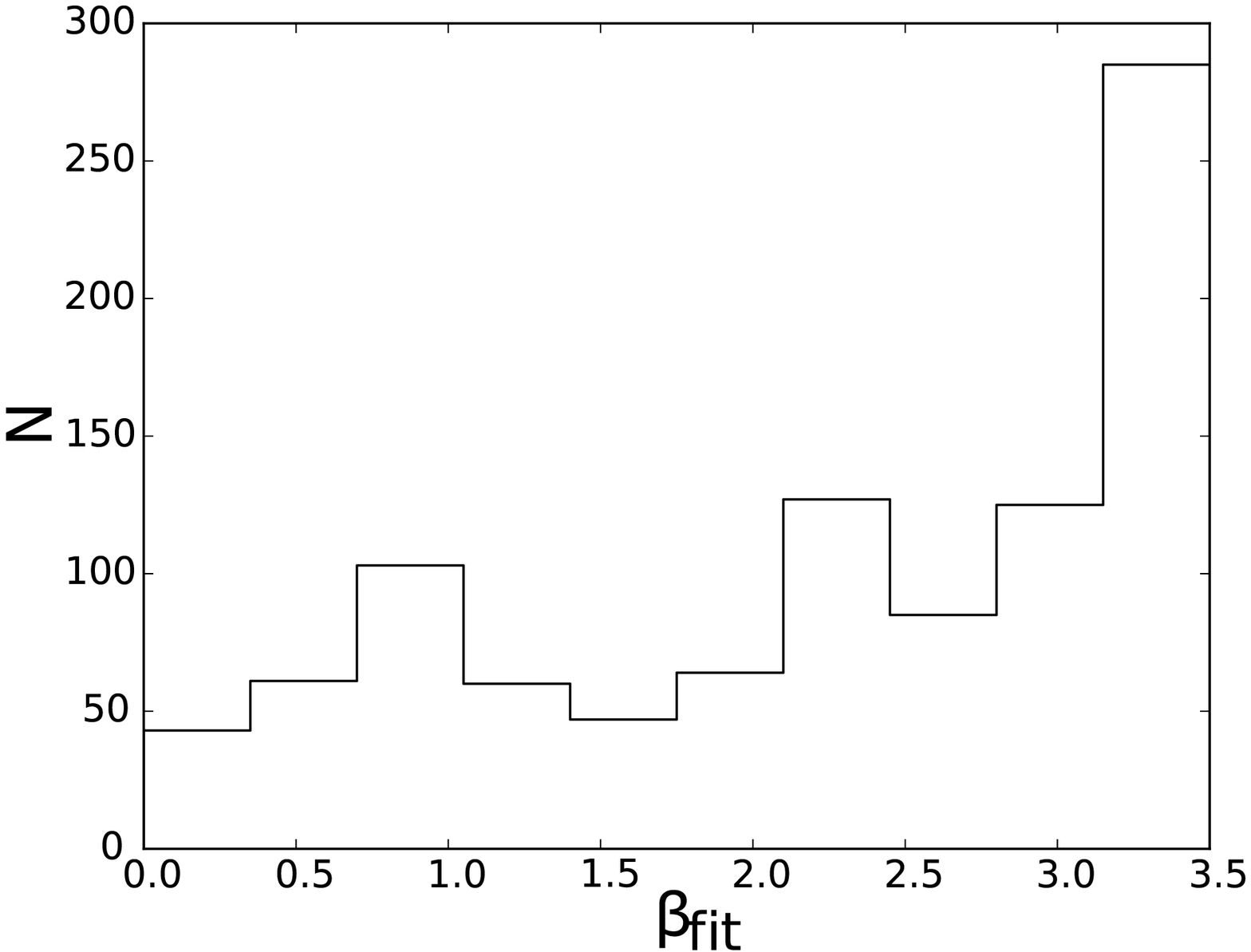}
\caption{Upper panel shows the OVRO data used to get the time sampling and the flux density uncertainties. Four lower panels are the distribution of best fit values for 1000 simulated light curves in each case. In each case this distribution gives an estimation of the error on the fit and is used to construct the confidence band. Top left is for $\beta_{\rm sim}  = 0.0$ and $\beta_{\rm fit} = 0.05^{+0.55}_{-0.05}$, top right is for $\beta_{\rm sim}  = 1.0$  and $\beta_{\rm fit} = 1.3^{+1.5}_{-0.55}$, lower left is for $\beta_{\rm sim}  = 2.0$  and $\beta_{\rm fit} = 1.9^{+0.6}_{-0.55}$, and lower right is for $\beta_{\rm sim}  = 3.0$  and $\beta_{\rm fit} = 3.0^{+0.4}_{-1.85}$. In the cases of $\beta_{\rm sim} = 0.0$ and $3.0$ we report the mode and dispersion about that value. All the other cases use the median and dispersion.}
\label{val_ex_2_fit}
\end{center}
\end{figure}

In this test we use the sampling pattern for the source J1653+3945 and error bars consistent with the noise in this source. The results of the fit for simulated data as a distribution of best fit values are shown in Figure \ref{val_ex_2_fit}. In this case, the large measurement errors make recovering the PSD exponent very hard and the fitting procedure fails to yield a meaningful constraint.

% PSD fit for simulated data for J1653+3945 sampling and noise
%\clearpage

\paragraph*{OVRO sampling pattern 2 and noise}

\begin{figure}
\begin{center}
\includegraphics[width=8cm, trim=40 0 40 0]{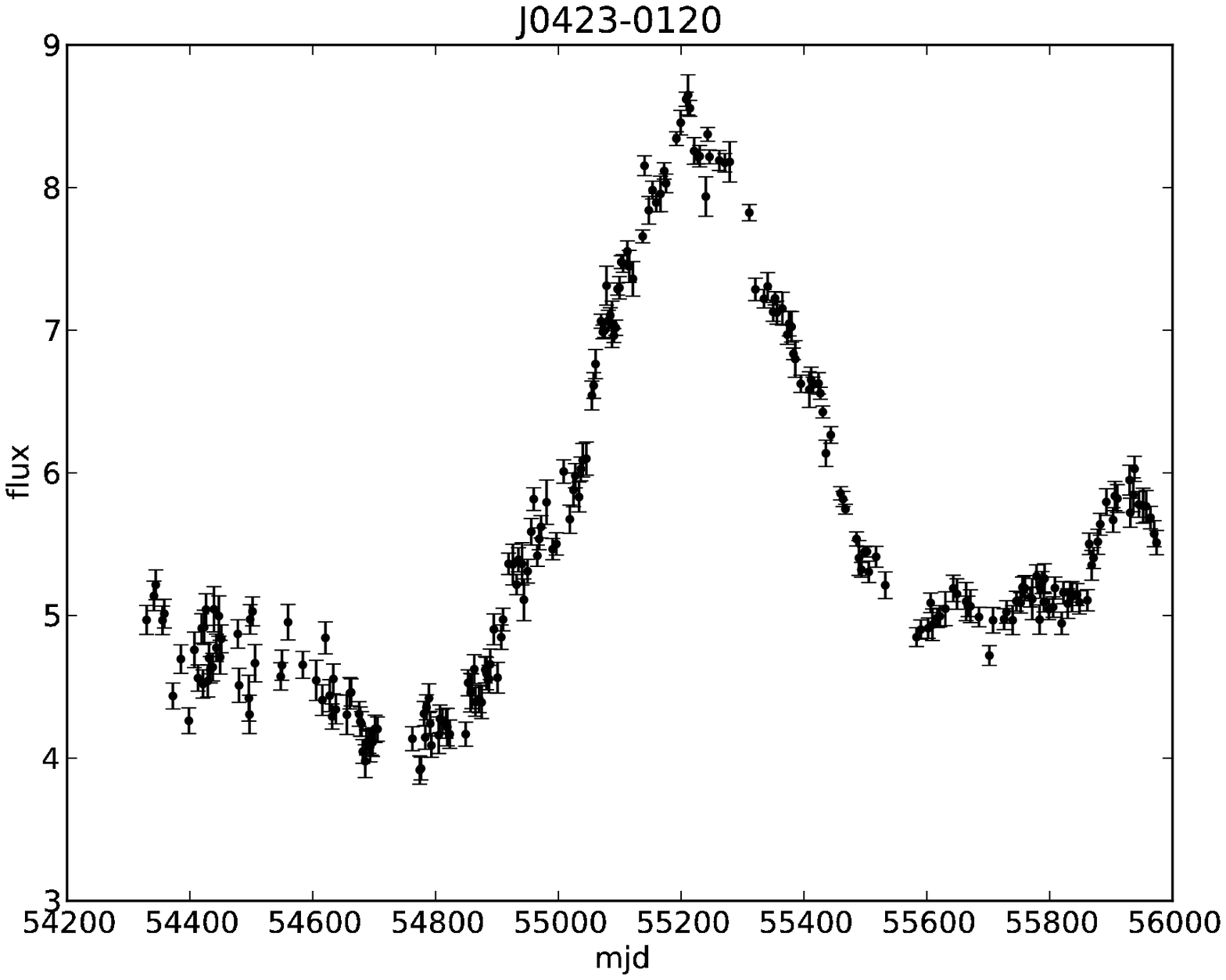}\\
\includegraphics[width=4cm, trim=30 5 20 0]{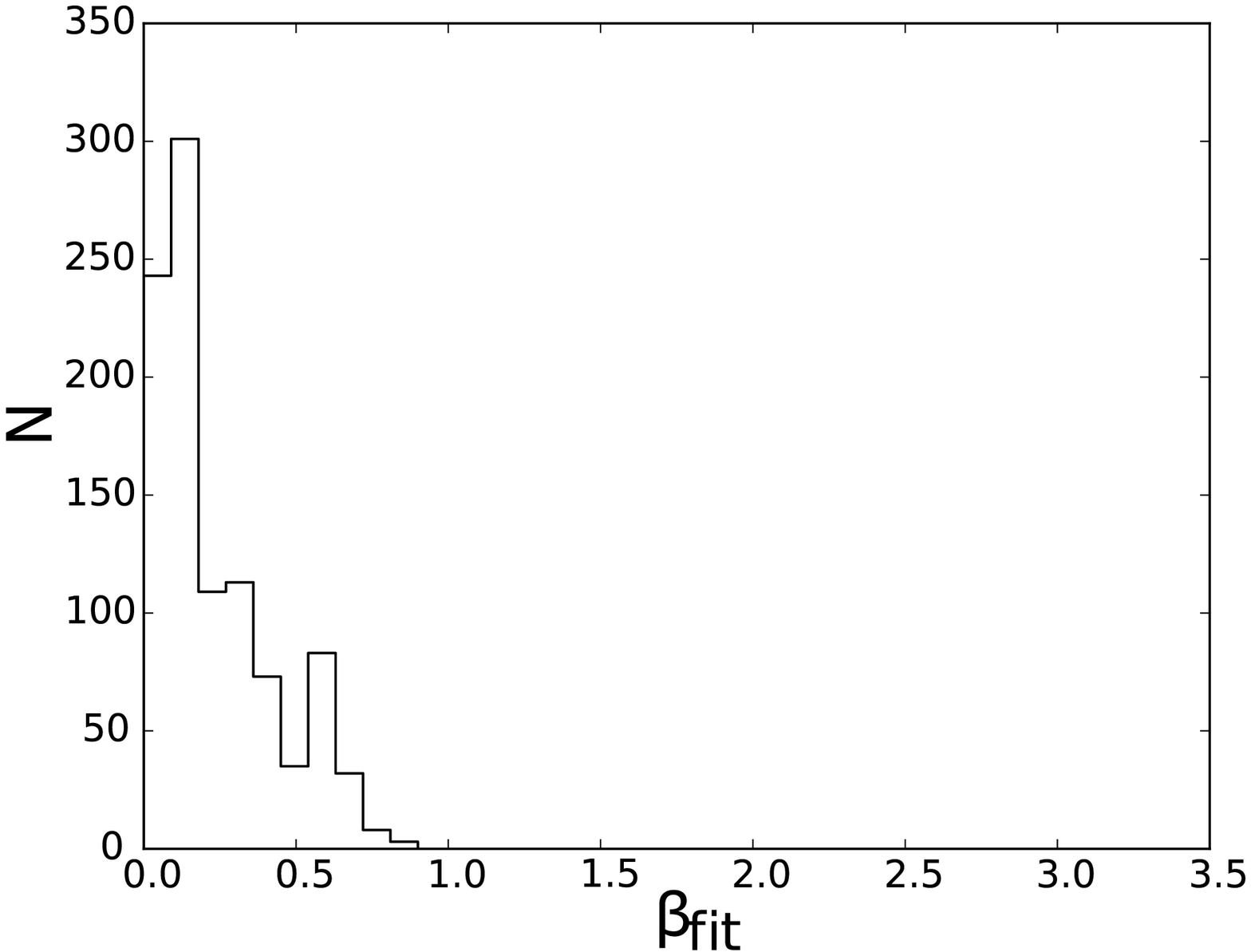}
\includegraphics[width=4cm, trim=30 5 20 0]{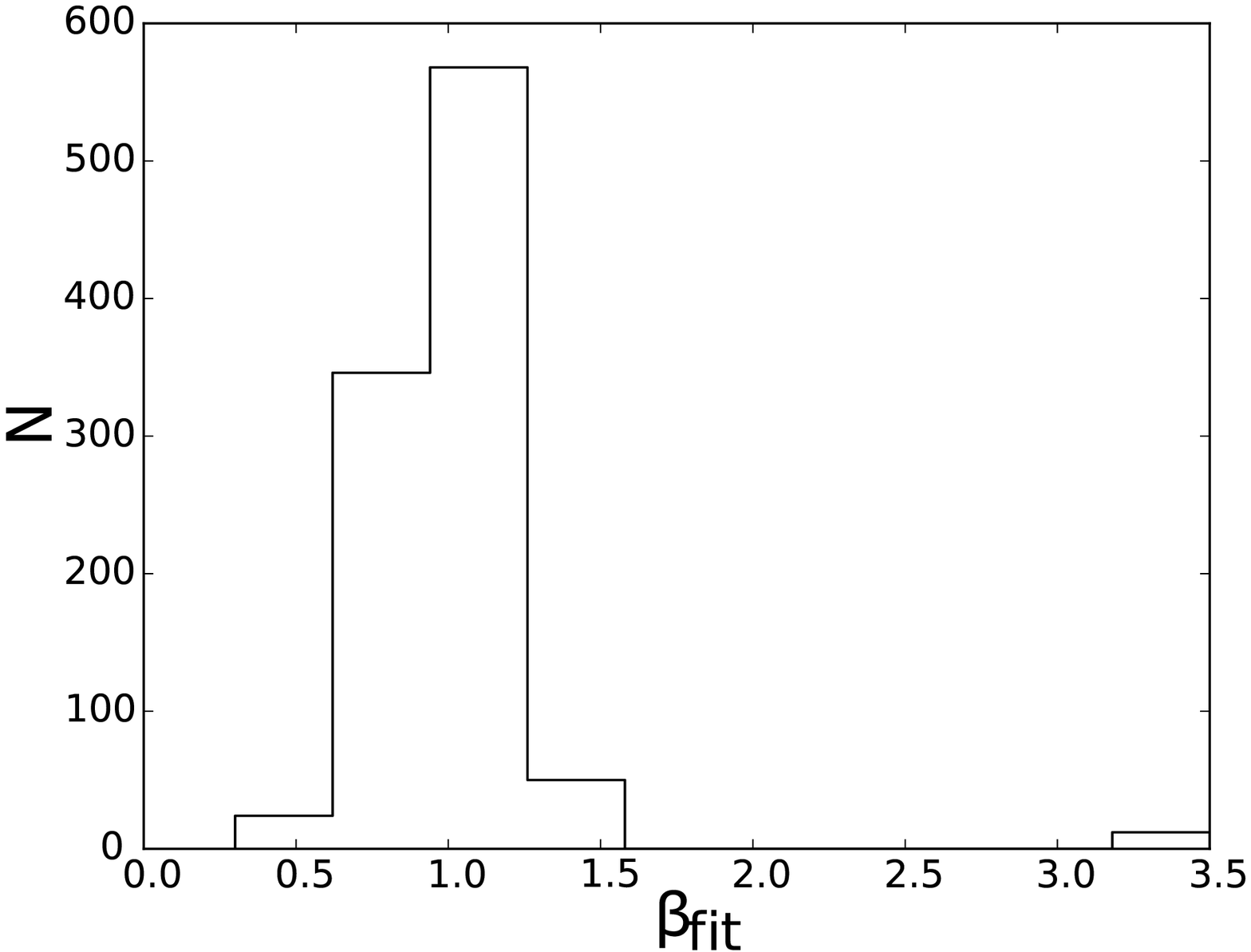}\\
\includegraphics[width=4cm, trim=30 5 20 0]{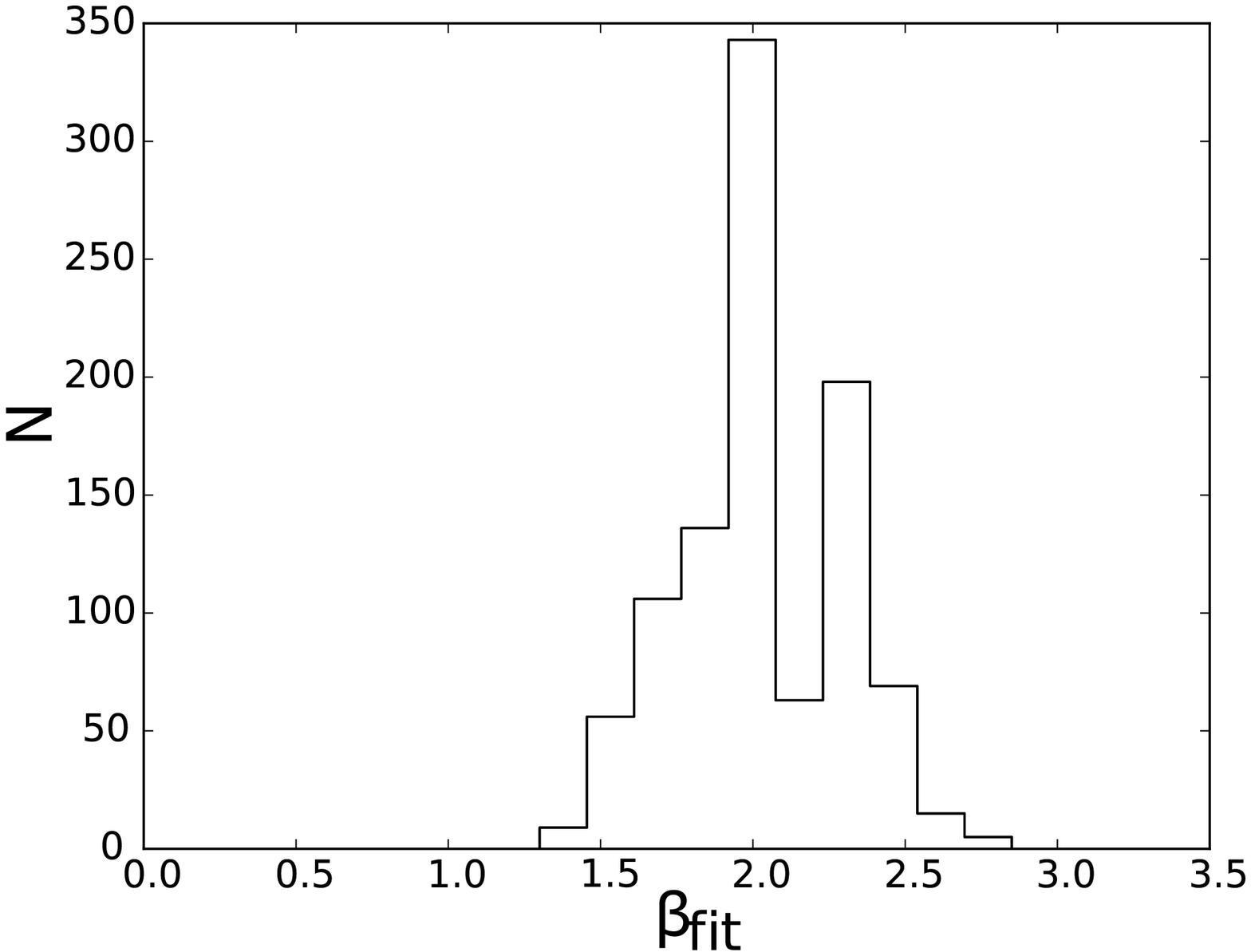}
\includegraphics[width=4cm, trim=30 5 20 0]{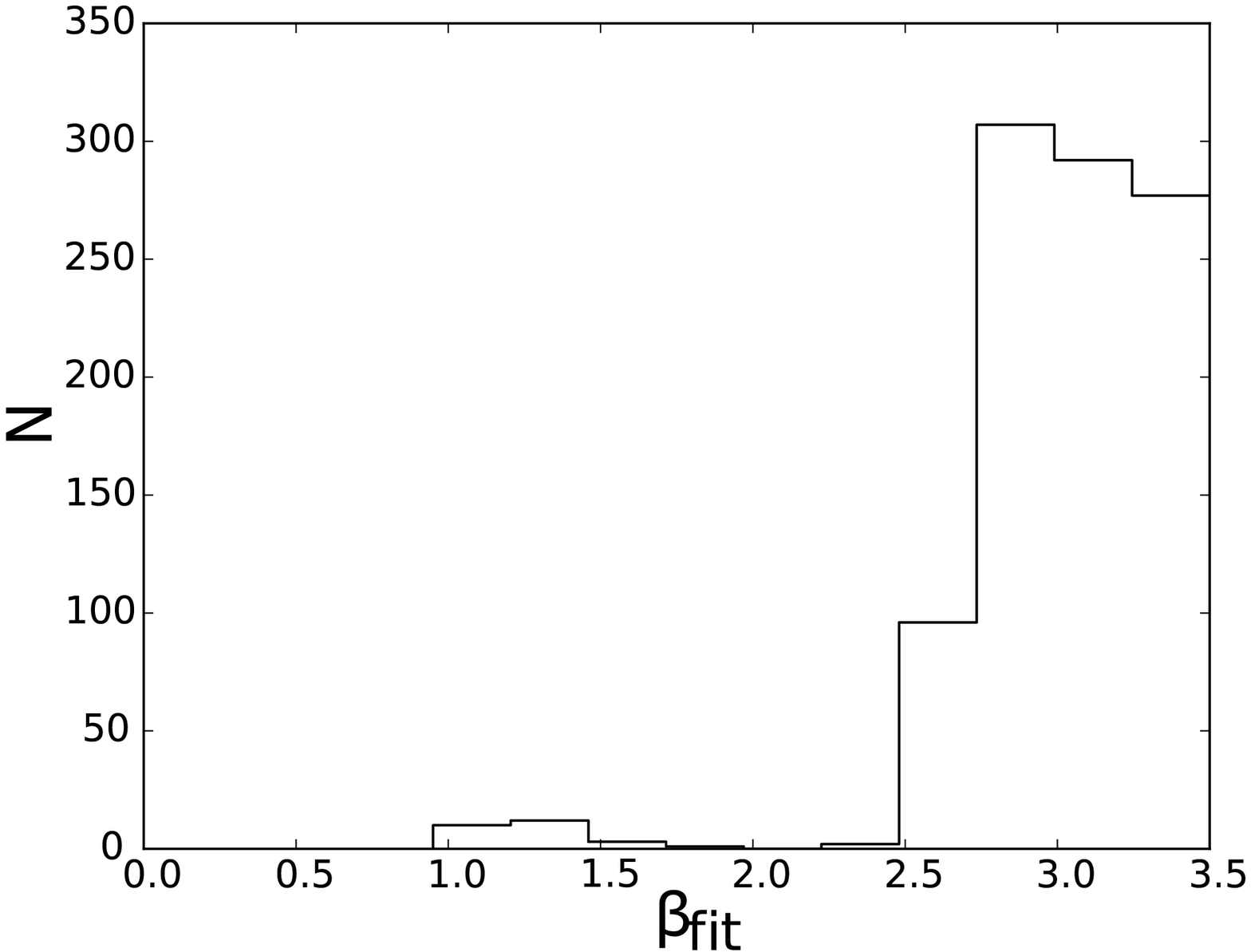}
\caption{Upper panel shows the OVRO data used to get the time sampling and the flux density uncertainties. Four lower panels are the distribution of best fit values for 1000 simulated light curves in each case. In each case this distribution gives an estimation of the error on the fit and is used to construct the confidence band. Top left is for $\beta_{\rm sim}  = 0.0$ and $\beta_{\rm fit} = 0.15^{+0.25}_{-0.1}$, top right is for $\beta_{\rm sim}  = 1.0$ and $\beta_{\rm fit} = 1.0 \pm 0.15$, lower left is for $\beta_{\rm sim}  = 2.0$ and $\beta_{\rm fit} = 2.0 \pm 0.25$, and lower right is for $\beta_{\rm sim}  = 3.0$ and $\beta_{\rm fit} = 3.05 \pm 0.3$. In the case of $\beta_{\rm sim} = 0.0$ we report the mode and dispersion about that value. All the other cases use the median and dispersion.}
\label{val_ex_5_fit}
\end{center}
\end{figure}

In this test we use the sampling pattern for the source J0423$-$0120 and error bars consistent with the noise in this source. The OVRO data are shown in Figure \ref{val_ex_5_fit}. The results of the fit for simulated data as a distribution of best fit values are shown in Figure \ref{val_ex_5_fit}. 

In this case the procedure also provides good constraints on $\beta$ except for the case of $\beta = 3.0$. If necessary this could be handled by the use of a different window function.

% PSD fit for simulated data for J0423-0120 sampling and noise
%\clearpage

\paragraph*{OVRO sampling pattern 3 and noise}

\begin{figure}
\begin{center}
\includegraphics[width=8cm, trim=40 0 40 0]{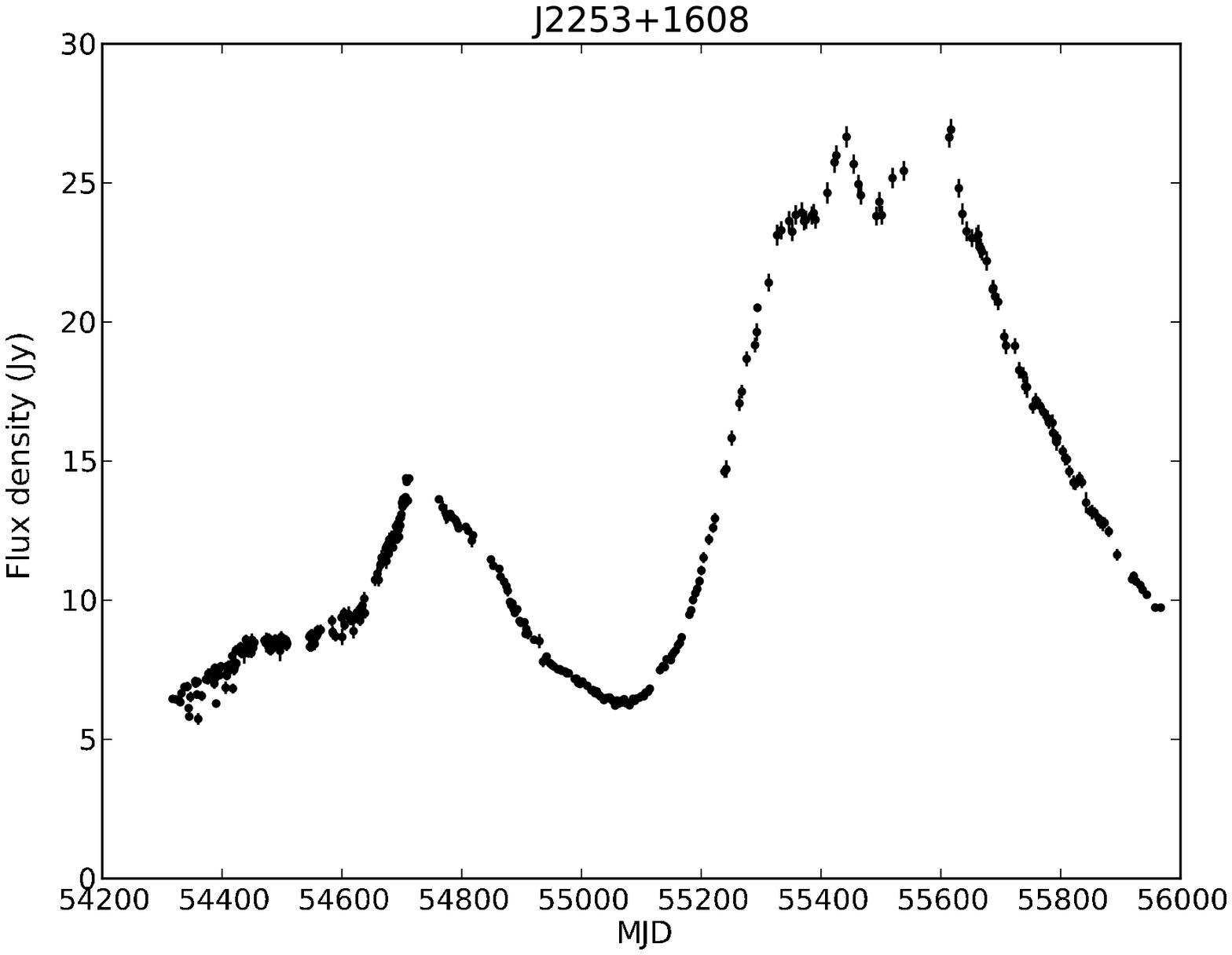}\\
\includegraphics[width=4cm, trim=30 5 20 0]{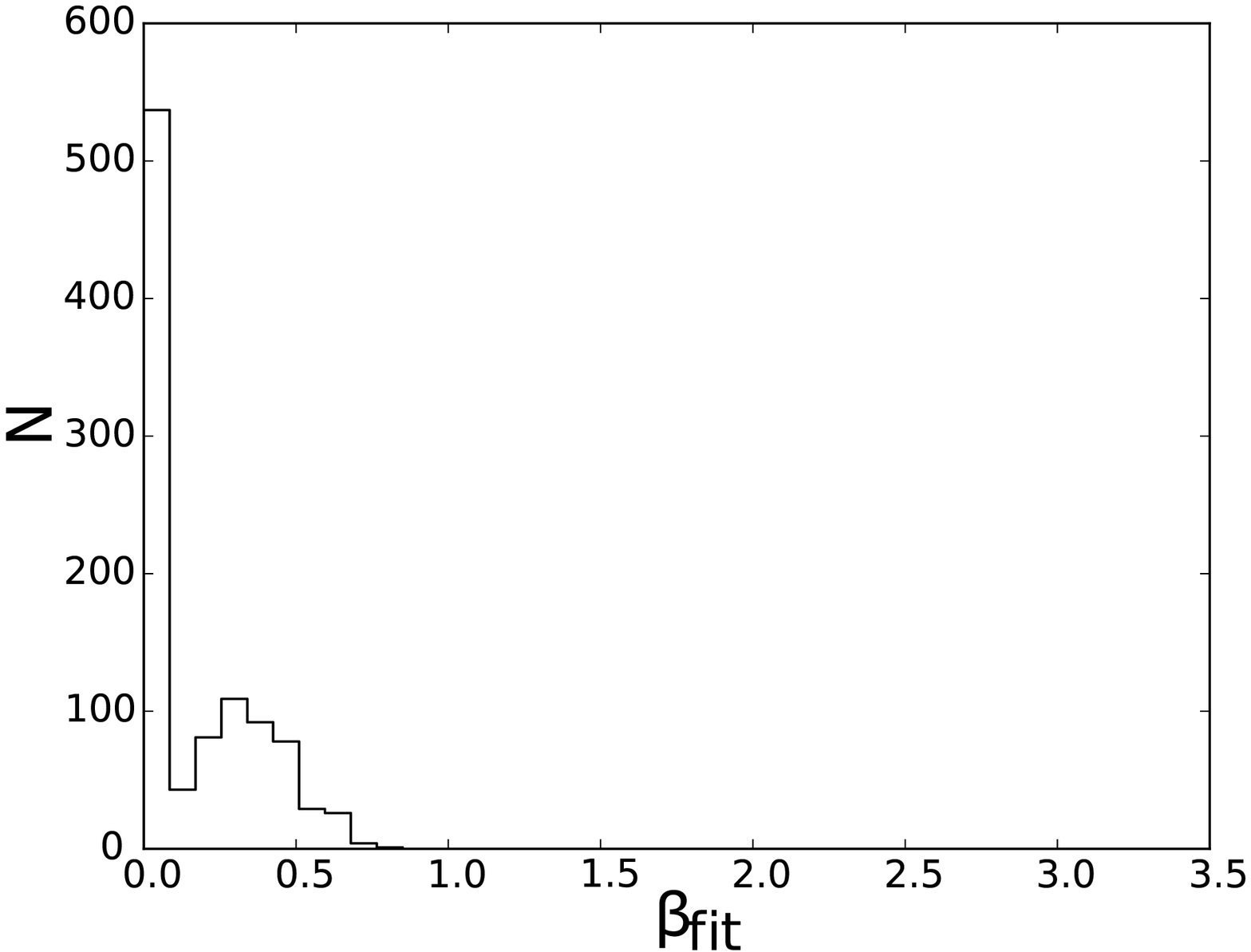}
\includegraphics[width=4cm, trim=30 5 20 0]{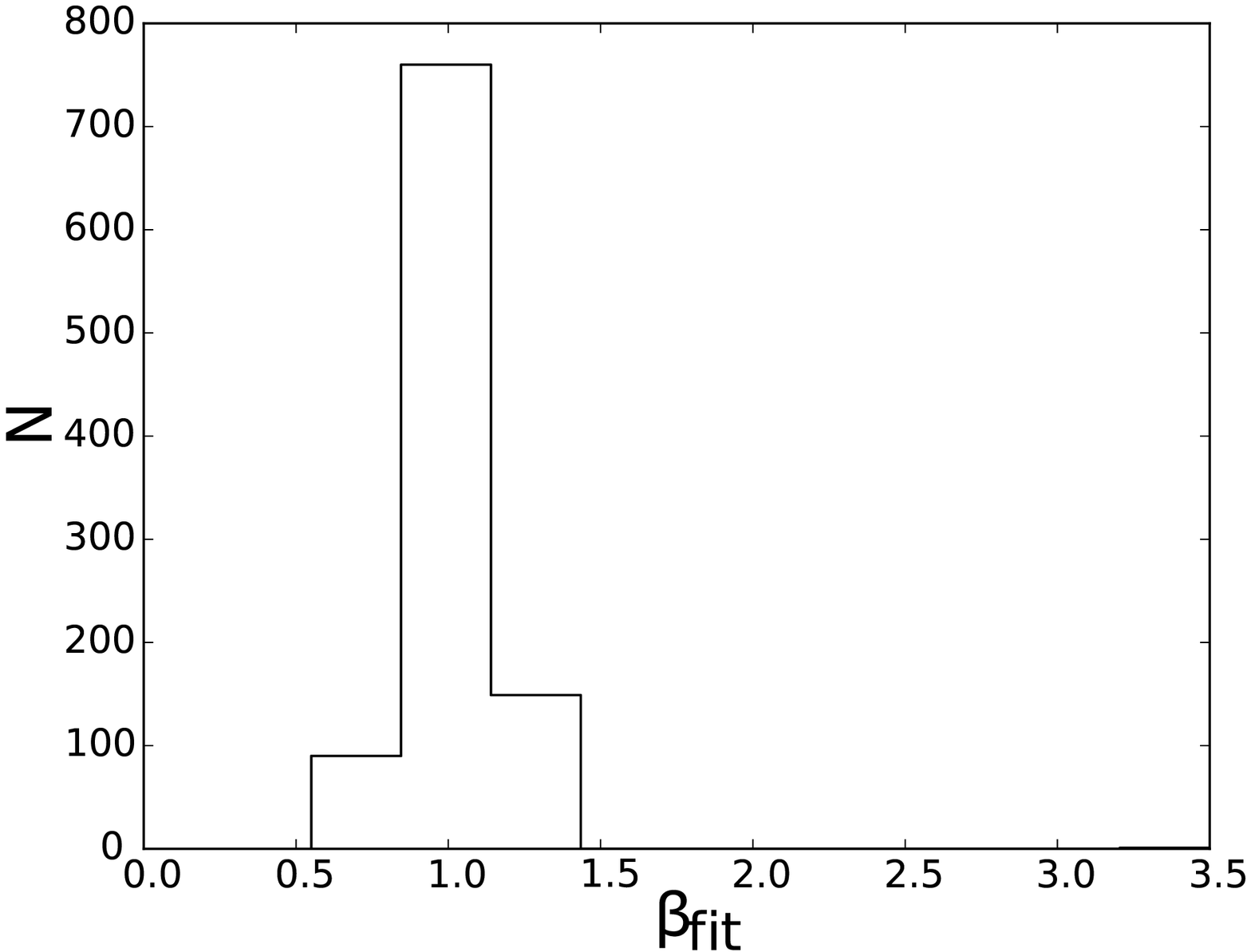}\\
\includegraphics[width=4cm, trim=30 5 20 0]{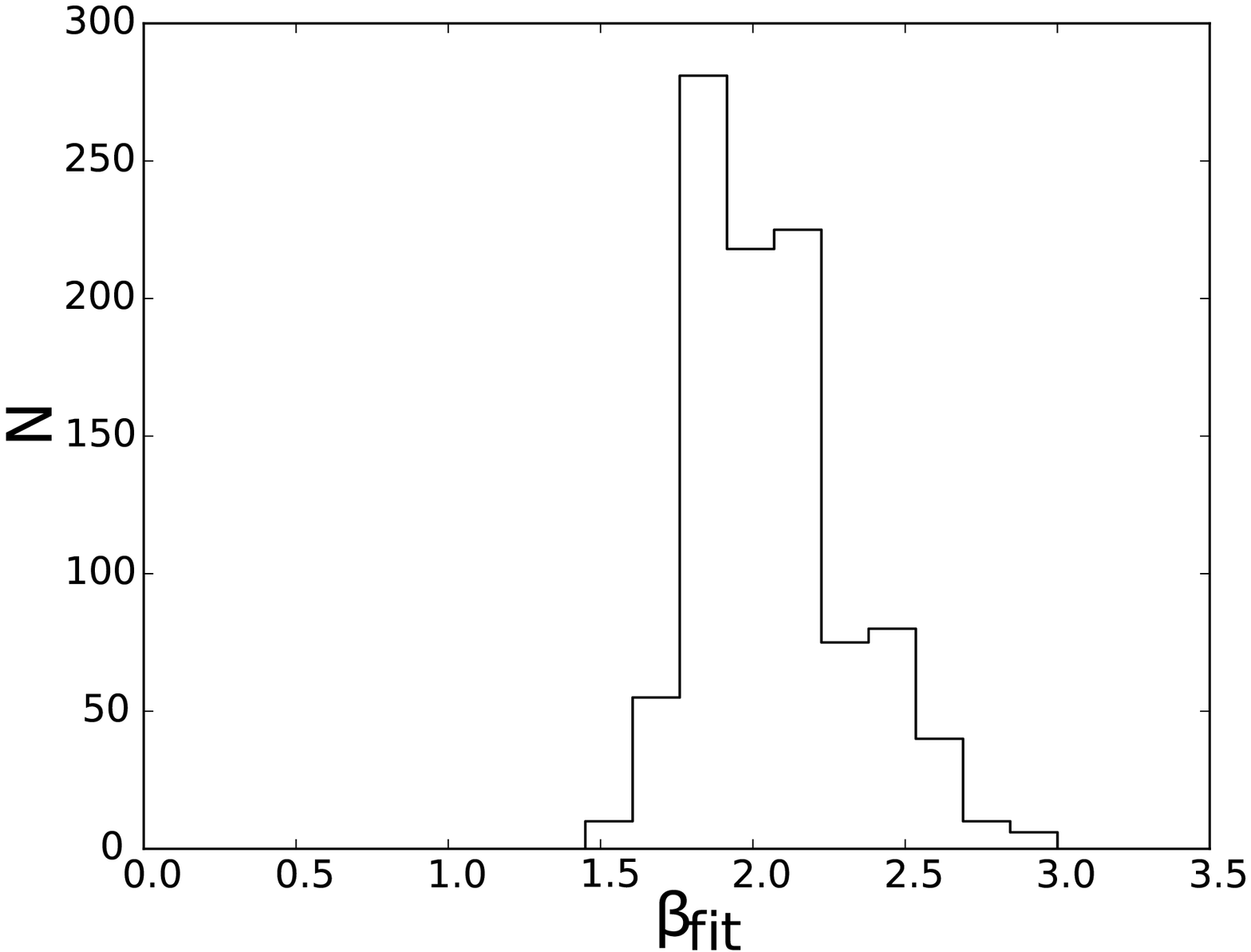}
\includegraphics[width=4cm, trim=30 5 20 0]{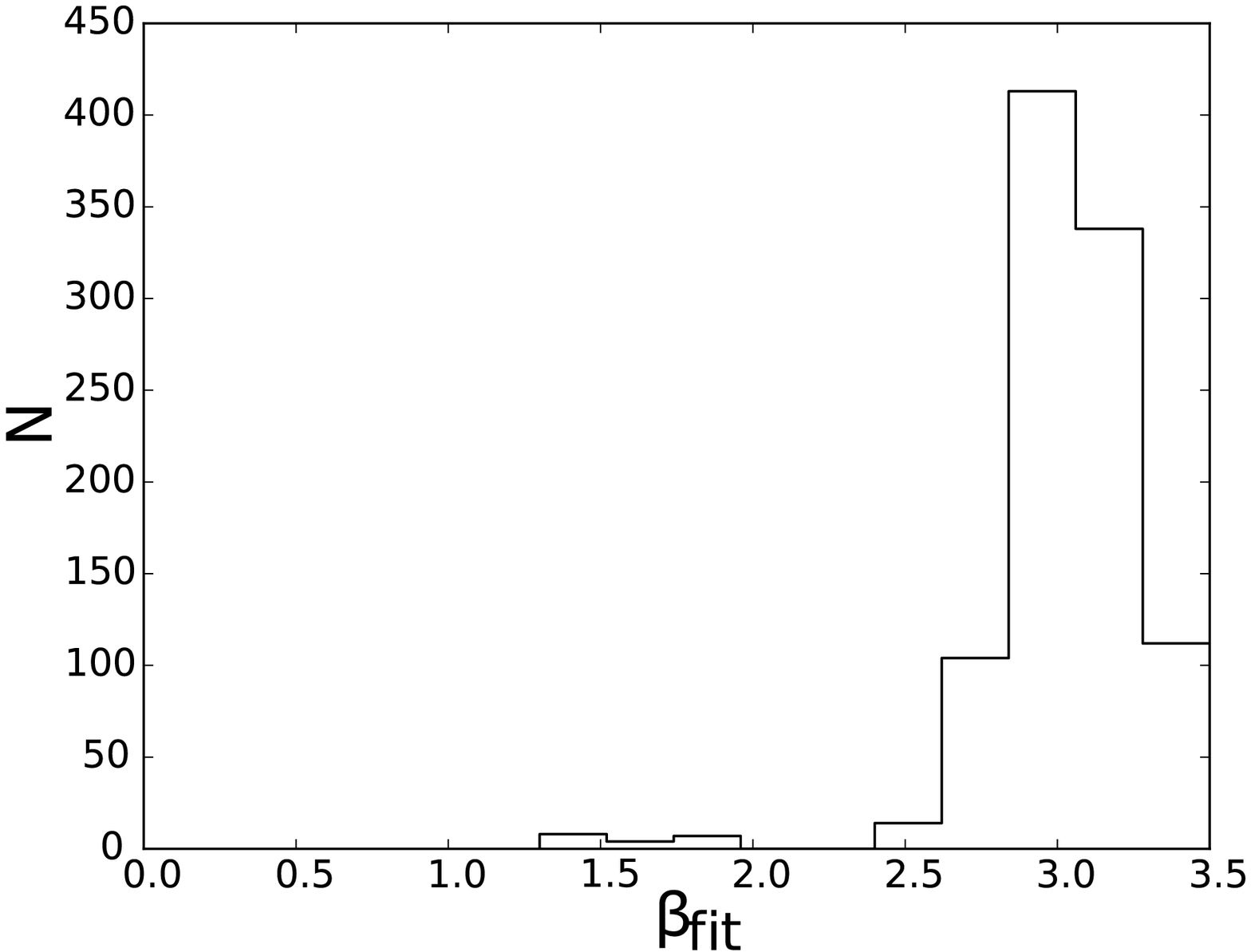}
\caption{Upper panel shows the OVRO data used to get the time sampling and the flux density uncertainties. Four lower panels are the distribution of best fit values for 1000 simulated light curves in each case. In each case this distribution gives an estimation of the error on the fit and is used to construct the confidence band. Top left is for $\beta_{\rm sim}  = 0.0$ and $\beta_{\rm fit} = 0.05^{+0.25}_{-0.05}$, top right is for $\beta_{\rm sim}  = 1.0$ and $\beta_{\rm fit} = 1.0^{+0.15}_{-0.1}$, lower left is for $\beta_{\rm sim}  = 2.0$ and $\beta_{\rm fit} = 2.05 \pm 0.25$, and lower right is for $\beta_{\rm sim}  = 3.0$ and $\beta_{\rm fit} = 3.0^{+0.2}_{-0.15}$. In the case of $\beta_{\rm sim} = 0.0$ we report the mode and dispersion about that value. All the other cases use the median and the 15.86 and 84.15 percentiles to measure the $1\sigma$ dispersion.}
\label{val_ex_7_fit}
\end{center}
\end{figure}

In this test we use the sampling pattern for the source J2253+1608  and error bars consistent with the noise in this source. The OVRO data are shown in Figure \ref{val_ex_7_fit}. The results of the fit for simulated data as a distribution of best fit values are shown in Figure \ref{val_ex_7_fit}. In this last case we are also able to constrain $\beta$ with an uncertainty of about 0.2.

% PSD fit for simulated data for J2253+1608 sampling and no noise
%\clearpage

\subsubsection{Effect of increasing the number of simulations \label{varying_N_and_repetability}}

Here we study the effect of varying the number of simulated light curves ($M$) on the repeatability of the results. We then establish a criterion to select $M$ for the data analysis and to get an idea of possible errors associated with that choice. We test this by fitting the same simulated data set used in Section \ref{example_of_psd_method}, 100 times using $M$=100, $M$=1,000 and $M$=10,000 simulated light curves at each trial power-law exponent of the PSD. The distribution of best fit values is used to estimate the repeatability of the fitting process. The second test does the same but in this case it fits the OVRO data for J0423$-$0120 shown in Figure \ref{val_ex_5_fit}, and this time incorporating the observational noise in the fit. The results are summarized in Table \ref{repeatability_table} which shows that the repeatability of the results increases as $M$ increases, as we would expect. The scatter is reduced by half when going from $M$=100 to $M$=10,000. We also note that in the case of the OVRO data we get a big increase (a factor of 1.9) in accuracy when going from $M$=100 to $M$=1,000, but a much smaller one (a factor of 1.2) by going to $M$=10,000. This shows that exceeding $M$=1,000 is not necessary for these data and can save a significant amount of time when studying large samples of sources. Similar tests can be performed for other data sets to determine the required number of simulations.
 
\begin{table}
\caption{Repeatability of fitted parameters as a function of number of simulated light curves}
\begin{center}
\begin{tabular}{c c c c}
\hline
Test & $\beta$ & $\beta$ & $\beta$ \\
 & $M$=100 & $M$=1,000 & $M$=10,000 \\ 
 \hline
Simulated, known PSD & $1.85 \pm 0.08$ & $1.85 \pm 0.05$ & $1.86 \pm 0.03$ \\
OVRO data with noise & $2.27 \pm 0.13$ & $2.30 \pm 0.07$ & $ 2.32 \pm 0.06$ \\
\hline
\end{tabular}
\end{center}
\label{repeatability_table}
\end{table}

%--------------------------------------------------------------------------------------------------------------------------------
% Significance of cross-correlations
%--------------------------------------------------------------------------------------------------------------------------------
\section{Significance of cross-correlations between two wavebands}
\label{cross_corr_method}

Here we deal with the statistical problem of quantifying the significance of the cross-correlation between two time series, in the case of uneven sampling and non-uniform measurement errors. The two time series are assumed to contain no upper or lower limits.

\subsection{The estimation of the cross-correlation function \label{crosscorr_alt}}

Our basic data sets are two time series we call A and B. These time series are time ordered sequences of triplets $(t_{ai}, a_i, \sigma_{ai})$ with $i=1,...,N$ and $(t_{bj}, b_j, \sigma_{bj})$ with $j=1,...,P$. In both cases $t_{xi}$ is the observation time, $x_i$ is the measured value of a quantity of interest (e.g. flux density, photon flux, etc.) and $\sigma_{xi}$ an estimate of the observational error associated with the measurement. 

Since the time interval between successive samples is not uniform and the A and B time series are not sampled simultaneously, we need to resort to some kind of time binning in order to measure the cross-correlation. The cross-correlation between two unevenly sampled time series can be measured using a number of different approaches. The usual approach is to generalize a standard method and use time binning to deal with the uneven sampling. Here we consider two methods that are commonly encountered in the literature: the discrete correlation function \citep{edelson_1988} and the local cross-correlation function \citep[e.g.,][]{welsh_1999}. A number of other alternatives have been used to handle the problem of measuring the correlation between unevenly sampled time series. Among them are the interpolated cross-correlation function \citep[ICCF;][]{gaskell+1987}, inverse Fourier transform of the cross-spectrum \citep[][]{scargle1989} and the $z$-transformed cross-correlation function \citep[][]{alexander1997}. We do not explore these alternative methods in this work. These methods provide a way of estimating the cross-correlation coefficients, but do not provide an estimate of the associated statistical significance, which is discussed in Section \ref{crosscorr_sig}.

The two most commonly found alternatives are presented below.

\subsubsection{The Discrete Correlation Function (DCF)}

The discrete correlation function was proposed by \citet{edelson_1988} and developed in the context of reverberation mapping studies. For two time series $a_i$ and $b_j$, we first calculate the unbinned discrete correlation for each of the pairs formed by taking one data point from each time series
\begin{equation}
{\rm UDCF}_{ij} = \frac{(a_i - \bar{a}) (b_j - \bar{b})}{\sigma_a \sigma_b}
\end{equation}
where $\bar{a}$ and $\bar{b}$ are the mean values for the time series, and $\sigma_a$ and $\sigma_b$ are the corresponding standard deviations. This particular value, ${\rm UDCF}_{ij}$, is associated with a time lag of $\Delta t_{ij} = t_{bj} - t_{ai}$. The discrete cross-correlation is estimated within independent time bins of width $\Delta t$, by averaging the unbinned values within each bin,
\begin{equation}
\label{dcf_mean_value}
{\rm DCF}(\tau) = \frac{1}{M} \sum {\rm UDCF}_{ij}
\end{equation}

The uncertainty in the binned discrete cross-correlation is given by the scatter in the unbinned values for each time bin, and is given by
\begin{equation}
\label{dcf_error}
\sigma_{\rm DCF}(\tau) = \frac{1}{M - 1} \left ( \sum [{\rm UDCF}_{ij} - {\rm DCF}(\tau)]^2 \right )^{1/2}
\end{equation}
In the expressions above the sum is over the $M$ pairs for which $\tau \leq \Delta t_{ij} < \tau + \Delta t$, where $\tau$ is the time lag, and all the bins have at least two data points in order to get a well-defined error. In practice it is recommended to choose $M$ much larger than 2 to reduce the effect of statistical fluctuations.

In this case, the mean and standard deviation use all the data points in a given time series, but the DCF for a given time lag only includes overlapping samples. This particular choice for normalization produces values of the DCF which are not restricted to the usual $[-1, 1]$ interval of standard correlation statistics. This immediately challenges the interpretation of the amplitude of the DCF as a valid measure of the cross-correlation and invalidates the use of standard statistical tests developed for other correlation statistics, forcing us to find alternative ways to estimate the significance of correlations. A modification that corrects this normalization problem but not the significance evaluation issue is described below.

\subsubsection{The Local Cross-Correlation Function (LCCF) \label{LCCF}}

Motivated by the normalization problems presented by the DCF, some authors have proposed a different prescription \citep[e.g.,][]{welsh_1999}. In this case, we only consider the samples that overlap with a certain coarse grain of the time delays, which is equivalent to the width of time bins $\Delta t$. Hence, we have
\begin{equation}
{\rm LCCF}(\tau) = \frac{1}{M} \frac{\sum (a_i - \bar{a}_{\tau}) (b_j - \bar{b}_{\tau})}
{{\sigma_a}_{\tau} {\sigma_b}_{\tau}}
\end{equation}
where the sum is over the $M$ pairs of indices $(i, j)$ such that  $\tau \leq \Delta t_{ij} < \tau + \Delta t$. The averages ($a_{\tau}$ and $b_{\tau}$) and standard deviations ($\sigma_{a\tau}$ and $\sigma_{b\tau}$) are also over the $M$ overlapping samples only.

The main motivation for using this expression instead of the DCF is that we recover cross-correlation coefficients that are bound to the $[-1, 1]$ interval. This latter property is a result of using only the overlapping samples to compute the means and standard deviations, which in effect reduces the problem to a standard cross-correlation bounded to $[-1, 1]$, as a consequence of the Cauchy-Schwarz inequality. Additionally, \citet[][]{welsh_1999} shows that the LCCF can determine time lags more accurately than the DCF in simulated data sets. These are certainly desirable properties, but as explained in Section \ref{crosscorr_sig}, they do not solve the estimation of significance problem.

\subsubsection{Relation between the DCF and LCCF}

In Section \ref{compare_methods} we perform a series of tests designed to help us compare the detection efficiency of the DCF and LCCF. In looking at these results, it is useful to consider the relation between those two correlation measures.

From our previous discussions, we can see that the only difference between the DCF and LCCF is in the values used for the means and standard deviations. In the case of the DCF, the mean and standard deviation are calculated from the complete time series ($\bar{a}, \bar{b}$ for the means and $\sigma_{a}, \sigma_{b}$ for the standard deviations), while for the LCCF only the overlapping samples at each time lag are used ($\bar{a_\tau}, \bar{b_\tau}$ for the means and $\sigma_{a \tau}, \sigma_{b \tau}$ for the standard deviations). It can be shown that the two are related at a given time lag by
\begin{equation}
{\rm DCF}(\tau) = {\rm LCCF}(\tau) \frac{\sigma_{a \tau} \sigma_{b \tau}}{\sigma_{a} \sigma_{b}} +                        \frac{(\bar{a_\tau} - \bar{a}) (\bar{b_\tau} - \bar{b})}{\sigma_{a} \sigma_{b}}.
\label{DCF_LCCF_linrel}
\end{equation}
This linear relation has coefficients that depend on the sampling pattern and the overlap between the two time series at different time lags. For long stationary time series, the means and variances of the overlapping and complete time series will be identical and the DCF will equal the LCCF. For short or non-stationary time series, the coefficients will make the DCF different from the LCCF. 

Deviations of the multiplicative coefficient $(\sigma_{a \tau} \sigma_{b \tau}) / (\sigma_{a} \sigma_{b})$ from 1 change the amplitude of the DCF, while deviations of the additive coefficient $((\bar{a_\tau} - \bar{a})(\bar{b_\tau} - \bar{b})) / (\sigma_{a} \sigma_{b})$ from 0 change the zero-point of the DCF. The combination of these variations explains why the DCF is not bounded to the $[-1, 1]$ interval as is the LCCF, and can also explain why they have different detection efficiencies.

\subsubsection{Estimation of the uncertainty in the location of the cross-correlation peak}

The standard method used by the reverberation mapping community \citep[][]{peterson+1998}, uses bootstrapping and randomization to generate slightly modified versions of the original data set, in order to quantify the uncertainty in the location of the cross-correlation peak. A modified data set is constructed by the application of two procedures. The first is ``random subset selection'', in which a bootstrapped light curve is constructed by randomly selecting with replacement samples from the original time series. In the second, we perturb the selected flux measurements by ``flux randomization'', in which normally distributed noise with a variance equal to the measured variance is added to the measured fluxes. Each of these modified data sets is cross-correlated using the method of choice and a value for the cross-correlation peak of interest is measured. By repeating this for many randomized data sets, a distribution of measured time lags for the cross-correlation peaks is obtained. This distribution is used to construct a confidence interval for the position of the peak.

\subsubsection{Light curve detrending}

There has been some discussion in the literature about the effects of detrending the light curves in order to improve the accuracy of the time lag estimates. \citet[][]{welsh_1999} strongly recommended removing at least a linear trend from the light curves. His results are based on simulations with even sampling and do not directly apply to uneven sampling as shown by \citet[][]{peterson+2004}. They find that detrending does not improve accuracy in unevenly sampled datasets, and produces large errors in some cases. Based on that finding, we have decided not to detrend our light curves. 

We emphasize that care must be taken when correlating time series where long term trends are present, as these are guaranteed to produce large values of the cross-correlation coefficient. Our studies are mostly concerned with the correlation between periods of high activity in different energy bands for light curves that appear to have a detectable ``quiescent" level. This is generally true for gamma-ray light curves, but is not always true for radio light curves. Radio light curves showing a single dominant increasing or decreasing linear trend should be analyzed with care, as they can produce spurious correlations. In our opinion, the best remedy for those cases is to collect longer light curves.

\subsection{The estimation of  the significance \label{crosscorr_sig}}

\begin{figure*}
\includegraphics[angle=0,width=14.0cm, trim=0 30 0 0]{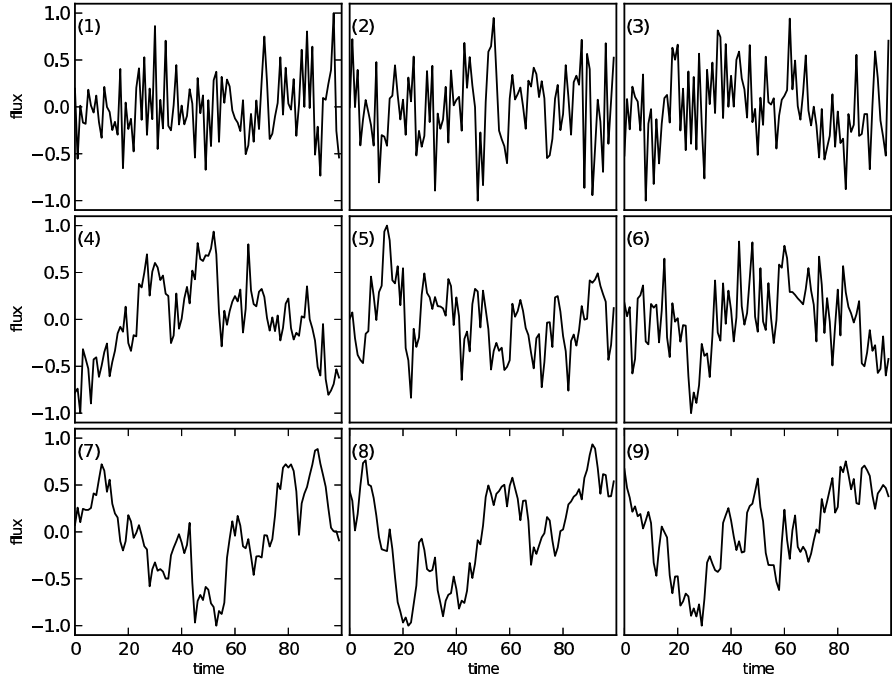}
\caption{Illustration of the time domain characteristic of simulated light curves with different power-law PSD. In all panels the horizontal axis is time and the vertical one is amplitude,  both in arbitrary units. Top panels 1, 2 and 3 for PSD $\propto 1/\nu^0$, central panels 4, 5 and 6 for $\propto 1/\nu^1$ and lower panels 7, 8 and 9 for $\propto 1/\nu^2$. The light curves with steeper PSD show more flare-like features that can induce high values of the cross-correlation coefficient as shown in Figure \ref{example_xcorr_simulated_light_curves}.}
\label{example_simulated_light_curves}
\end{figure*}

A complete quantification of the cross-correlation needs an estimate of its statistical significance. In our case, we need to consider the intrinsic correlation between adjacent samples of a given time series, which are produced by the presence of flare-like features; a distinctive characteristic of blazar light curves. This behavior can be modeled statistically by red-noise stochastic processes (e.g., \citet[][]{hufnagel+1992} in the radio and optical, \citet[][]{lawrence_1993} in the X-rays, and \citet[][]{abdo_variability_2010} in gamma-rays). Red-noise processes are characterized by their PSD, show variability at all time scales, and appear as time series in which flare-like features are a common phenomenon. The frequent appearance of flares means that high correlation coefficients between any two energy bands are to be expected, even in the absence of any relation between the processes responsible for their production. To illustrate this point Figure \ref{example_simulated_light_curves} shows simulated light curves with power-law power spectral densities (PSD $\propto 1/\nu^{\beta}$).

% Example light curves

In fact, every time we cross-correlate two time series, each of which has a flare, we will get  a peak in the cross-correlation at some time lag. Then quantifying the chances of such peak being just a random occurrence is of critical importance. The problem is further complicated by the uneven sampling and non-uniform errors, so the only feasible method is to use Monte Carlo simulations. Standard methods are not suitable for this analysis, as they assume that the individual data points are uncorrelated. The effect of ignoring the correlations will lead to an overestimate of the significance of the cross-correlations and to an erroneous physical interpretation.

\begin{figure*}
\includegraphics[angle=0, width=14.0cm, trim=0 20 0 0]{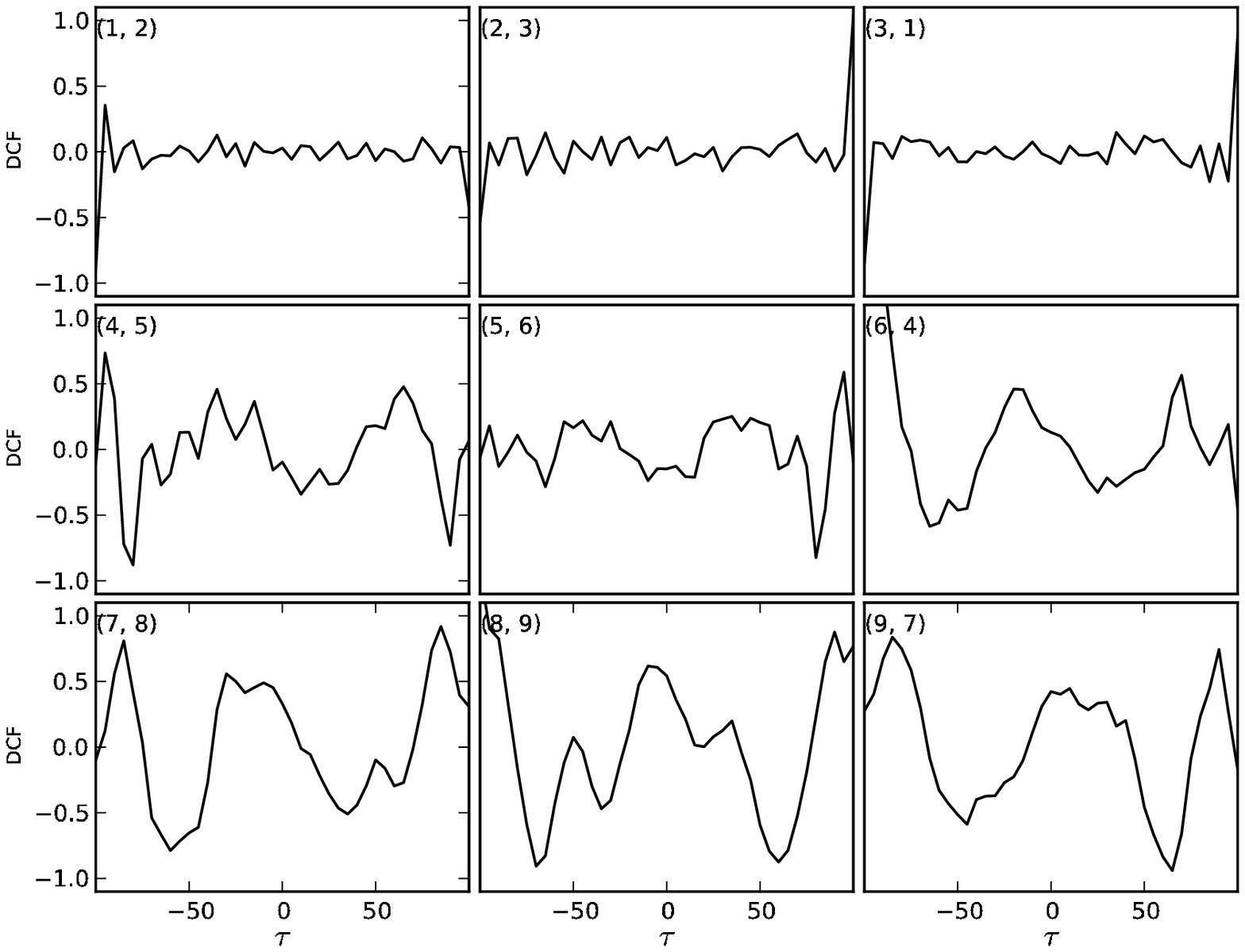}
\includegraphics[angle=0,width=14.0cm, trim=0 30 0 0]{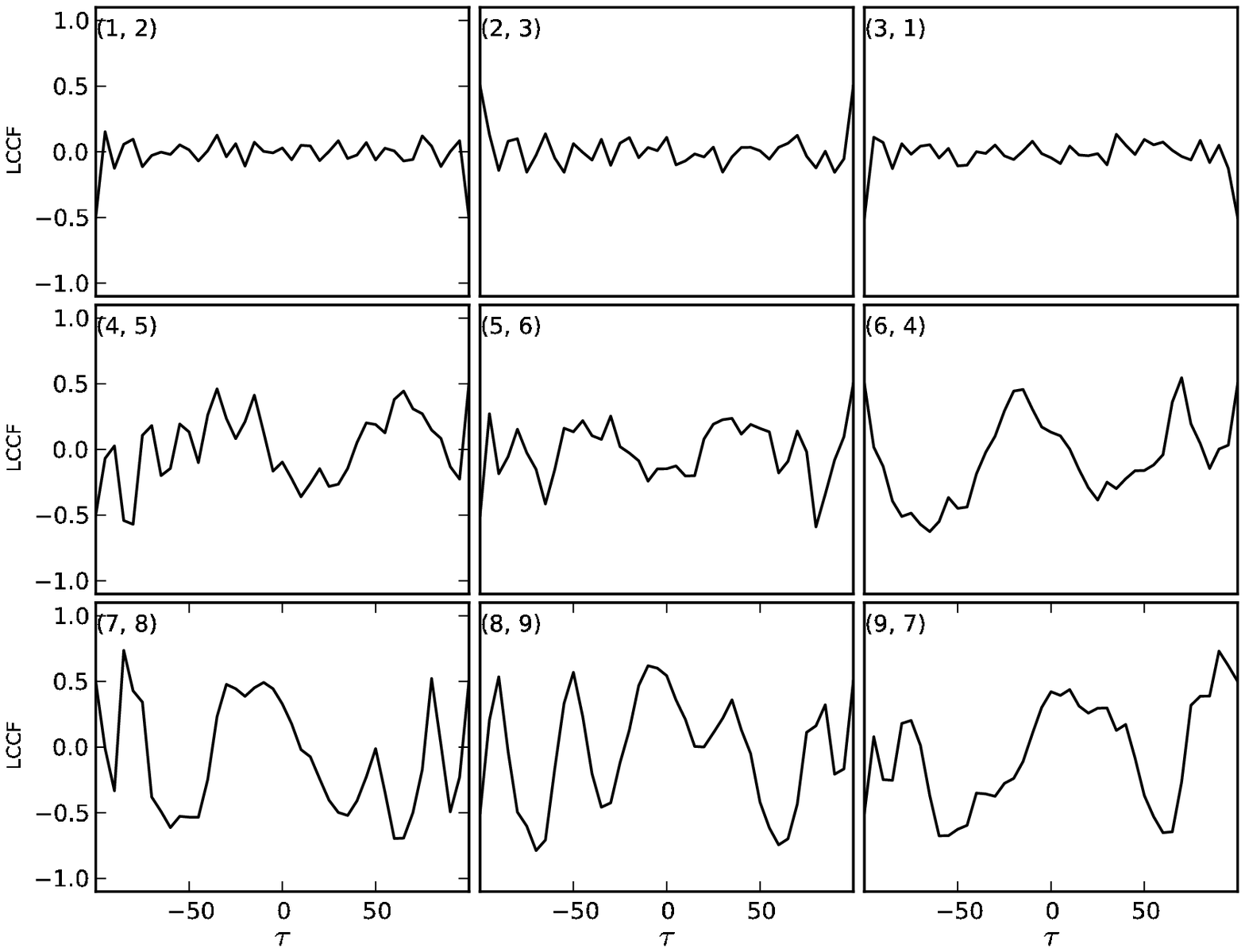}
\caption{Examples of the cross-correlation of simulated light curves shown in Figure \ref{example_simulated_light_curves} using the DCF (upper figure) and LCCF (lower figure). In all panels the horizontal axis is time lag in arbitrary units and the vertical one is the amplitude of the cross-correlation. Upper panels, cross-correlation of independent $\beta=0.0$ light curves. Central panels, cross-correlation of independent $\beta=1.0$ light curves. Lower panels, cross-correlation of independent $\beta=2.0$ light curves. The pair of numbers on the upper left corner of each panel are the light curve numbers from Figure \ref{example_simulated_light_curves} which are correlated in each case. The light curve pairs have been simulated independently and yet show large peaks in the discrete cross-correlation function for the cases of $\beta=1.0$ and $2.0$. The existence and amplitude of peaks in the cross-correlation appears to increase for steeper power spectral densities, independently of the method used.}
\label{example_xcorr_simulated_light_curves}
\end{figure*}

In Figure \ref{example_xcorr_simulated_light_curves}, we show the results of cross-correlating the independently simulated light curves from Figure \ref{example_simulated_light_curves}, which have different values of the power-law exponent for the PSD. It can be seen that correlating light curves with steep PSD, which show frequent flare-like features, can result in high cross-correlation coefficients that have nothing to do with a physical relation between the light curve pairs. The results illustrate how common it is to get high cross-correlations for unrelated light curves with steep PSDs and the dangers of interpreting them as signs of a physical connection. Standard statistical tests that assume uncorrelated data are equivalent to the case of white noise time series (PSD $\propto 1/\nu^0$), which is illustrated in the upper panels of Figure \ref{example_xcorr_simulated_light_curves}. Since blazar light curves are more similar to simulated light curves with steep PSDs, it is easy to see how misleading it is to use statistical tests that ignore the long term correlations in the individual time series.

% Example cross-correlation of simulated light curves with DCF and LCCF

\subsubsection{Monte Carlo procedure for the estimation of the significance}

To estimate the significance of the cross-correlation coefficients, we use a Monte Carlo method to estimate the distribution of random cross-correlations, that uses simulated time series with statistical properties similar to the observations. These and related ideas have been applied by several authors \citep[e.g.,][]{edelson+1995, uttley+2003, arevalo+2008, chatterjee_2008}. The details of the procedure vary from author to author, so we provide a detailed description of our implementation to enable others to evaluate and reproduce our analysis.

The algorithmic description of the method we use to measure the significance of the time lags is as follows:

\begin{enumerate}
\item
We calculate the cross-correlation coefficients between the unevenly sampled time series using one of the methods described in Section \ref{crosscorr_alt}.

\item
Using an appropriate model for the PSDs at each energy band, we simulate time series with the given noise properties and sampled exactly as the data. The resulting flux densities are perturbed by adding noise according to the observational errors. We calculate the cross-correlation coefficients of the simulated light curve pairs using the same method as for the real data.

\item
We repeat the previous step for a large number of radio/gamma-ray simulated light curve pairs and accumulate the resulting cross-correlation coefficients for each time lag.

\item
For each time lag bin, the distribution of the simulated cross-correlation coefficients is used to estimate the significance levels of the data cross-correlation coefficients.
\end{enumerate}

\begin{figure*}
\begin{center}
\includegraphics[width=7.5cm, trim=30 10 30 0]{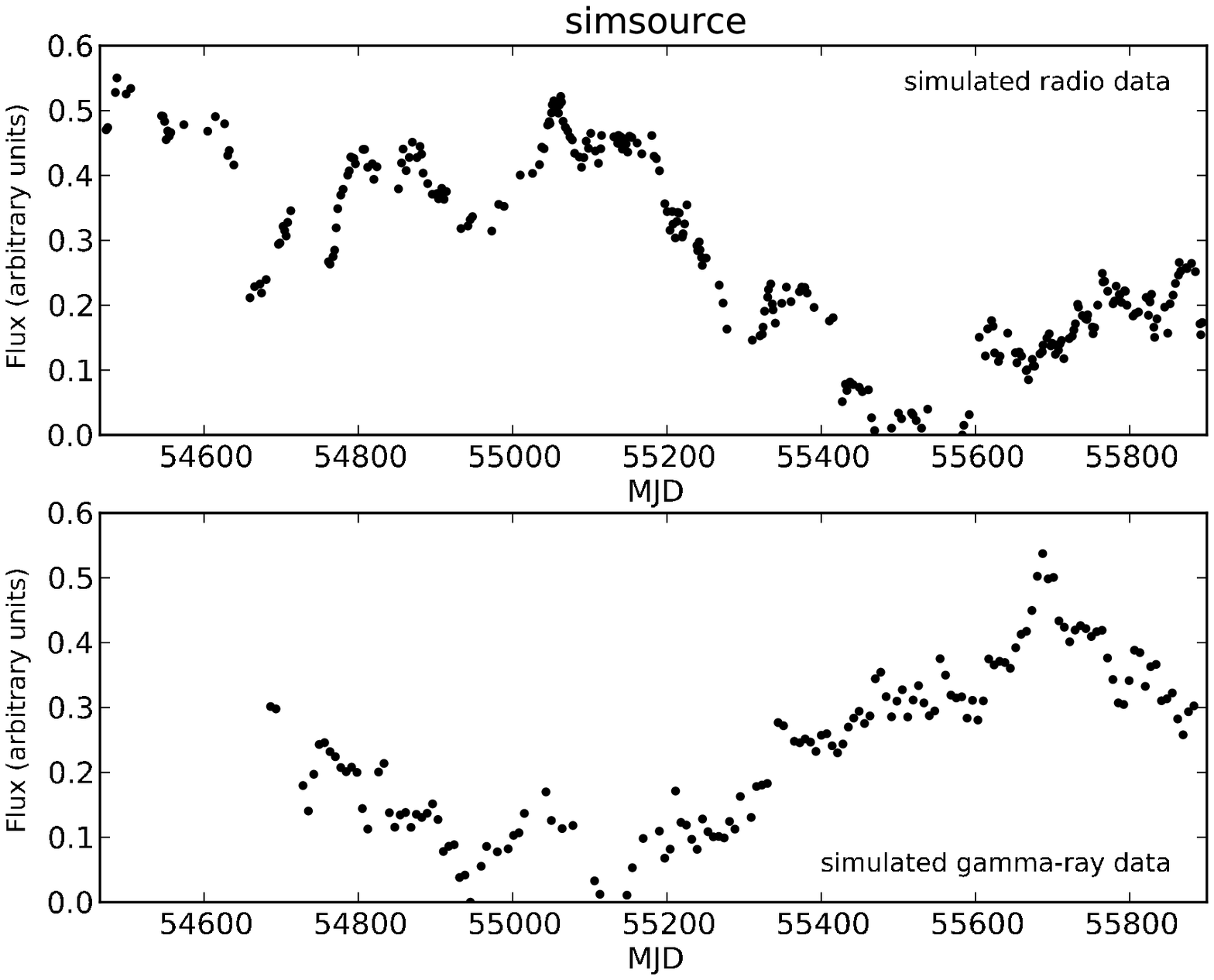}\\
\includegraphics[width=7.5cm, trim=30 0 20 0]{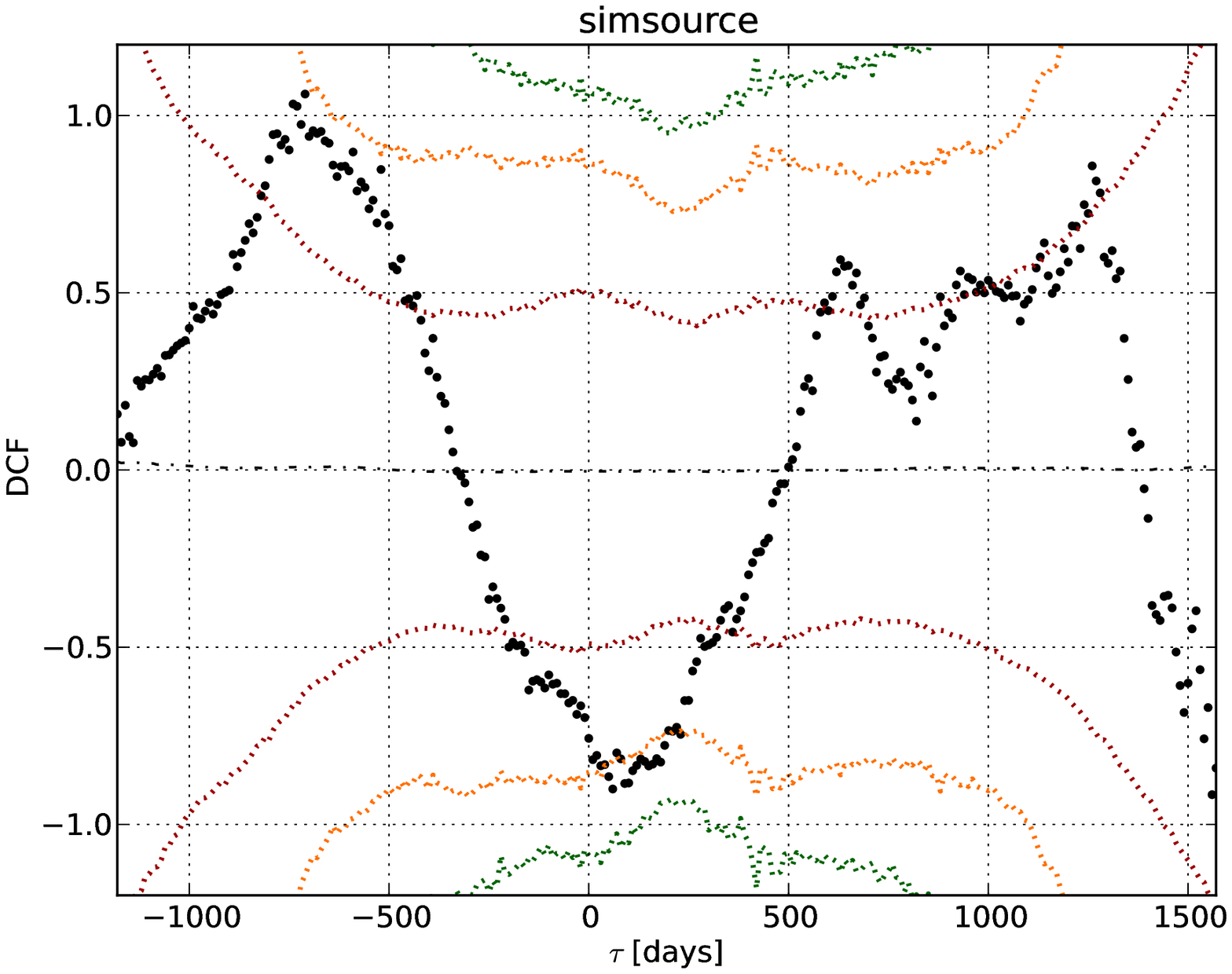}
\includegraphics[width=7.5cm, trim=30 0 20 0]{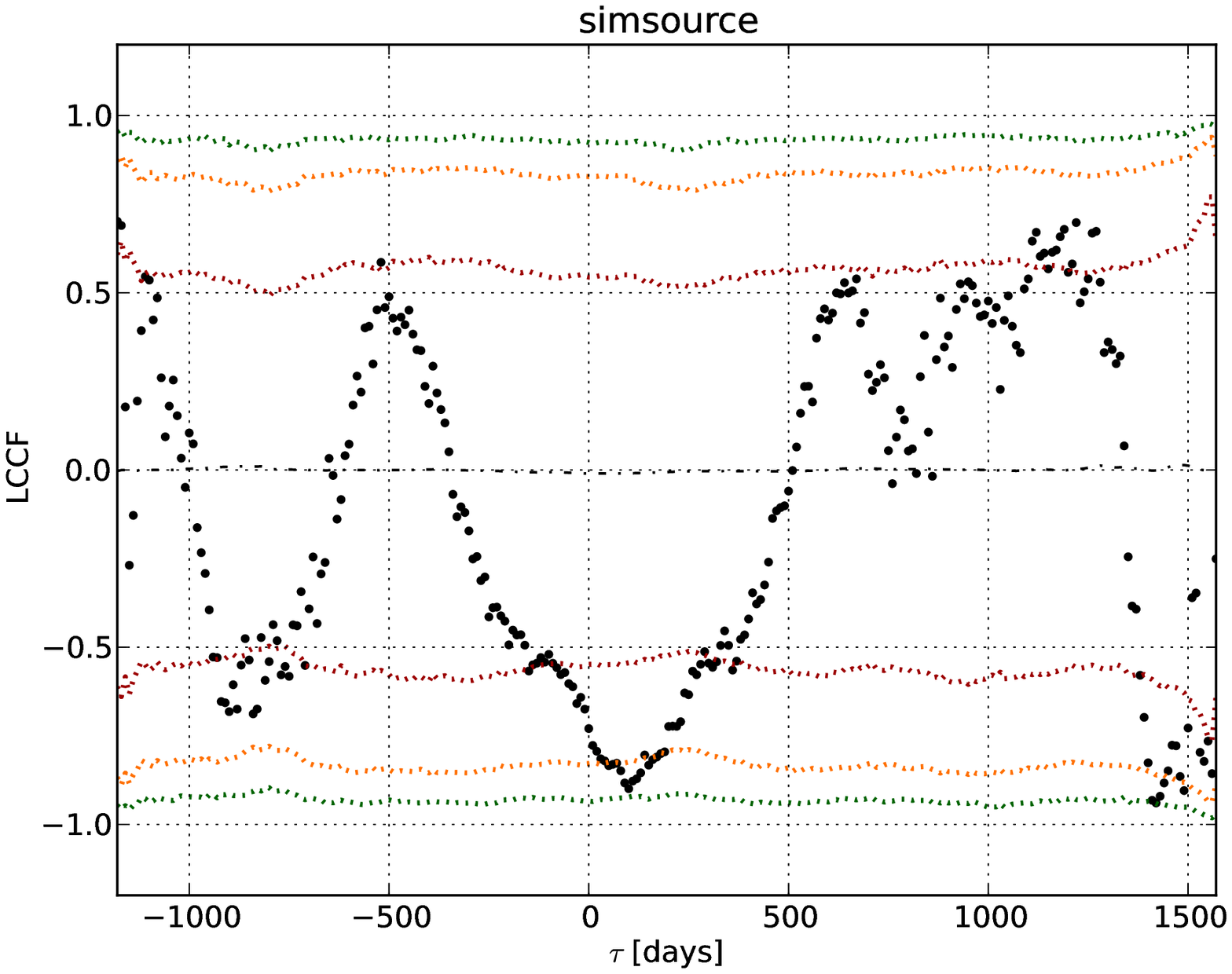}
\caption{Example of cross-correlation significance results. Upper panel shows simulated radio data with arbitrary flux units (top) and simulated gamma-ray data with arbitrary units (bottom). Both light curves use a typical sampling pattern from our monitoring program. Lower left panel is for the DCF and lower right panel for the LCCF. The black dots represent the cross-correlation for the data, while the color contours show the distribution of random cross-correlations obtained by the Monte Carlo simulation with $1\sigma$ (red), $2\sigma$ (orange) and for $3\sigma$ significance (green). A time lag $\tau > 0$ indicates the gamma-ray emission lags the radio and $\tau < 0$ the opposite.}
\label{example_xcorr_data}
\end{center}
\end{figure*}

An additional detail is that the gamma-ray time series are the result of long integrations, so each simulated data point is generated by averaging the required number of samples to replicate the time binning. For the radio light curves, the integrations are so short that the closest sample can be chosen. Figure \ref{example_xcorr_data} shows the application of the method for an example using simulated data with the sampling pattern from our monitoring program. We use $\beta = 2$ in both bands for the DCF and the LCCF. In both cases, the cross-correlation coefficient at each time lag is represented by the black dots and the distribution of random cross-correlations by the colored dotted lines. A time lag $\tau > 0$ means the gamma-ray emission lags the radio, while $\tau < 0$ represents the opposite. The red lines contain 68.27\% of the random cross-correlations, so we refer to them as the $1\sigma$ lines, the orange lines contains 95.45\% ($2\sigma$), and the green lines contains 99.73\% ($3\sigma$)\footnote{In what follows we refer to them as the 1, 2 and 3$\sigma$ lines or significance levels}. The colored contours provide a quick way to evaluate the significance of the cross-correlation and are used for this purpose throughout this paper. In this case, although the amplitudes are relatively high for both the DCF and LCCF, the significance is not even 2$\sigma$ indicating only marginal evidence of a correlation.

% Significance analysis cross-corr

\subsection{Comparison of the DCF and the LCCF \label{compare_methods}}

We use both DCF and LCCF in our tests, to determine quantitatively which is the best for the problem of detecting significant correlations between two time series. The comparison is made in terms of detection efficiency of correlations, at a given significance level, and a maximum time lag error. For the tests, we simulate a time series with a very fine time resolution and make two copies, one for each band, in which the only difference is a known time lag and the different sampling pattern, which is taken from example light curves from our monitoring program.

In all the cases, we bin the cross-correlation with $\Delta t = 10$ days and model the time series with a PSD $\propto 1/\nu^2$, which is also used for the Monte Carlo evaluation of the significance. We use $M=1000$ uncorrelated time series to estimate the distribution of random cross-correlations and significance, and use these results to estimate the significance of cross-correlation for 1000 correlated time series. This enables us to determine the significance of the correlations and the error in the recovered time lag.

This corresponds to the ideal case of a perfect intrinsic correlation, which is only distorted by the time lag and different sampling of the two time series. The case is also ideal with respect to the significance evaluation, as we perfectly know the model for the light curves. It is important to keep these points in mind and to realize that the actual detection efficiencies could be much lower than what we find through these tests.

\subsubsection{Uniform and identical sampling for both time series, with zero lag and no noise.}

\begin{figure}
\begin{center}
\includegraphics[width=9cm]{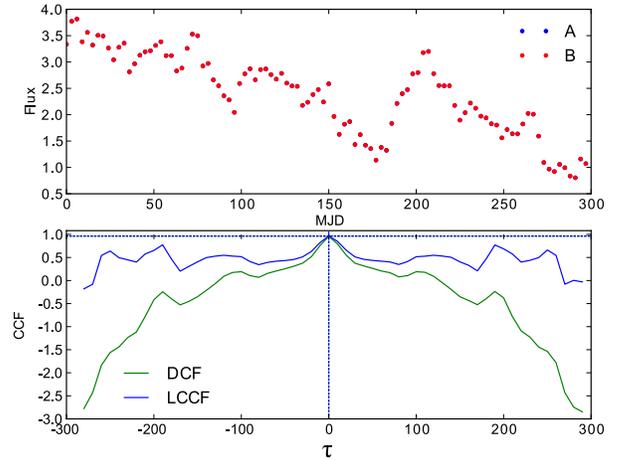}
\caption{Example of simulated data with PSD $\propto 1/\nu^2$, uniform and identical sampling, zero time lag and no noise. The upper panel shows the two time series which overlap perfectly in this case. The lower panel has the results of the DCF and LCCF for this case. The vertical lines show the position of the most significant peak with color corresponding to the method used. Horizontal color lines mark the amplitude of the most significant peak for each method. The most striking difference between the two methods is the normalization which is not restricted to the $[-1, 1]$ interval in the case of the DCF.}
\label{example_sim_data_perfect}
\end{center}
\end{figure}

\begin{figure}
\begin{center}
\includegraphics[width=9cm, trim=0 0 0 0]{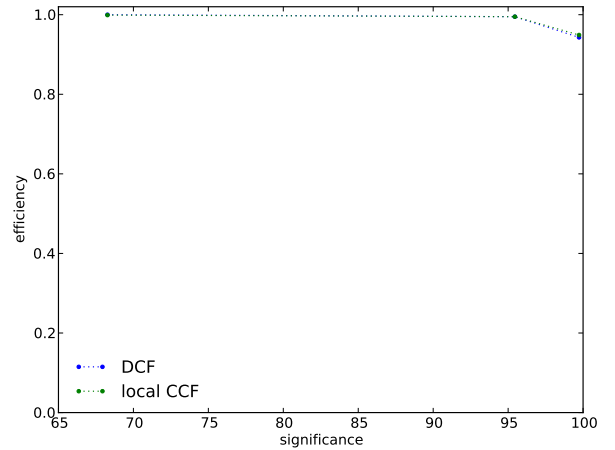}
\caption{Detection efficiency versus significance for both methods for the case of \emph{uniform and identical sampling for both time series, zero lag and no noise}. In this case close to 95\% of the lags are recovered at the right value and $3\sigma$ significance.}
\label{efficiency_perfect}
\end{center}
\end{figure}

As a check of the method and to help the reader understand the results, we first test our ability to detect correlations in a very simple case. In this case a time series with a uniform sampling period of 3 days is correlated with a copy of itself without any delay or noise. An example of the simulated data set along with the results for the DCF and LCCF is shown in Figure \ref{example_sim_data_perfect}. The same procedure is repeated for all simulated time series with known time lag and correlation properties, and the fraction of detected lags at the known lag ($\pm \Delta t$) with a given significance level is reported as an efficiency in Figure \ref{efficiency_perfect}.

% Example test light curve

% Efficiency

\begin{figure}
\begin{center}
\includegraphics[width=9cm, trim=0 0 0 0]{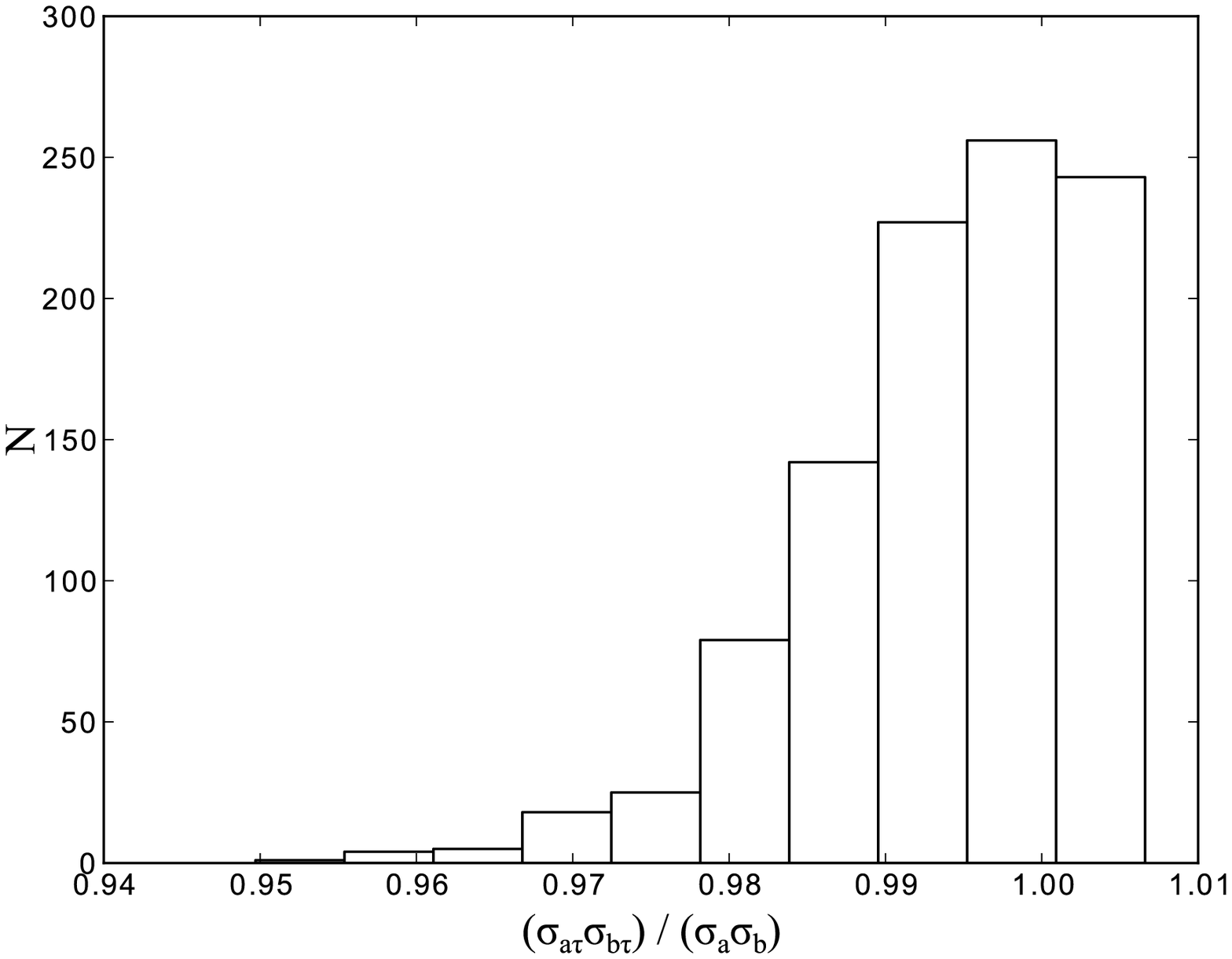}
\includegraphics[width=9cm, trim=0 0 0 0]{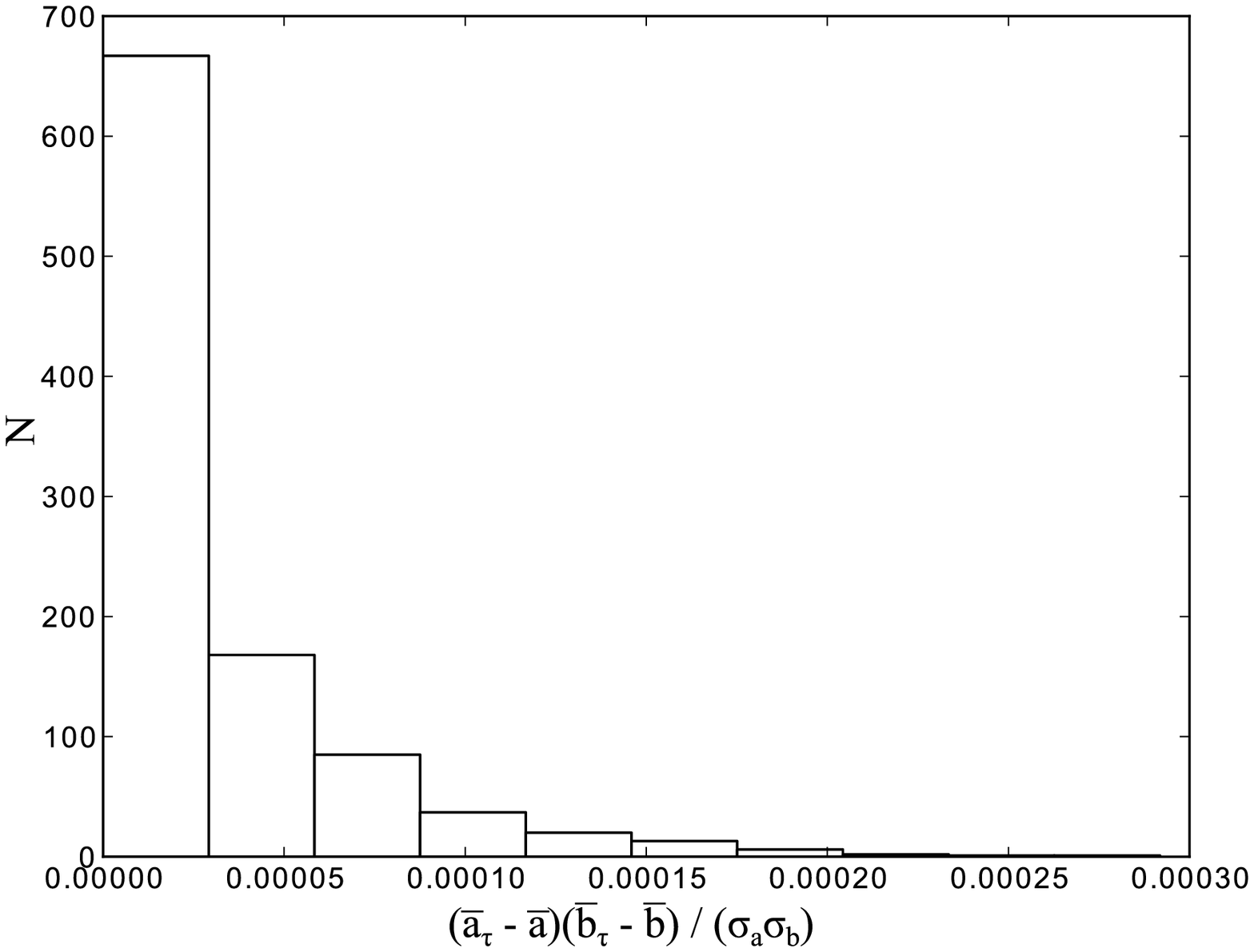}
\caption{Distribution of coefficients of the linear relation between DCF and LCCF for $\tau = 0$ day, for the case of \emph{uniform and identical sampling for both time series, zero lag and no noise}. Upper panel is the multiplicative factor, which is very close to 1 in most cases. Lower panel is the additive constant which is very close to 0. These values make DCF $\approx$ LCCF which makes the results of both methods very similar as can be seen in Figure \ref{efficiency_perfect}.}
\label{coeff_linear_perfect}
\end{center}
\end{figure}

In this case, we recover most of the time lags at the right value and the behaviors of the DCF and LCCF are very similar. The values of the coefficients of the linear relation for $\tau = 0$ (Equation \ref{DCF_LCCF_linrel}), are very close to the case when the DCF and LCCF are equal (Figure \ref{coeff_linear_perfect}).

% Coefficients linear relation

\subsubsection{Data sampling case 1, ``short data set": 2 years of OVRO and 1 year of \emph{Fermi}-LAT. \label{test_short_data_set}}

We now study a case with sampling taken from the OVRO 40 meter blazar monitoring program and \emph{Fermi}-LAT data set. Again we add no noise to the simulations and have zero lag between the two light curves, so the only difference is in the sampling pattern. In this case, a source was observed for two years with the OVRO 40 meter telescope at 15 GHz with a nearly twice per week sampling \citep[][]{richards+2011}. The gamma-ray data for the same source has one observation per week and a one year time duration \citep[][]{abdo_variability_2010}.

\begin{figure}
\begin{center}
\includegraphics[width=9cm]{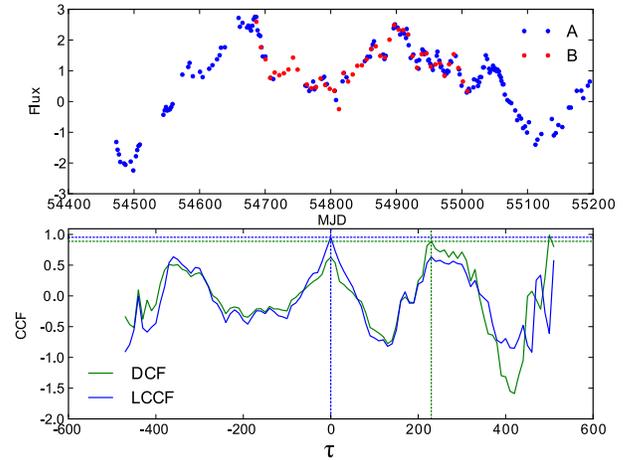}
\caption{Example of simulated data  with PSD $\propto 1/\nu^2$, for the case ``short data set". Upper panel shows the two time series, which have some small differences produced by the different sampling at each waveband. Lower panel has the results of the DCF and LCCF for this case. The vertical lines show the position of the most significant peak with color corresponding to the method used. Horizontal color lines mark the amplitude of the most significant peak for each method. In this example the LCCF recovers the right time lag, but the DCF finds a spurious time lag.}
\label{example_sim_data_short}
\end{center}
\end{figure}

An example of simulated data with this sampling is shown in Figure \ref{example_sim_data_short} (upper panel), along with the results for the cross-correlation (lower panel). In this case the radio sampling (blue dots) covers a longer time span than the gamma-ray one (red dots).

% Example test light curve

% Efficiency
\begin{figure}
\begin{center}
\includegraphics[width=9cm, trim=0 0 0 0]{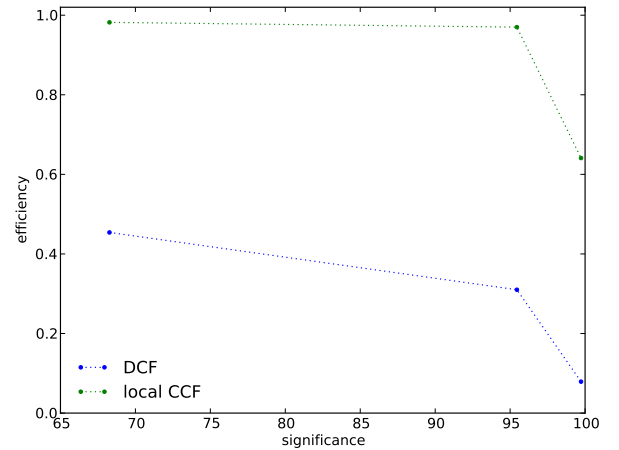}
\caption{Detection efficiency versus significance for both methods for the case of the ``short data set". In this case the efficiencies differ significantly between both methods, with the LCCF being the more efficient.}
\label{efficiency_short}
\end{center}
\end{figure}

\begin{figure}
\begin{center}
\includegraphics[width=9cm, trim=0 0 0 0]{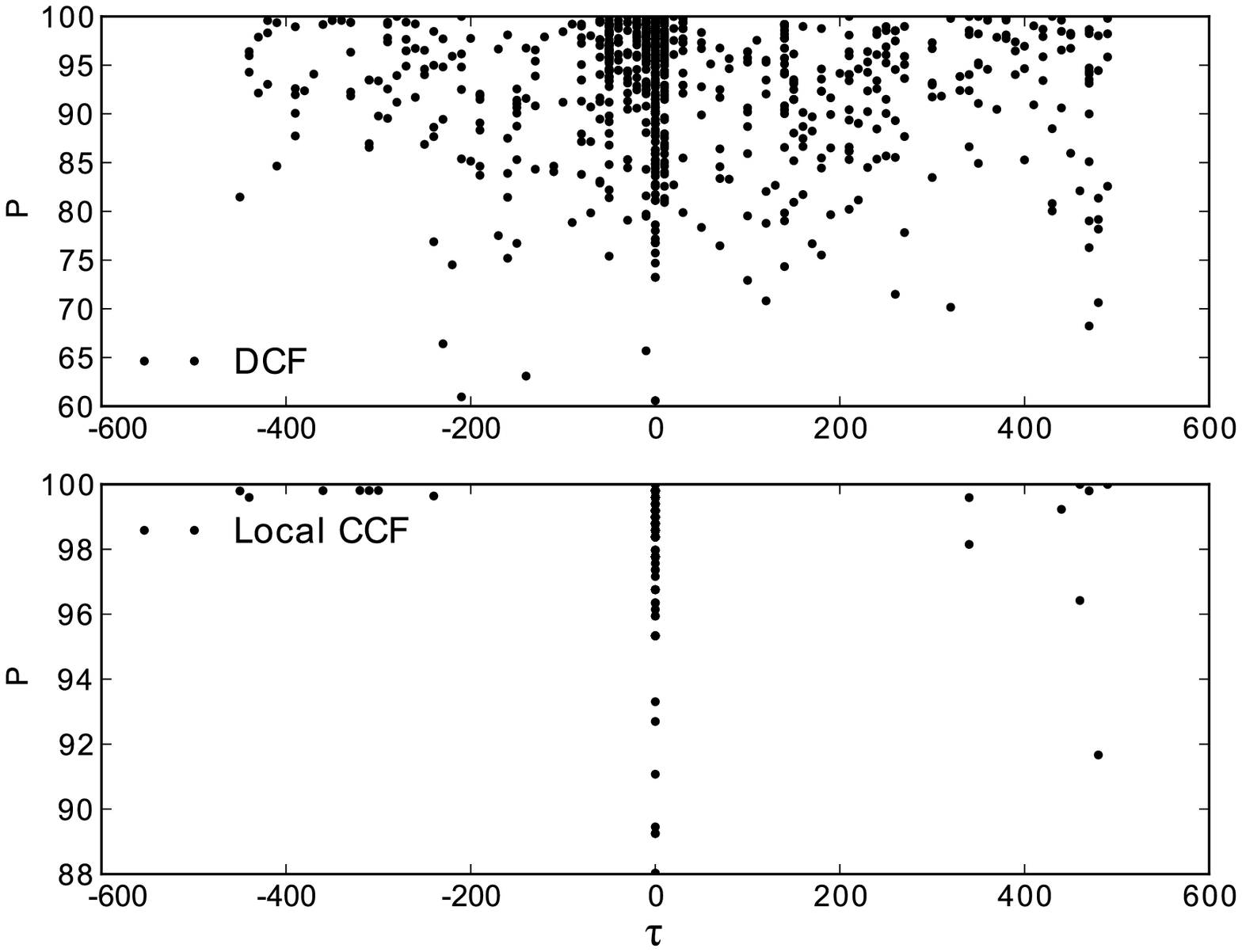}
\includegraphics[width=9cm, trim=0 0 0 0]{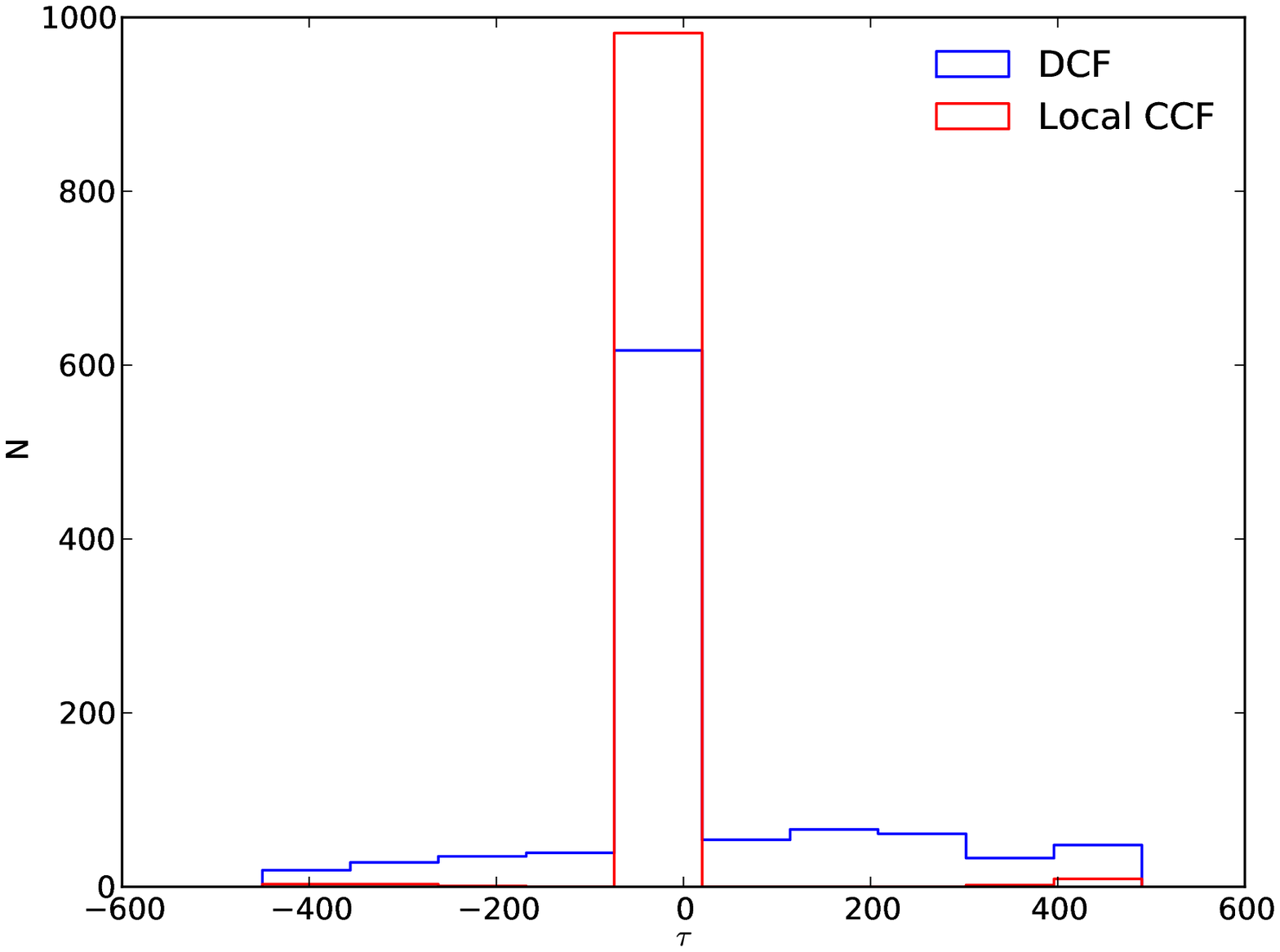}
\caption{Distribution of most significant peaks in the correlation for both methods for the case of the ``short data set". Upper panels show the lag and significance of the most significant peak for both methods. The lower panel is a histogram for the distribution of lags for the most significant peak.}
\label{peak_dist_short}
\end{center}
\end{figure}

Figure \ref{efficiency_short} shows that in this case we only recover a fraction of the time lags at a $3\sigma$ significance. This is because the DCF often finds the most significant peak at a lag different from zero (Fig. \ref{peak_dist_short}). Moreover some of those spurious lags are of high statistical significance. We still get some significant peaks at lags different from zero for the LCCF, but at a much smaller rate.
 
% Peak distribution

\begin{figure}
\begin{center}
\includegraphics[width=9cm, trim=0 0 0 0]{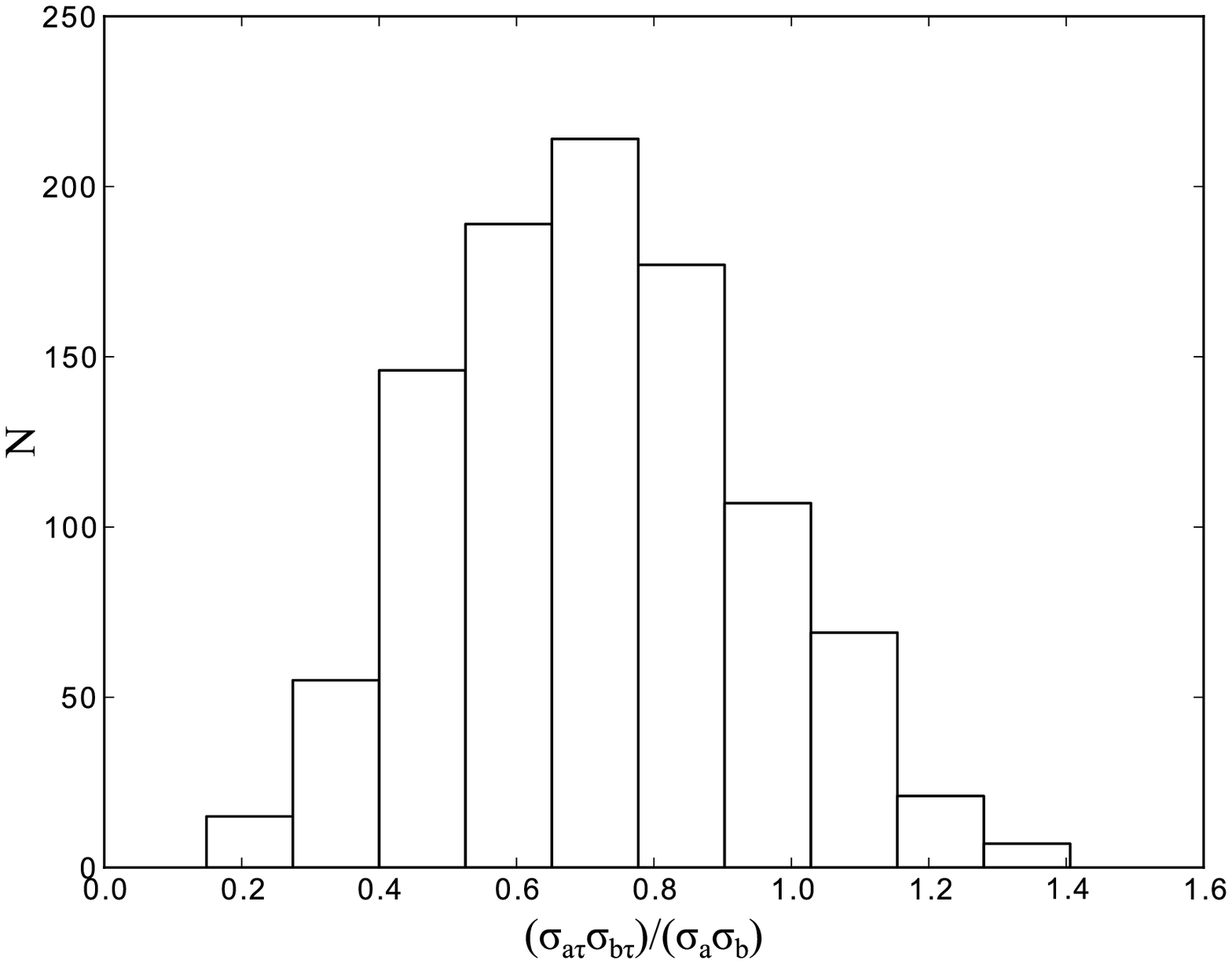}
\includegraphics[width=9cm, trim=0 0 0 0]{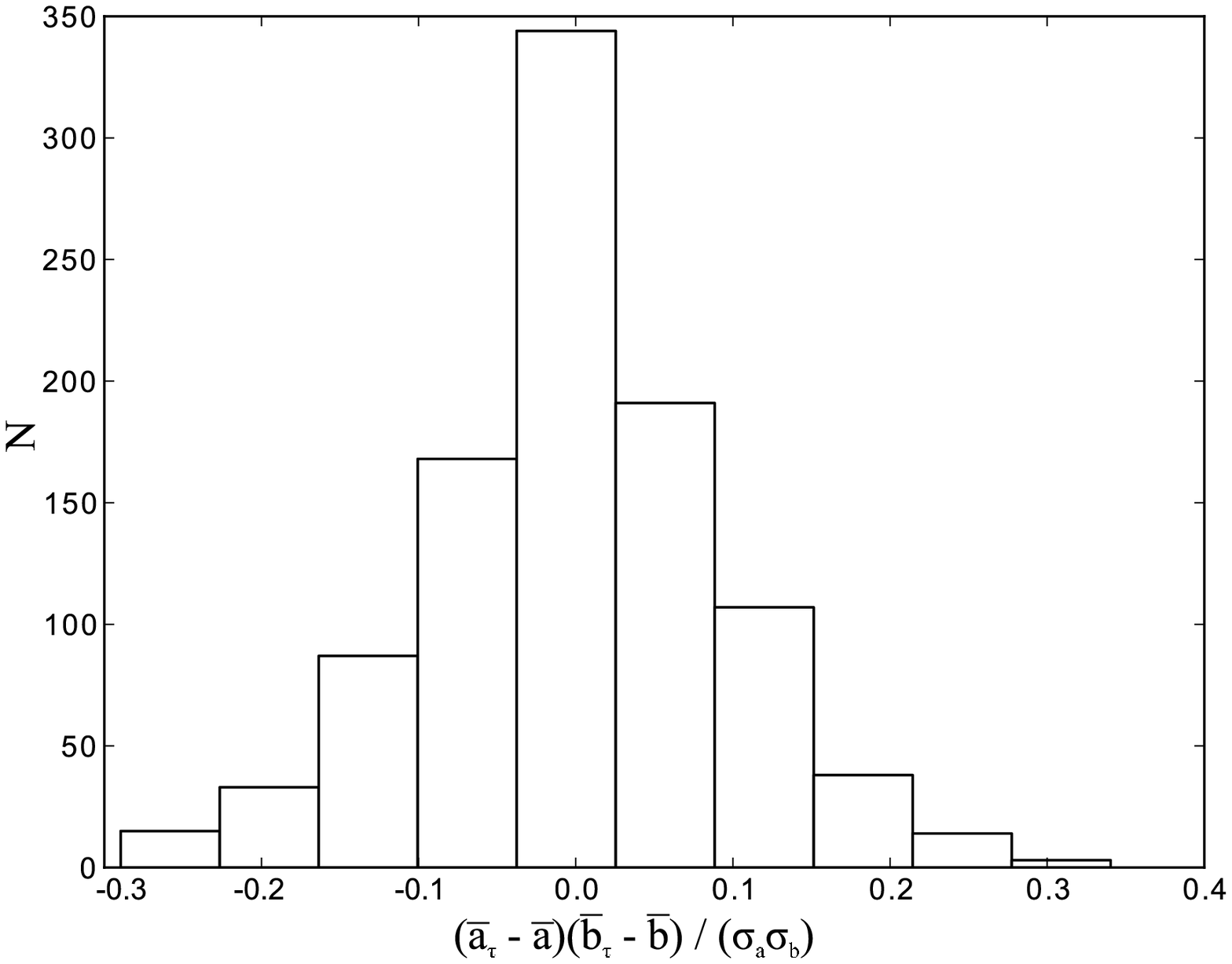}
\caption{Distribution of coefficients of the linear relation between DCF and LCCF for $\tau = 0$ day, for the case of the ``short data set". Upper panel is the multiplicative factor, which has a very broad distribution, different from 1 in most cases. Lower panel is the additive constant which also has a very broad distribution, different from the ideal case of 0. These values show the DCF to be different from the LCCF and have a role in producing spurious highly significant peaks in the correlation.}
\label{coeff_linear_short}
\end{center}
\end{figure}

\begin{figure}
\begin{center}
\includegraphics[width=9cm]{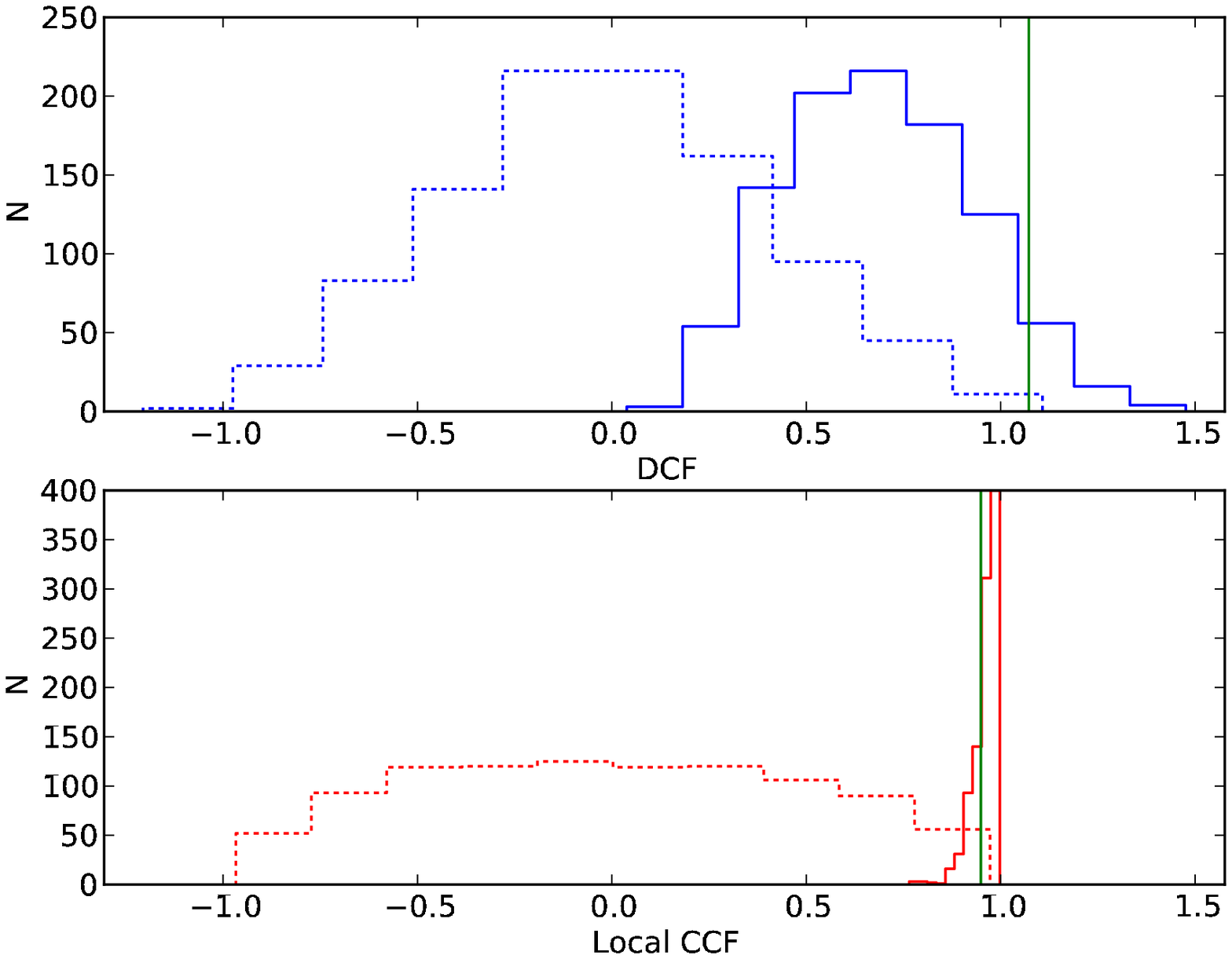}
\caption{Distribution of the cross-correlation coefficient for both methods at $\tau = 0$ day, for the case of the ``short data set". Upper panel is for the DCF and lower panel for the LCCF. Both panels show the distribution of random cross-correlations with a dotted line and for correlated data with a solid line. Points with cross-correlation coefficient to the right of the vertical green line have a significance of at least $3\sigma$.}
\label{dist_dcf_and_lccf_short}
\end{center}
\end{figure}

To understand how we can get small values of the DCF at zero lag while still having large values of the LCCF, we can take a look at the distributions of the coefficients of the linear relation (Equation \ref{DCF_LCCF_linrel}), shown in Figure \ref{coeff_linear_short}. The multiplicative coefficient should be one in the ideal case, but instead it has a broad distribution (upper panel). The additive coefficient should be zero in the ideal case, but it also has a broad distribution (lower panel). This can effectively reduce the value of the correlation coefficient or make its distribution broader, either way reducing its discriminating power. This effect is seen in Figure \ref{dist_dcf_and_lccf_short}, which shows the distribution of cross-correlation coefficients at $\tau = 0$ days. In the figure, the distribution of random cross-correlations is represented with a dotted line and the one for correlated data with a solid line. The upper panel is for the DCF and the lower panel for the LCCF. The vertical green line represents the $3\sigma$ significance threshold amplitude for cross-correlation coefficients. The fraction of cross-correlations for correlated data (solid line) that is to the right of the green line is approximately equal to the detection efficiency \footnote{The equality is only approximate because a peak with larger significance might have appeared in a lag different than $\tau = 0$. These cases are not excluded from the histogram.}. It can be seen that this fraction is much larger for the LCCF, as a result of increased scatter in the distribution of the DCF when compared to the LCCF of correlated data, for the reasons presented earlier.

% Coefficients linear relation

% Distribution of DCF and LCCF at zero lag

\subsubsection{Data sampling case 2, ``long data set": 4 years of OVRO and 3 years of \emph{Fermi}-LAT. \label{test_long_data_set}}

\begin{figure}
\begin{center}
\includegraphics[width=9cm]{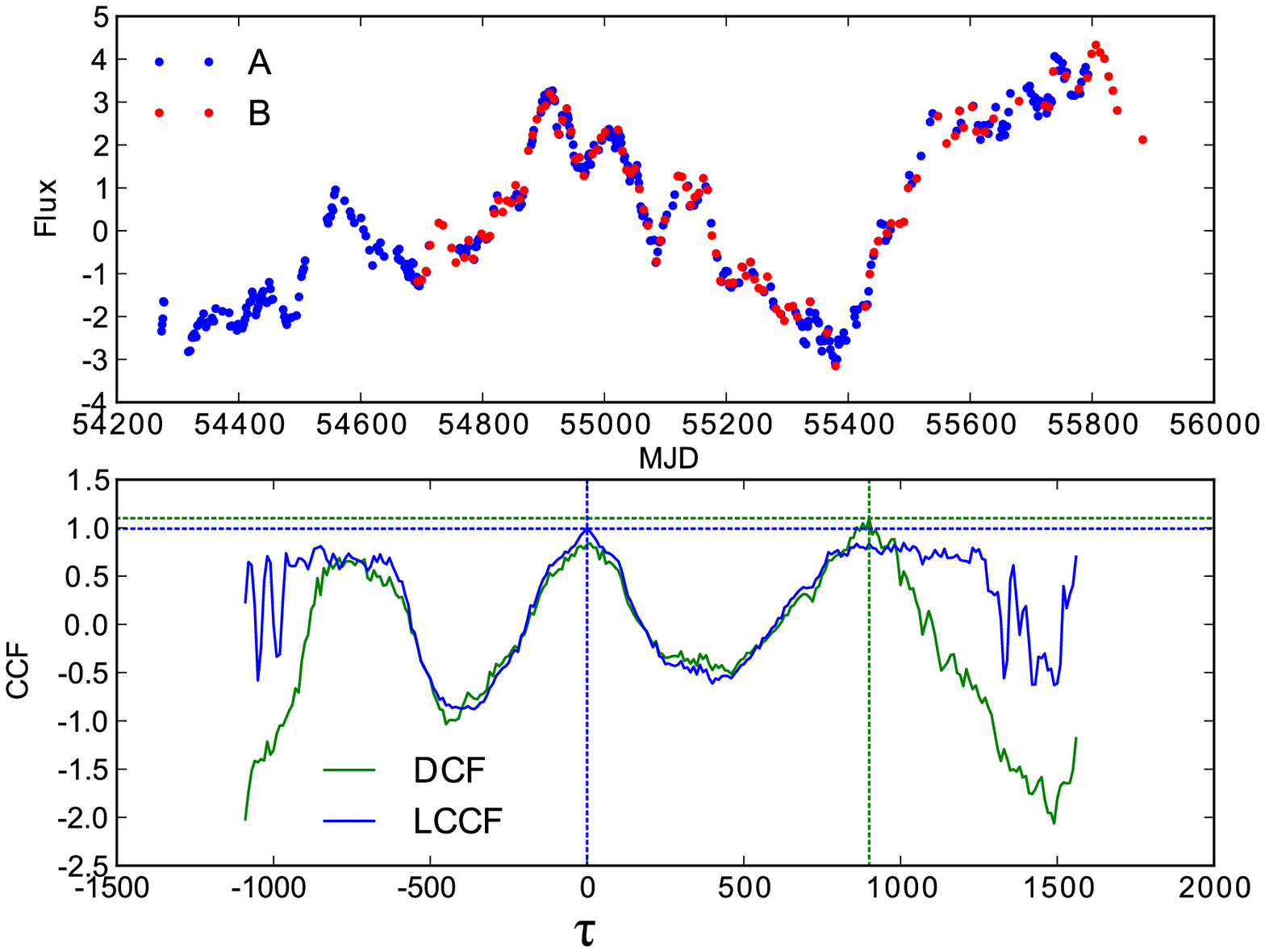}
\caption{Example of simulated data for the ``long data set". Upper panel shows the two time series, which have some small differences produced by the different sampling at each waveband. Lower panel has the results of the DCF and LCCF for this case. The vertical lines show the position of the most significant peak. In this example the LCCF recovers the right time lag, but the DCF finds an spurious time lag.}
\label{example_sim_data_long}
\end{center}
\end{figure}

We make the same comparison using a data set with radio light curves of 4 years time duration sampled about twice a week, and gamma-ray light curves with a 3 year time duration and weekly sampling. We again consider the case with no noise and zero lag between the two light curves, so the only difference is in the sampling pattern. An example of a simulated data set with this sampling is shown in Figure \ref{example_sim_data_long} (upper panel), along with the results for the cross-correlation (lower panel). Comparison of the results of this section with the shorter dataset test (Section \ref{test_short_data_set}), can give us an idea of the variation of the relative power to detect correlations in different data sets.

% Example test light curve

% Efficiency
\begin{figure}
\begin{center}
\includegraphics[width=9cm, trim=0 0 0 0]{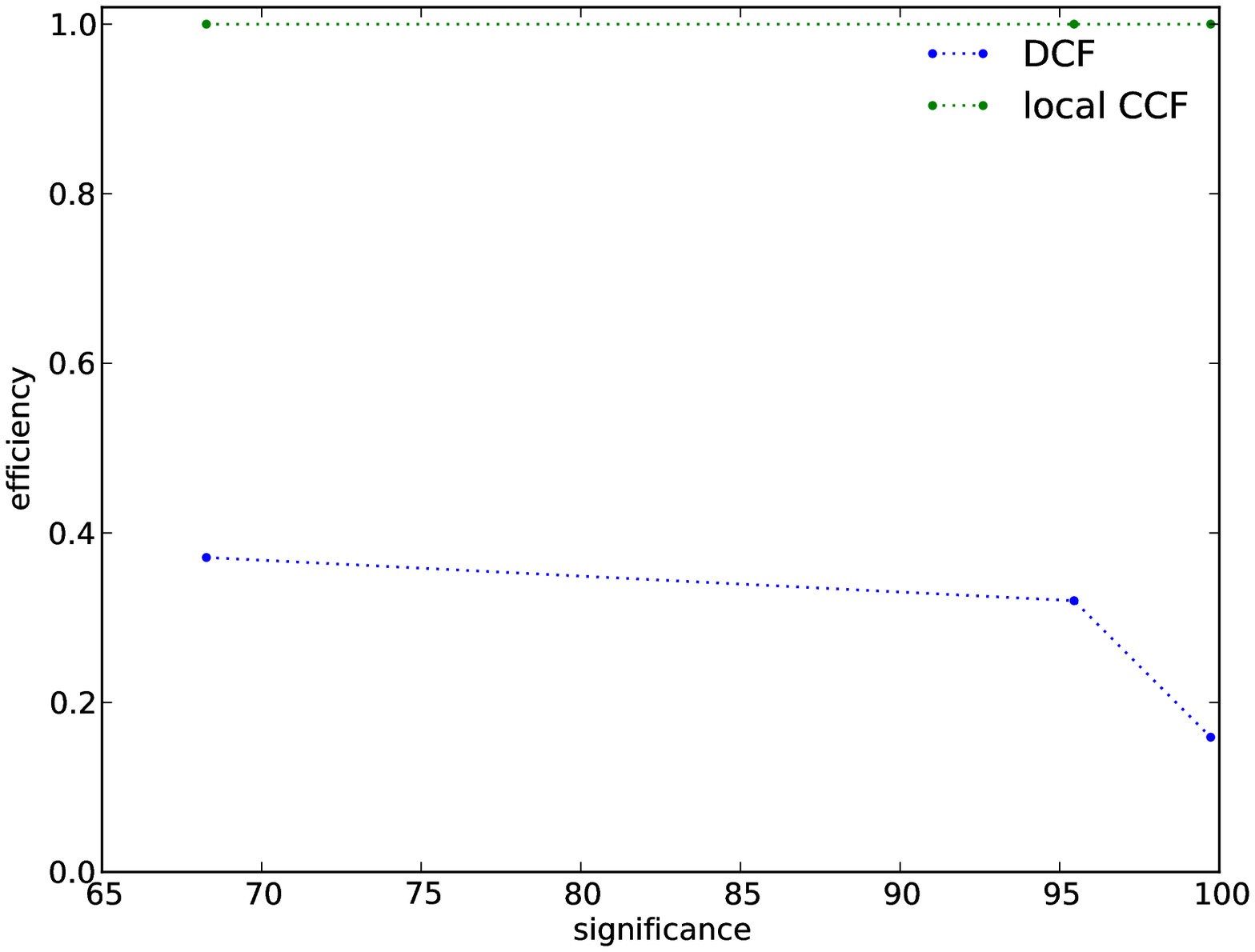}
\caption{Detection efficiency versus significance for both methods, for the case of the ``long data set". In this case the efficiencies differ significantly between both methods, with the LCCF being the more efficient.}
\label{efficiency_long}
\end{center}
\end{figure}

\begin{figure}
\begin{center}
\includegraphics[width=9cm, trim=0 0 0 0]{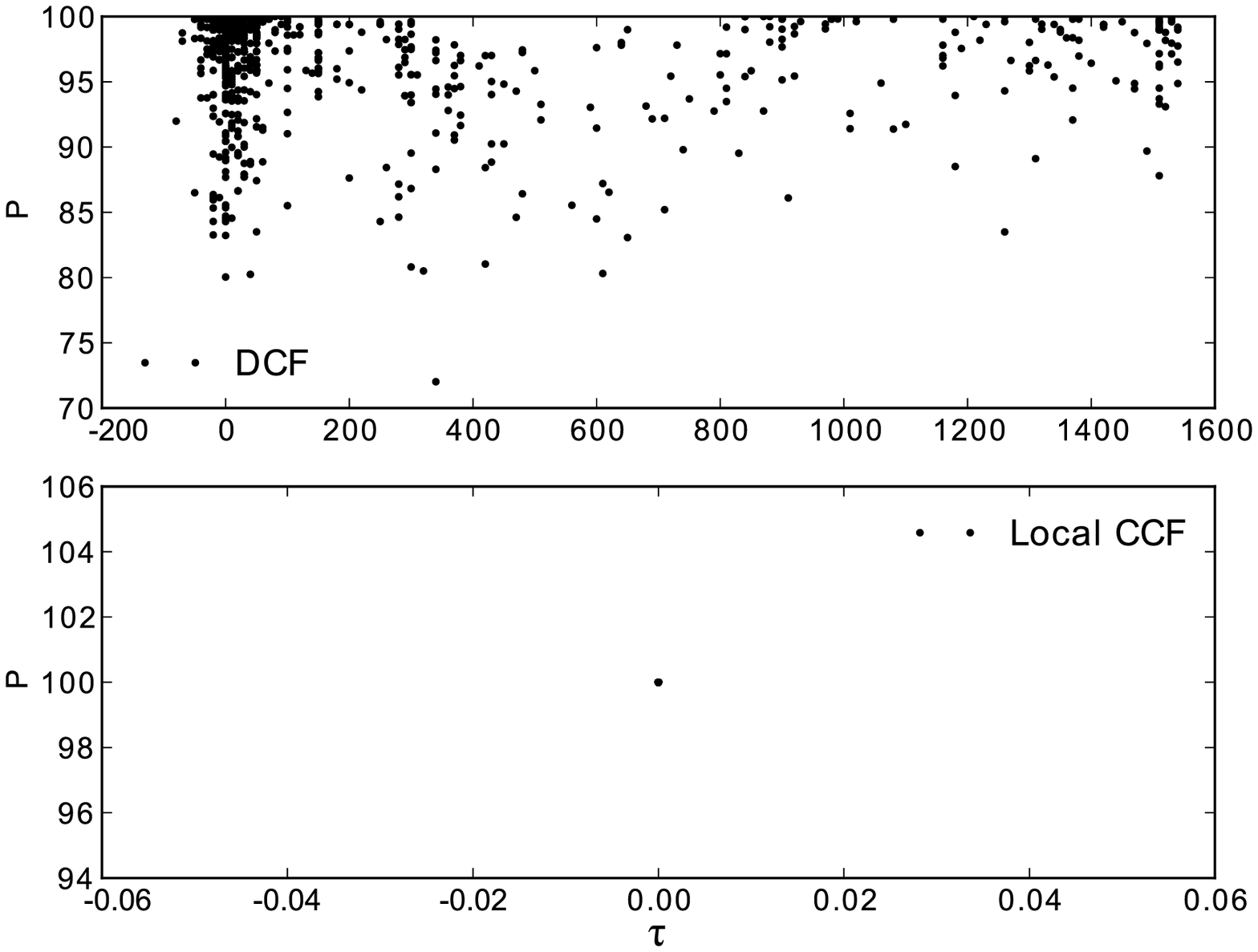}
\includegraphics[width=9cm, trim=0 0 0 0]{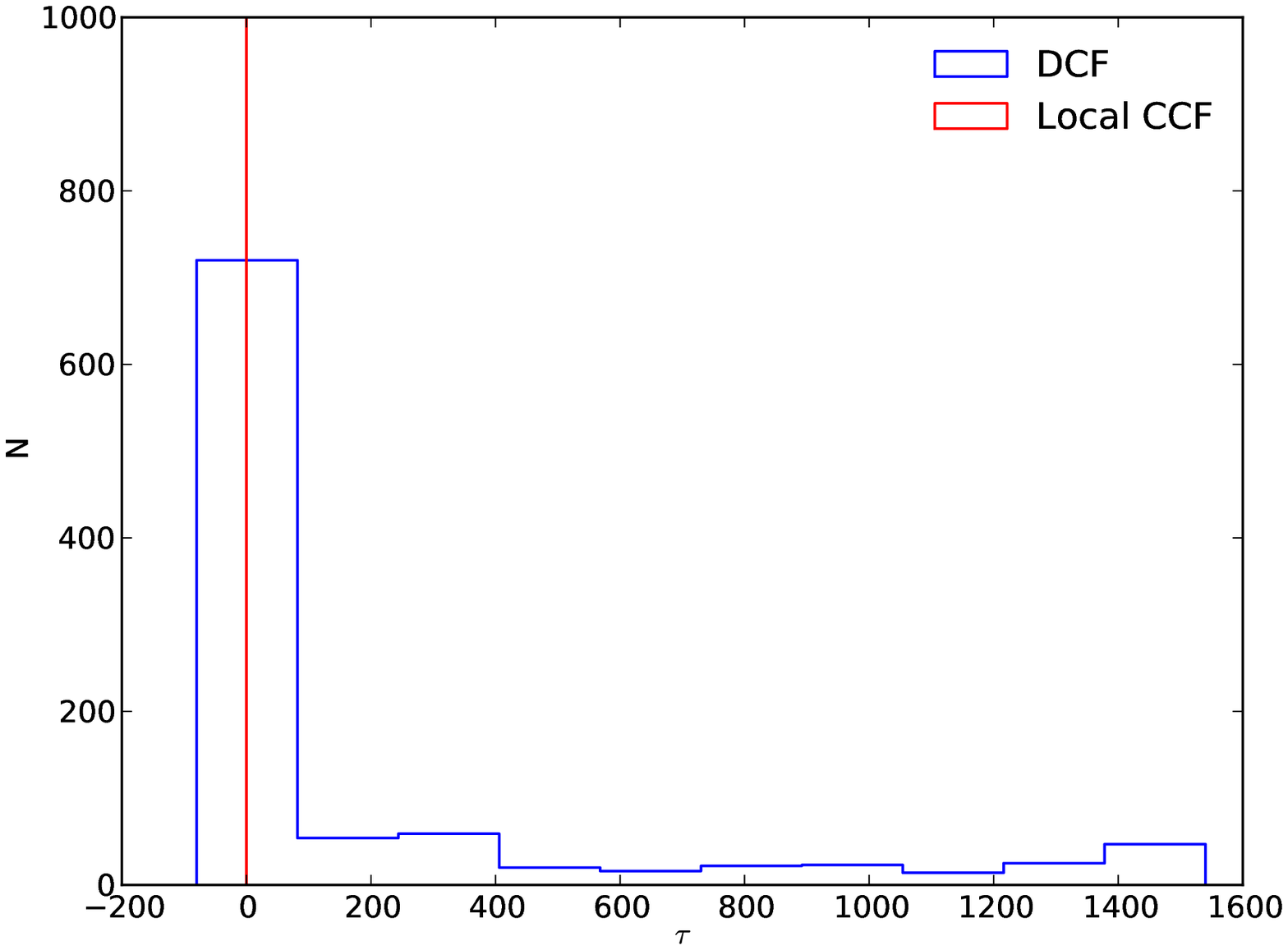}
\caption{Distribution of most significant peaks in the correlation for both methods, for the case of the ``long data set". Upper panels show the lag and significance of the most significant peak for both methods. The lower panel is a histogram for the distribution of lags for the most significant peak.}
\label{peak_dist_long}
\end{center}
\end{figure}

As in the case of the ``short data set", we find that the efficiency of detection strongly depends on the method used. Figure \ref{efficiency_long} shows that the LCCF recovers the right time lag at high significance for all the cases, while the DCF does so in only about 15\% of the cases. An examination of Figure \ref{peak_dist_long} shows that the DCF produces spurious correlation peaks with a wide distribution. As in the case of the ``short data set", some of those spurious peaks have high statistical significance.

A comparison of Figures \ref{efficiency_short}  and \ref{efficiency_long} shows that the performance of both methods improves as expected when using longer time series. However, as can be seen from Figure \ref{peak_dist_long}, the DCF produces a large fraction of spurious statistically significant correlation peaks, while the LCCF recovers a significant correlation at $\tau = 0$ in all cases.

% Peak distribution

\begin{figure}
\begin{center}
\includegraphics[width=9cm, trim=0 0 0 0]{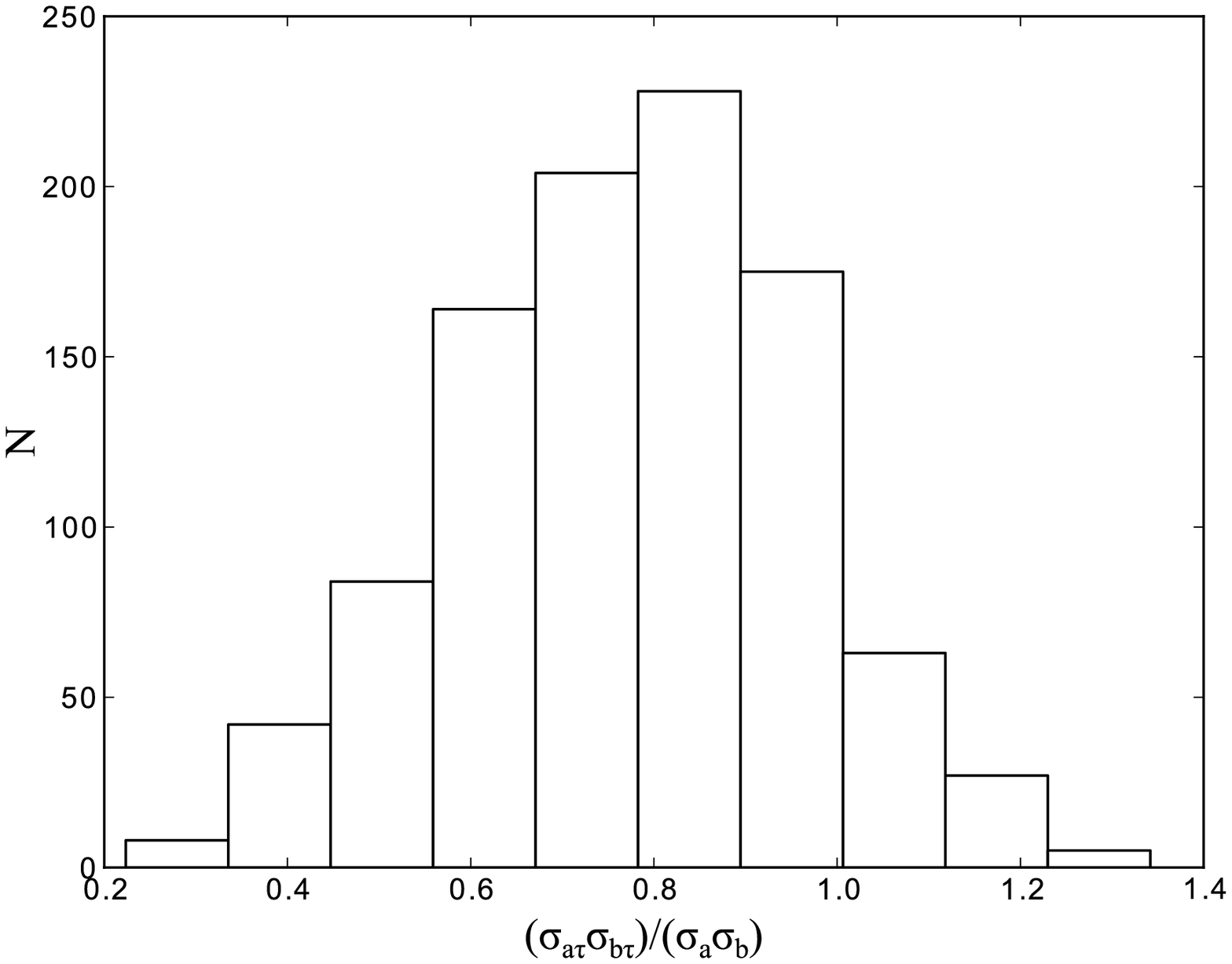}
\includegraphics[width=9cm, trim=0 0 0 0]{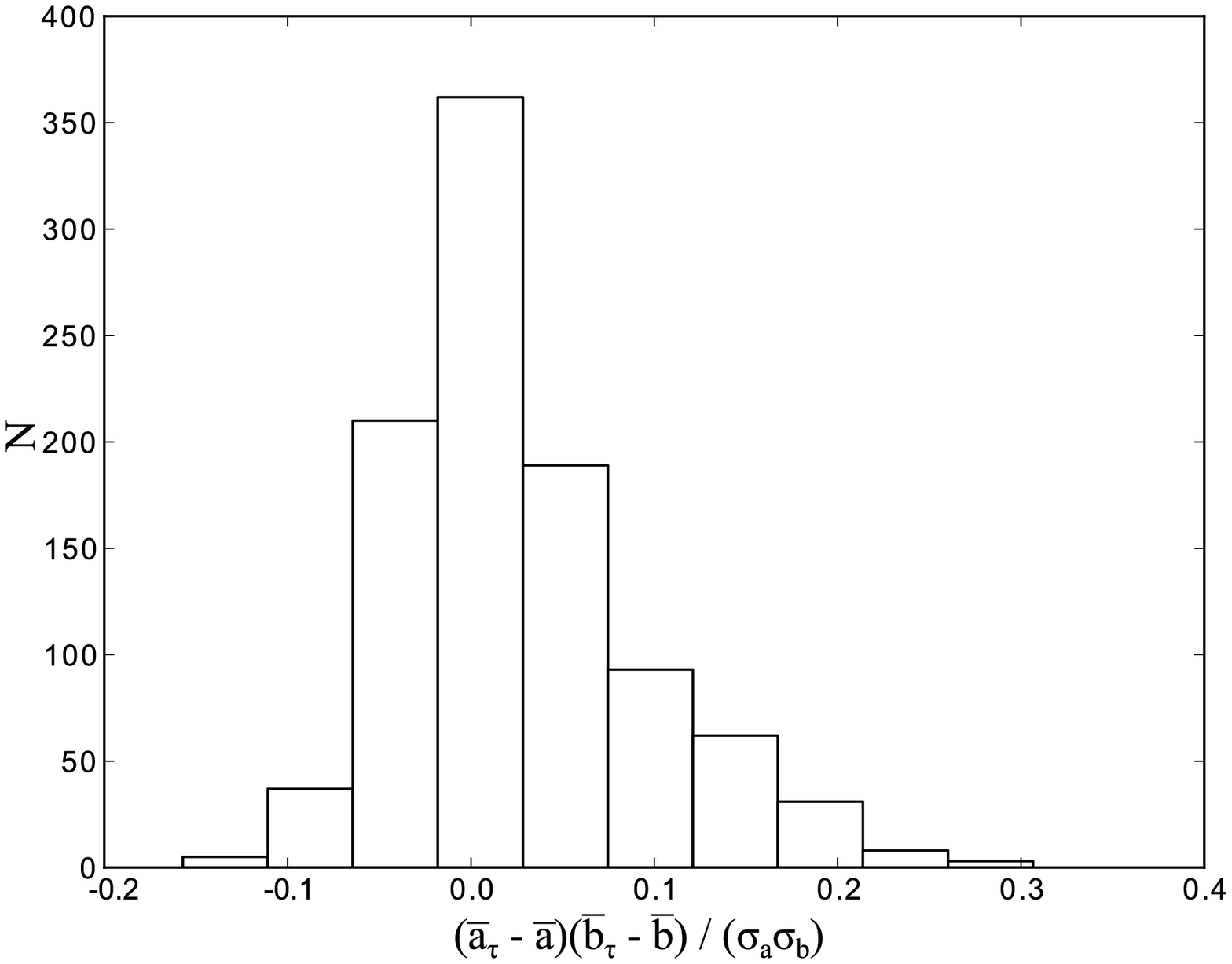}
\caption{Distribution of coefficients of the linear relation between DCF and LCCF for $\tau = 0$ day,  for the case of the ``long data set". Upper panel is the multiplicative factor, which has a very broad distribution, different from 1 in most cases. Lower panel is the additive constant which also has a very broad distribution, different from the ideal case of 0. These values show the DCF to be different from the LCCF and have a role in producing spurious highly significant peaks in the correlation.}
\label{coeff_linear_long}
\end{center}
\end{figure}

\begin{figure}
\begin{center}
\includegraphics[width=9cm]{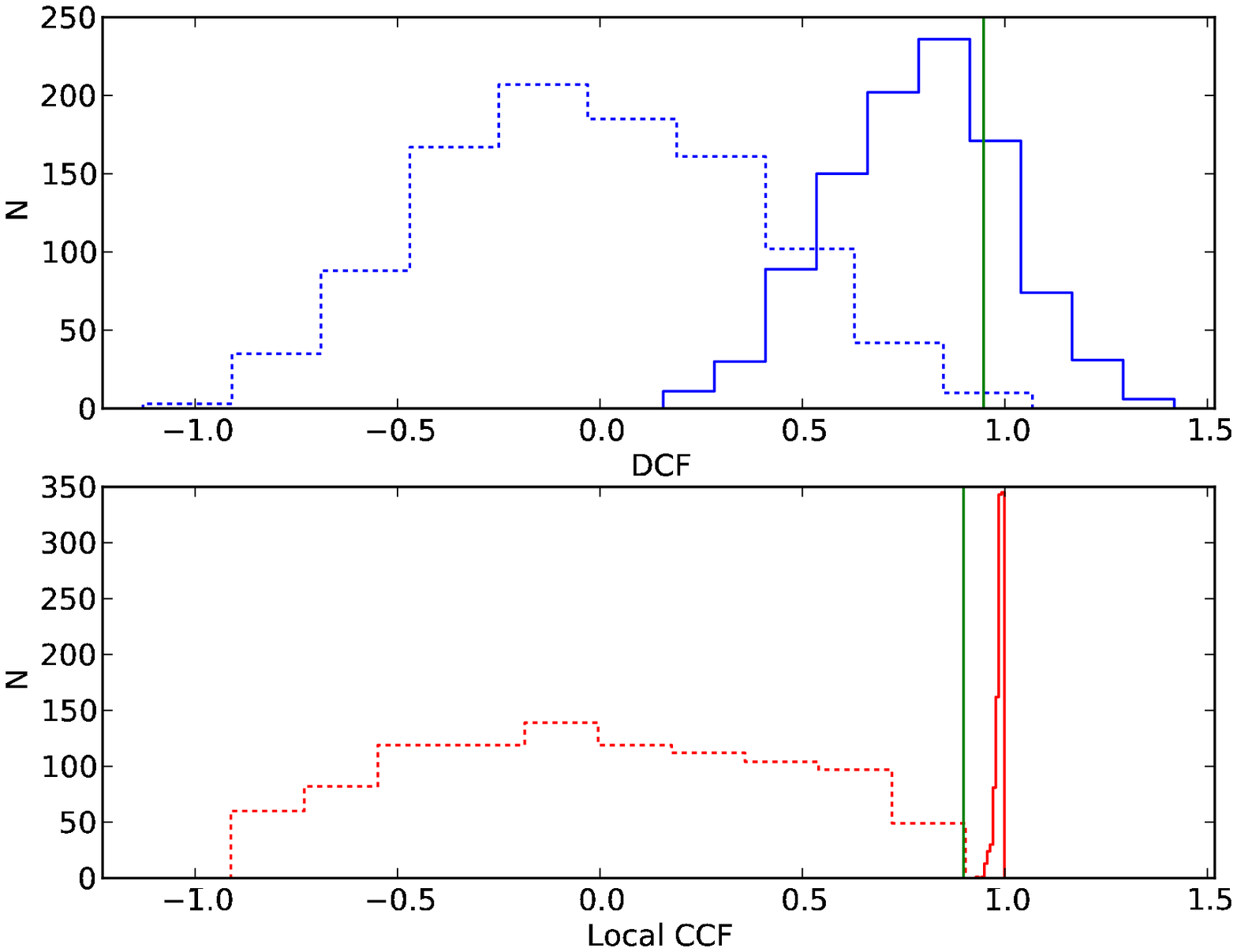}
\caption{Distribution of the cross-correlation coefficient for both methods at $\tau = 0$ day, for the case of the ``long data set". Both panels show the distribution of random cross-correlations with dotted line and the one for correlated data with solid line. Points with cross-correlation coefficient to the right of the vertical green line have a significance of at least $3\sigma$. Upper panel is for the DCF and lower panel for the LCCF.}
\label{dist_dcf_and_lccf_long}
\end{center}
\end{figure}

Figure \ref{coeff_linear_long} shows the distribution of the coefficients for the linear relation between the DCF and LCCF (Equation \ref{DCF_LCCF_linrel}). We again see that they significantly differ from the ideal case of a stationary time series. This provides an explanation for the difference between these two estimators of the correlation. As for the case of the ``short data set'', we also look at the distribution of cross-correlation coefficients for the uncorrelated and correlated data sets at $\tau = 0$ (Figure \ref{dist_dcf_and_lccf_long}). We again see the broad distribution of correlation coefficients for the DCF of correlated data sets, while a much narrower distribution for the LCCF, demonstrating the better discriminating power of the LCCF.

% Coefficients linear relation
%\clearpage

% Distribution of DCF and LCCF at zero lag
%\clearpage

\subsubsection{Additional tests}

Additional tests were performed introducing various time lags for the time series and measuring the efficiency of detection for the DCF and LCCF. They all show the same qualitative information and are thus not included here. In all cases the LCCF outperforms the DCF and the efficiency of detection improves when using a longer time duration dataset. These results demonstrate that the LCCF is the more efficient method for recovering time lags with high significance.

\subsection{Further considerations}

In this section, we describe some additional issues that should be considered when estimating the significance of cross-correlations using the Monte Carlo test we have devised, or similar methods. The error on the significance estimate has been mostly ignored in the literature, while the dependence of the significance estimate on the model light curves - when not fully appreciated - can lead to significance tests that are not consistent with the basic statistical properties of blazar light curves.

Another effect not considered here has recently been raised by \citet[][]{emmanoulopoulos+2013}. In their paper, they propose a method to simulate light curves that reproduces not only the power spectral density, but also the power density function of the flux measurements. This method is an improvement on \citet[][]{timmer+1995}, that produces Gaussian distributed fluxes and can provide a better approximation to light curves that have non-Gaussian probability density functions. For our data, this can be the case for gamma-ray light curves but it is not of much concern for the radio light curves.

\subsubsection{The dependence of the significance estimate on the model light curves \label{sig_and_psds}}

\begin{figure}
\begin{center}
\includegraphics[angle=0,width=9.0cm, trim=0 10 0 0]{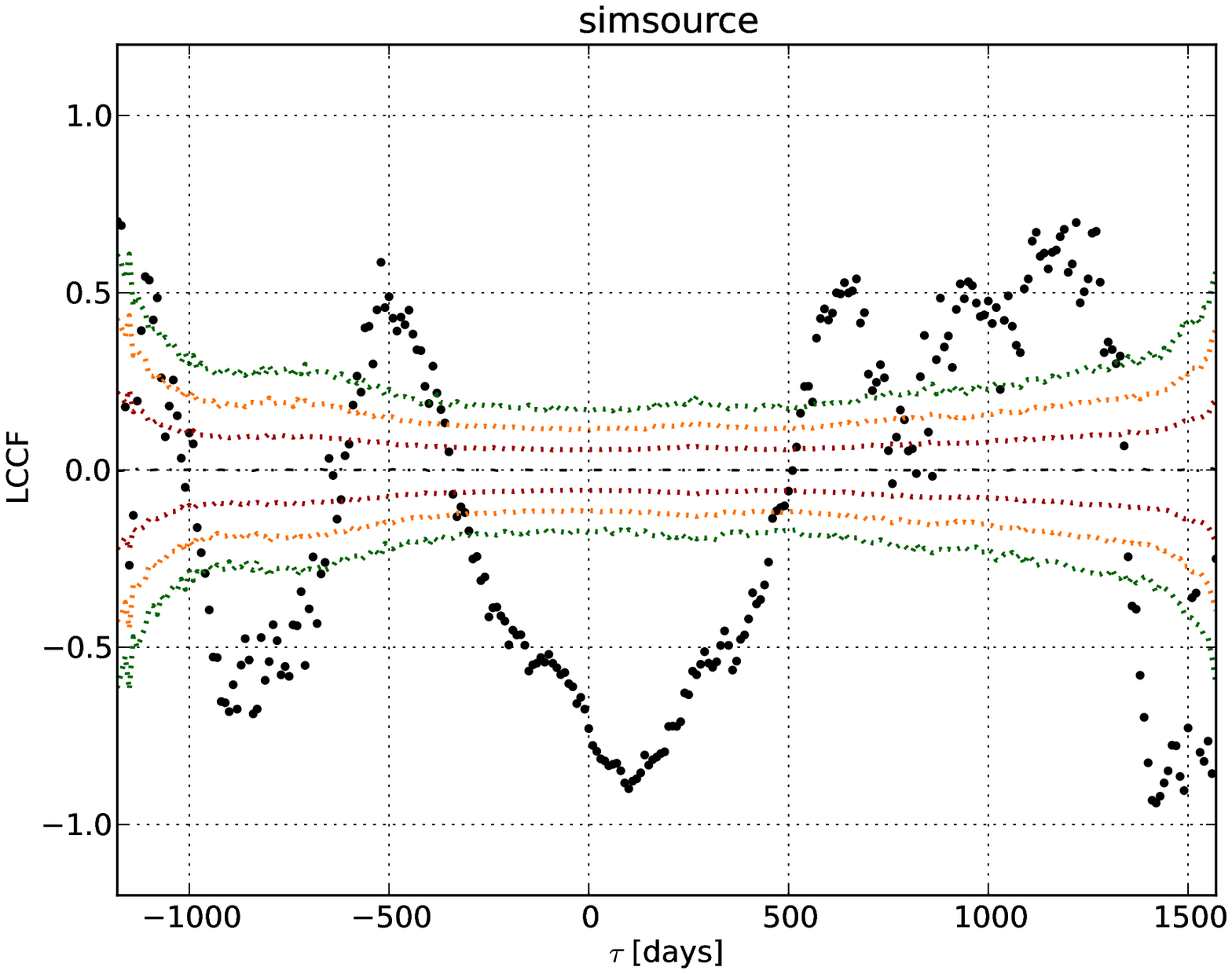}
\includegraphics[angle=0,width=9.0cm, trim=0 10 0 0]{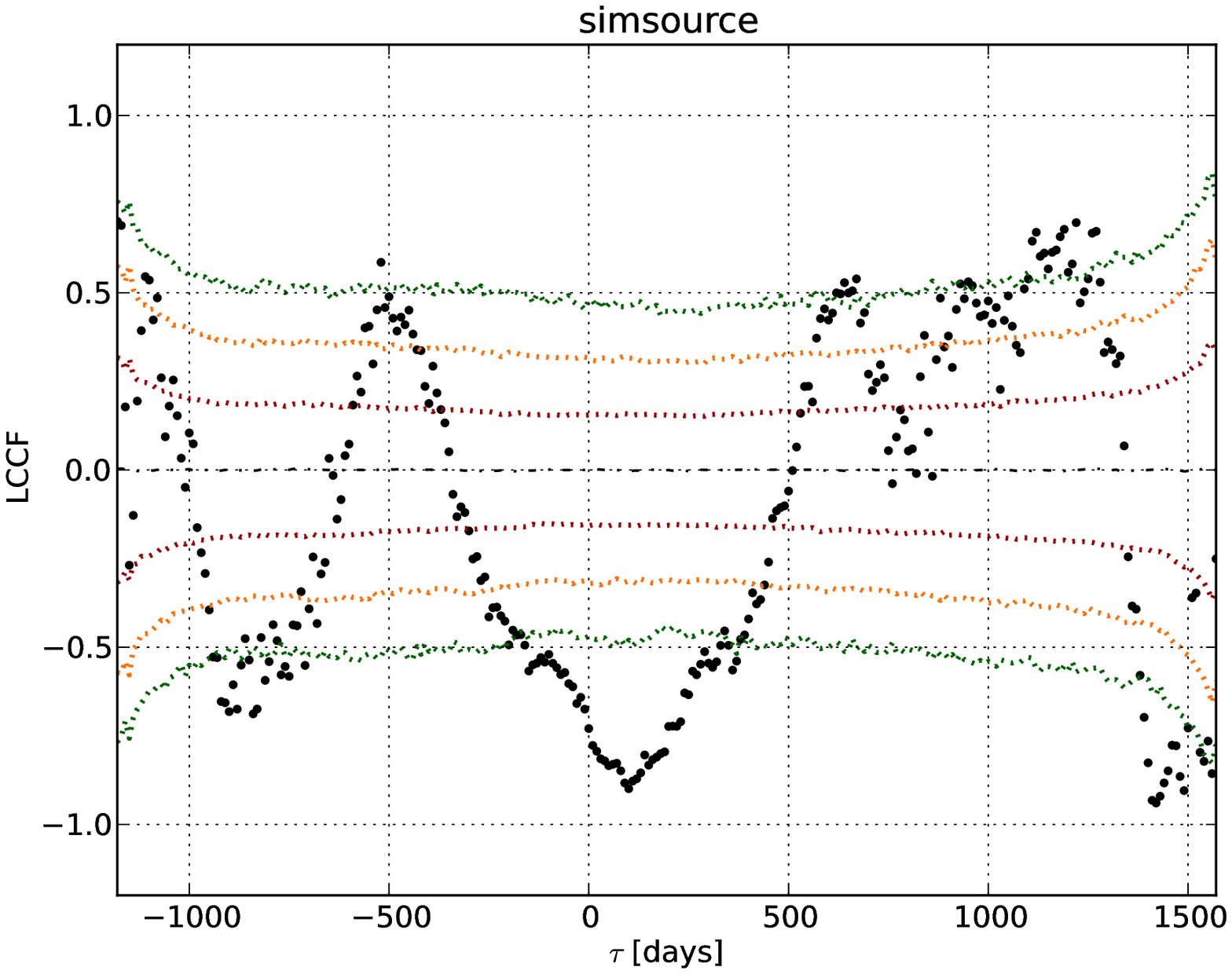}
\includegraphics[angle=0,width=9.0cm, trim=0 10 0 0]{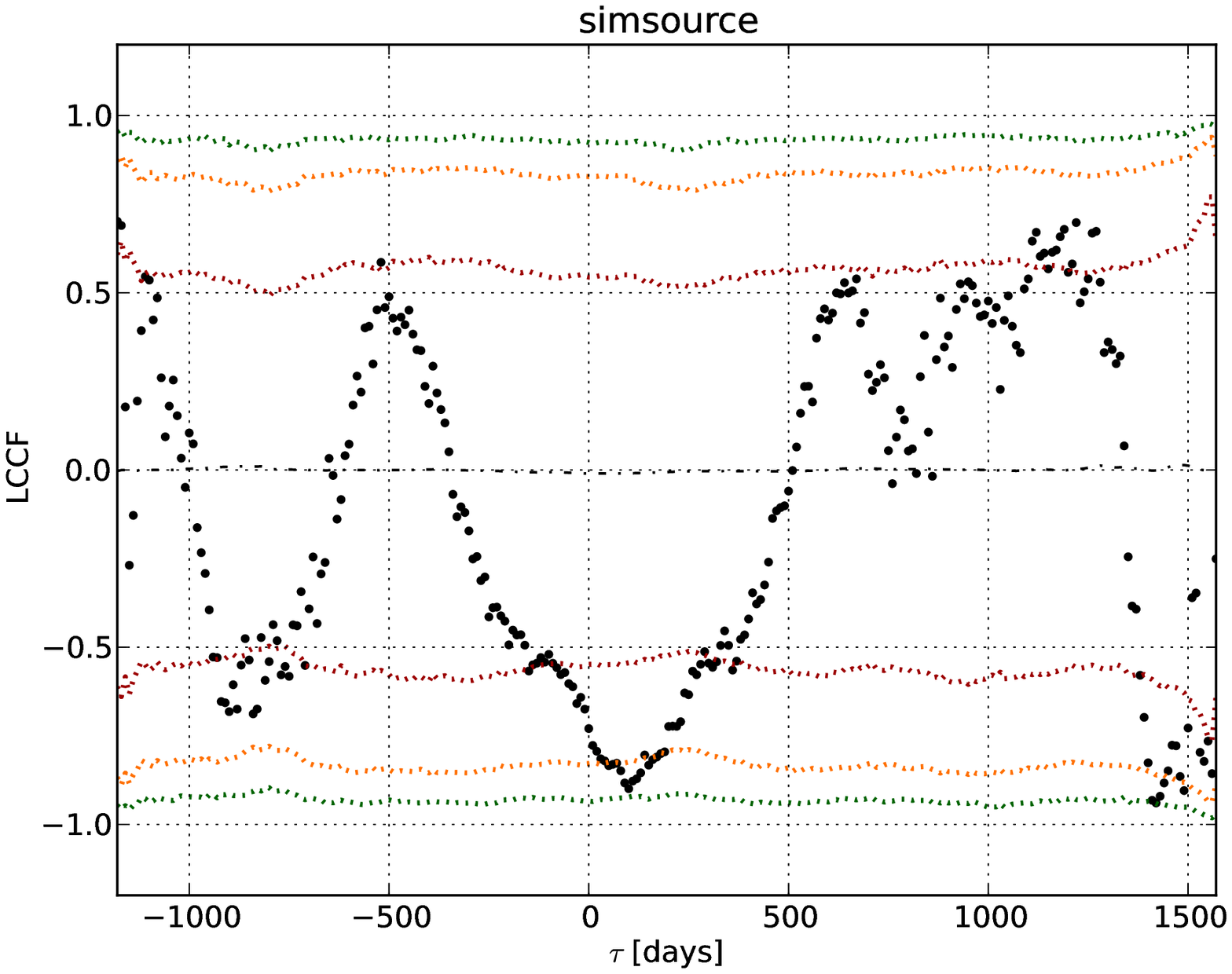}
\caption{Example of cross-correlation significance results for simulated data using the typical sampling from our monitoring program (same light curves as Figure \ref{example_xcorr_data}). We use $\beta = 0$ (upper panel), $\beta = 1$ (central panel) and $\beta = 2$ (lower panel). The black dots represent the LCCF for the data, while the color contours the distribution of random cross-correlations obtained by the Monte Carlo simulation with red for $1\sigma$, orange for $2\sigma$ and green for $3\sigma$. The increased amplitude of random cross-correlations is evident for steeper PSDs.}
\label{significance_various_models}
\end{center}
\end{figure}

As illustrated in Figure \ref{example_xcorr_simulated_light_curves}, the distribution of random cross-correlation coefficients will depend on the model used for the simulated light curves. In order to better appreciate that dependence, we have estimated the significance of the cross-correlation for an example using simulated data with the sampling pattern from our monitoring program (same as Figure \ref{example_xcorr_data}). We have used 10,000 simulated light curves with PSD $\propto 1/\nu^\beta$ for $\beta = 0$, $1$ and $2$. Figure \ref{significance_various_models} presents the results in the form introduced in Figure \ref{example_xcorr_data}. As in Figure \ref{example_xcorr_simulated_light_curves}, we observe an increase in the amplitude of the random cross-correlation when steeper power spectral densities are used in the simulations. This  manifests as increased scatter in the distribution of random cross-correlations and a lower significance estimate for the cross-correlations. The dependence of the results on the particular model of the light curves illustrates the importance of a proper characterization of the light curves variability, a subject we discussed in section \ref{psd_estimation_method}.

% Example cross-correlation significance	

\subsubsection{Error on the significance estimate and minimum number of simulations \label{sig_error_estimate}}

\begin{figure}
\begin{center}
\includegraphics[angle=0,width=9.0cm]{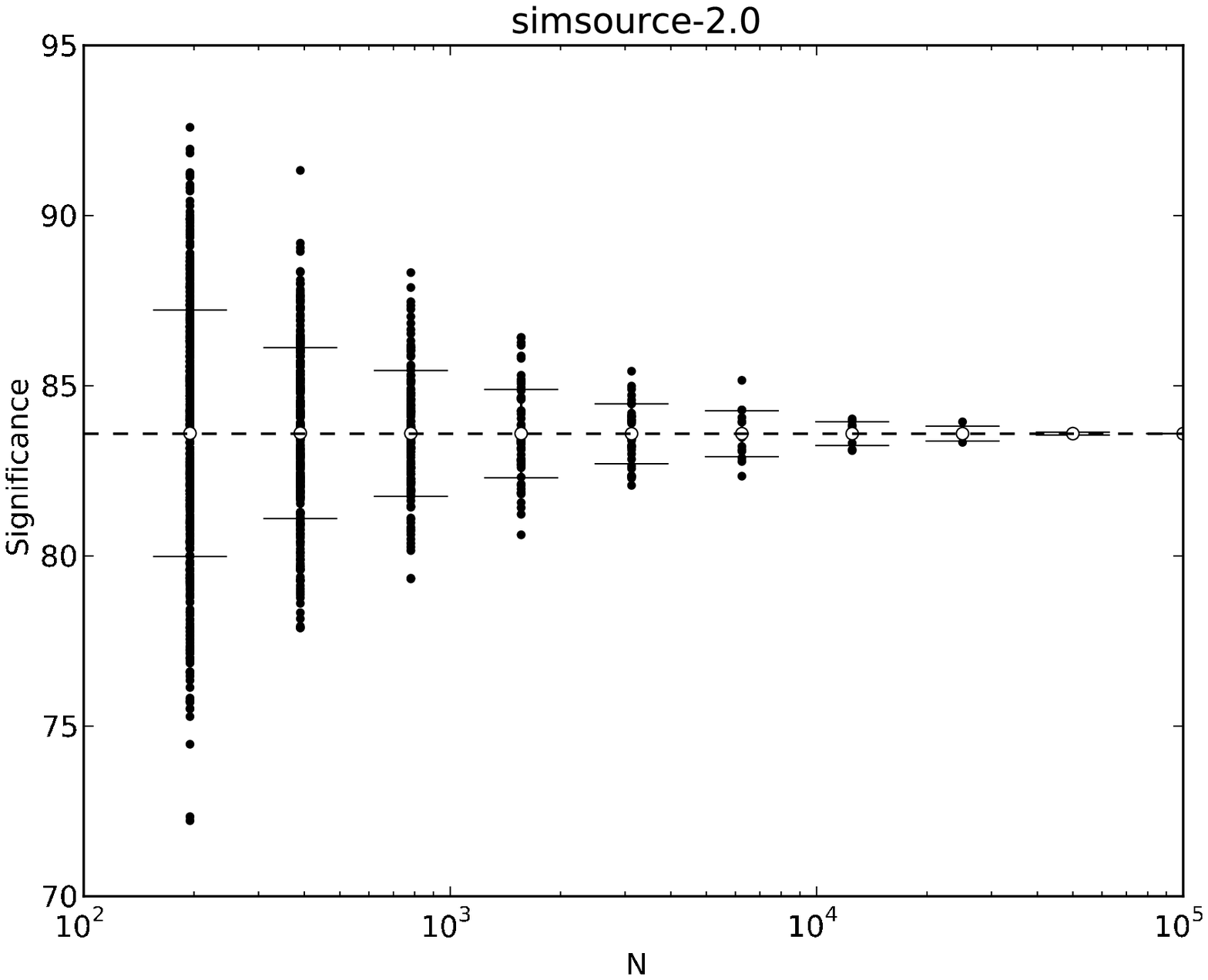}
\caption{Example of scatter in the significance estimate for independent subsets of the full simulation using different numbers of light curve pairs. The horizontal axis shows the number of simulations used to get each estimate and the vertical the significance. Black dots represent each of the independent subsets of the full simulation. The empty circles and error bars represent the mean and standard deviation for subsets of a given number of simulations. The horizontal segmented line corresponds to the results using the whole simulation. As expected, the scatter of the estimates obtained using smaller number of simulations is larger.}
\label{example_crosscorr_error}
\end{center}
\end{figure}

It is expected that the precision of the significance estimates will increase as the number of simulated light curve pairs increases. In order to get an estimate on the expected error in our significance estimate, due to the finite number of simulations, we have divided a full simulation with 100,000 simulated light curve pairs into independent subsets, and provide independent estimates for each of them. The idea is to observe the scatter when a small number of simulations is used and compare its variation as more simulations are included. The original simulation is divided in two halves which are subsequently divided into two. The process is repeated until the number of simulations in each subset is small enough that results have a very large scatter, and do not give us reliable significance estimates. For all sources we find that the results of a test with smaller number of simulations is less precise than the one using all the simulations. In all cases, the average gives the result of the complete simulation, an expected result since together they encode the same information. As expected, the scatter is much smaller when a large number of simulations is used. An example is presented in Figure \ref{example_crosscorr_error}, which clearly shows the reduction in the scatter as the number of simulated light curve pairs is increased. With less than 1,000 simulations the scatter is of a few percentage points, and gets to about 0.4\% for more than 10,000 simulations.

% Scatter as a function of number of simulations

\begin{figure}
\begin{center}
\includegraphics[angle=0,width=9.0cm]{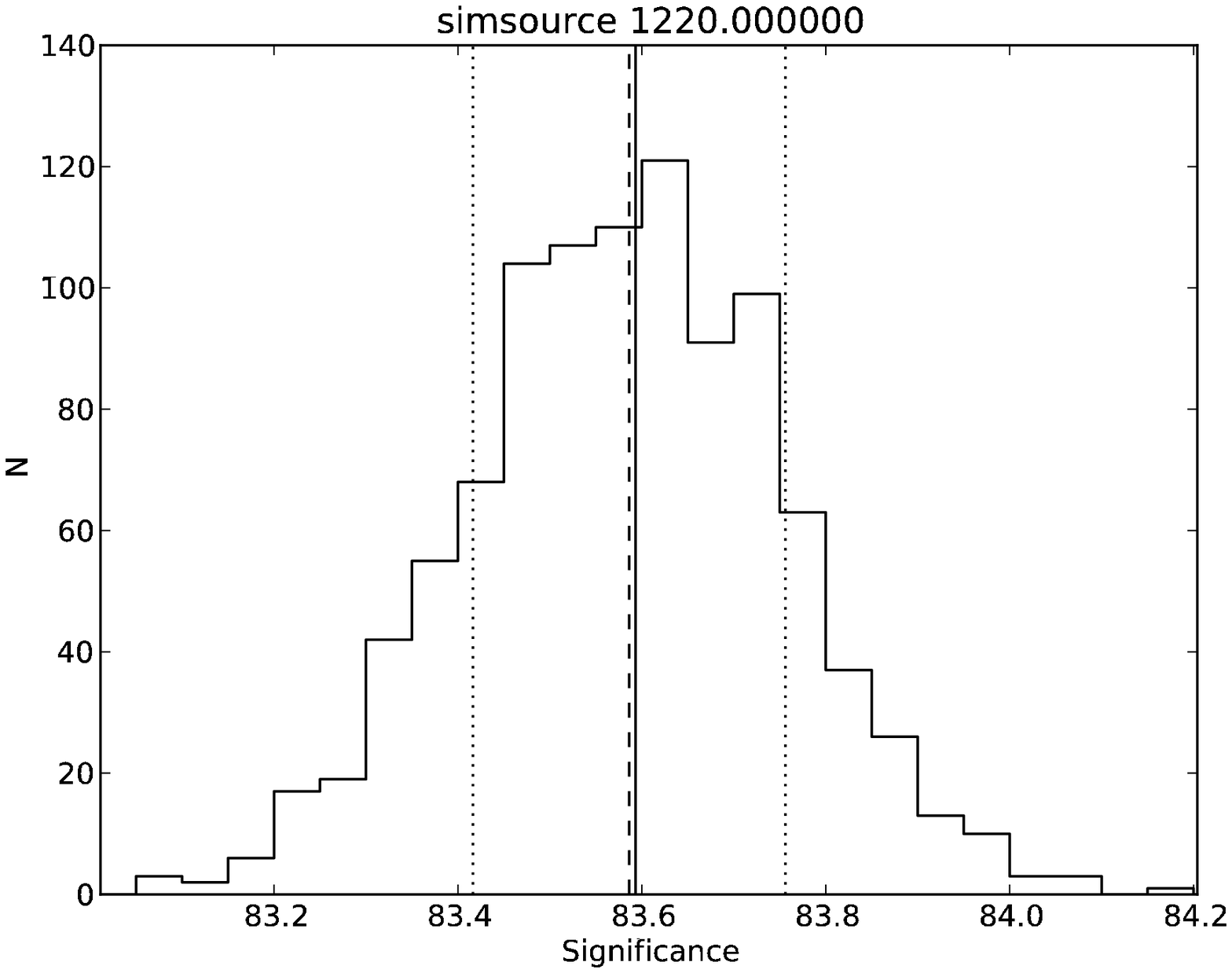}
\caption{The distribution of the significance estimates for the bootstrap samples is represented as a histogram. The solid line represents the value obtained using the whole simulation. The segmented line is the mean of the distribution, with the dotted lines the one standard deviation upper and lower limits.}
\label{example_crosscorr_error_bootstrap}
\end{center}
\end{figure}

The process described above could in principle be used to obtain an error estimate, but instead we compute a more conventional bootstrap estimate of the standard error, following the procedure described below \citep[this is applied in][]{max-moerbeck+2014}. For the time lag of interests, we have $N$ values of the random cross-correlations obtained from the $N$ simulated light curves. From these $N$ random cross-correlations 1,000 bootstrap samples are obtained, each one giving a different significance estimate. The sample standard deviation of these bootstrap replications is used as the error in the significance estimate. An example of the distribution of bootstrapped estimates is shown in Figure \ref{example_crosscorr_error_bootstrap}. We think this error estimate is a required step of any Monte Carlo estimate of the significance, and we recommend the adoption of this or equivalent procedures - an issue that has surprisingly been up to now ignored by all authors.

% Scatter as a function of number of simulations

\section{Summary}
\label{conclusions}

We presented a description of a Monte Carlo method to estimate the significance of cross-correlations between two unevenly sampled time series. We demonstrated the dependence of the significance estimates on the model of the light curves, and presented a method based on \cite{uttley+2002}, that allow us to determine the best fit for a simple power-law power spectral density model for a light curve. An improved way of dealing with the effects of red-noise leakage is implemented. This method uses interpolation and windowing with a Hanning window, and provides the ability to fit steep PSDs like those found in our data sets. We demonstrated that windowing is essential to obtain an upper limit on the value of the PSD power-law index. An upper limit is required for meaningful cross-correlation significance estimates, which depend on the model used for the light curves. The method used for error estimation of the best fit was modified for one which decouples the goodness of fit estimate from the estimation of confidence intervals, and that can indicate the presence of biases in the fitting procedure. The method was evaluated using simulated data sets and found to be accurate with a typical error in $\beta$ of less than $\pm 0.3$, for cases in which the signal power is large compared to observational noise. The performance of the method is degraded when fitting time series in which the signal power is comparable to the observational noise. In these cases, the procedure fails to provide a reliable constraint on the shape of the PSD, a situation we can consider when analysing our data set by using the Neyman construction to obtain confidence intervals. We also checked the repeatability of the best fit value when running the procedure multiple times, and find that it improves when using a large number of simulated light curves ($M$). For an example using the OVRO data set, we find that big improvements are expected when going from $M$=100 to $M$=1,000, but any further increase provides a small improvement, and might not be worth the increased computational time.

Finally, we described the problem of estimating the cross-correlation for unevenly sampled time series. We have shown that high values of the cross-correlation coefficients for red-noise time series are ubiquitous, and that any method that aims at quantifying the significance of correlation coefficients for light curves having flare-like features needs to take this into account. We have described a general Monte Carlo method to estimate the significance of cross-correlation coefficients between two wavebands. A number of tests aimed at measuring the effectiveness of a particular cross-correlation method have been performed to compare the LCCF and the DCF. Given the absence of a physical model for the expected correlations, the method cannot be used to give a definitive value of the detection efficiency, but it can be used to compare different alternatives. The main result is that the LCCF has a much larger detection efficiency than the DCF when trying to recover a linear correlation. The DCF has the additional problem of producing a large fraction of spurious high significance time correlations, which could be mistaken as real correlations. This problem is less important for the LCCF especially when long time series are used.

The origin of the difference, and the lack of discriminating power for the DCF, seems to originate in the short duration or non-stationarity of the time series involved. In conclusion, we recommend the use of the LCCF as a tool to search for correlations.

We also show that the significance of the cross-correlation coefficients is strongly dependent on the power-law slope of the PSD,  which makes characterization of the light curves critical. We investigate the error on the estimated significance by repeating the analysis using different numbers of simulations. Especially in cases where high significances are claimed, we suggest using a bootstrap estimate of the error on the significance and reporting its value as part of the analysis results. The results of the application of this method to a data set combining data from the OVRO monitoring program and \emph{Fermi} Large Area Telescope are presented in \citet[][]{max-moerbeck+2014}.

%--------------------------------------------------------------------------------------------------------------------------------
% Acknowledgements
%--------------------------------------------------------------------------------------------------------------------------------
\section*{Acknowledgments}
The OVRO program is supported in part by NASA grants NNX08AW31G and NNX11A043G and NSF grants AST-0808050 and AST-1109911. Support from MPIfR for upgrading the OVRO 40-m telescope receiver is acknowledged. W.M. thanks Jeffrey Scargle, James Chiang, Iossif Papadakis and Glenn Jones for discussions. The National Radio Astronomy Observatory is a facility of the National Science Foundation operated under cooperative agreement by Associated Universities Inc. TH was supported in part by the Jenny and Antti Wihuri foundation and by the Academy of Finland project number 267324. We thank the anonymous referee for constructive comments that greatly improved the presentation of some sections of this paper.

%--------------------------------------------------------------------------------------------------------------------------------
% Bibliography
%--------------------------------------------------------------------------------------------------------------------------------

\bsp % ``This paper has been produced using the ...''
\label{lastpage}

\end{document}